\documentclass[twocolumn,twocolappendix]{aastex631}

\submitjournal{PSJ}

\usepackage{amssymb}

\usepackage{rotating}
 \usepackage{mathtools}
 
\newcommand*\patchAmsMathEnvironmentForLineno[1]{%
  \expandafter\let\csname old#1\expandafter\endcsname\csname #1\endcsname
  \expandafter\let\csname oldend#1\expandafter\endcsname\csname end#1\endcsname
  \renewenvironment{#1}%
     {\linenomath\csname old#1\endcsname}%
     {\csname oldend#1\endcsname\endlinenomath}}% 
\newcommand*\patchBothAmsMathEnvironmentsForLineno[1]{%
  \patchAmsMathEnvironmentForLineno{#1}%
  \patchAmsMathEnvironmentForLineno{#1*}}%
\AtBeginDocument{%
\patchBothAmsMathEnvironmentsForLineno{equation}%
\patchBothAmsMathEnvironmentsForLineno{align}%
\patchBothAmsMathEnvironmentsForLineno{flalign}%
\patchBothAmsMathEnvironmentsForLineno{alignat}%
\patchBothAmsMathEnvironmentsForLineno{gather}%
\patchBothAmsMathEnvironmentsForLineno{multline}%
}

\def\CO2{CO$_2$}
\def\H20{H$_2$O}
\def\H2{H$_2$}

\def\co2{CO$_2$}
\def\h2o{H$_2$O}
\def\h2{H$_2$}

\def\r{$\mathcal{R}$}

 \shorttitle{Volatile loss depends on surface conditions}
\shortauthors{Lock \& Stewart}

\begin{document}
% \sloppy

\title{Atmospheric loss in giant impacts depends on pre-impact surface conditions}
\author[0000-0001-5365-9616]{Simon J. Lock}
\affiliation{School of Earth Sciences, University of Bristol, Bristol, BS8 1RJ, UK}
\email{s.lock@bristol.ac.uk}
\correspondingauthor{Simon J. Lock}

\author[0000-0001-9606-1593]{Sarah T. Stewart}
\affiliation{Department of Earth and Planetary Sciences, U. California Davis, Davis, CA 95616, USA}

%\date{}							% Activate to display a given date or no date

%xxxxxxxxxxxxxxxxxxxxxxxxxxxxxxxxxxxxxxxxxxxxxxxxxxxxxxxxxxxxxxxxxxxxxxxxxxxxxxxxxxxxxxxxxxxxxxxxxxxxxxxxxxxxxxxxxxxxxxxxxxxxxxxxxxxxxxxx
%xxxxxxxxxxxxxxxxxxxxxxxxxxxxxxxxxxxxxxxxxxxxxxxxxxxxxxxxxxxxxxxxxxxxxxxxxxxxxxxxxxxxxxxxxxxxxxxxxxxxxxxxxxxxxxxxxxxxxxxxxxxxxxxxxxxxxxxx
%xxxxxxxxxxxxxxxxxxxxxxxxxxxxxxxxxxxxxxxxxxxxxxxxxxxxxxxxxxxxxxxxxxxxxxxxxxxxxxxxxxxxxxxxxxxxxxxxxxxxxxxxxxxxxxxxxxxxxxxxxxxxxxxxxxxxxxxx
\begin{abstract}

Earth likely acquired much of its inventory of volatile elements during the main stage of its formation. Some of Earth's proto-atmosphere must therefore have survived the giant impacts, collisions between planet-sized bodies, that dominate the latter phases of accretion. Here we use a suite of 1D hydrodynamic simulations and impedance match calculations to quantify the effect that pre-impact surface conditions (such as atmospheric pressure and presence of an ocean) have on the efficiency of atmospheric and ocean loss from proto-planets during giant impacts. We find that -- in the absence of an ocean -- lighter, hotter, and lower-pressure atmospheres are more easily lost. The presence of an ocean can significantly increase the efficiency of atmospheric loss compared to the no-ocean case, with a rapid transition between low and high loss regimes as the mass ratio of atmosphere to ocean decreases. However, contrary to previous thinking, the presence of an ocean can also reduce atmospheric loss if the ocean is not sufficiently massive, typically less than a few times the atmospheric mass. Volatile loss due to giant impacts is thus highly sensitive to the surface conditions on the colliding bodies. To allow our results to be combined with 3D impact simulations, we have developed scaling laws that relate loss to the ground velocity and surface conditions. Our results demonstrate that the final volatile budgets of planets are critically dependent on the exact timing and sequence of impacts experienced by their precursor planetary embryos, making atmospheric properties a highly stochastic outcome of accretion.

\end{abstract}

% \keywords{Giant impacts}

%xxxxxxxxxxxxxxxxxxxxxxxxxxxxxxxxxxxxxxxxxxxxxxxxxxxxxxxxxxxxxxxxxxxxxxxxxxxxxxxxxxxxxxxxxxxxxxxxxxxxxxxxxxxxxxxxxxxxxxxxxxxxxxxxxxxxxxxx
%xxxxxxxxxxxxxxxxxxxxxxxxxxxxxxxxxxxxxxxxxxxxxxxxxxxxxxxxxxxxxxxxxxxxxxxxxxxxxxxxxxxxxxxxxxxxxxxxxxxxxxxxxxxxxxxxxxxxxxxxxxxxxxxxxxxxxxxx
%xxxxxxxxxxxxxxxxxxxxxxxxxxxxxxxxxxxxxxxxxxxxxxxxxxxxxxxxxxxxxxxxxxxxxxxxxxxxxxxxxxxxxxxxxxxxxxxxxxxxxxxxxxxxxxxxxxxxxxxxxxxxxxxxxxxxxxxx
\section{Introduction}

\begin{figure*}[t]
    \centering
    \includegraphics[height=0.96\textheight]{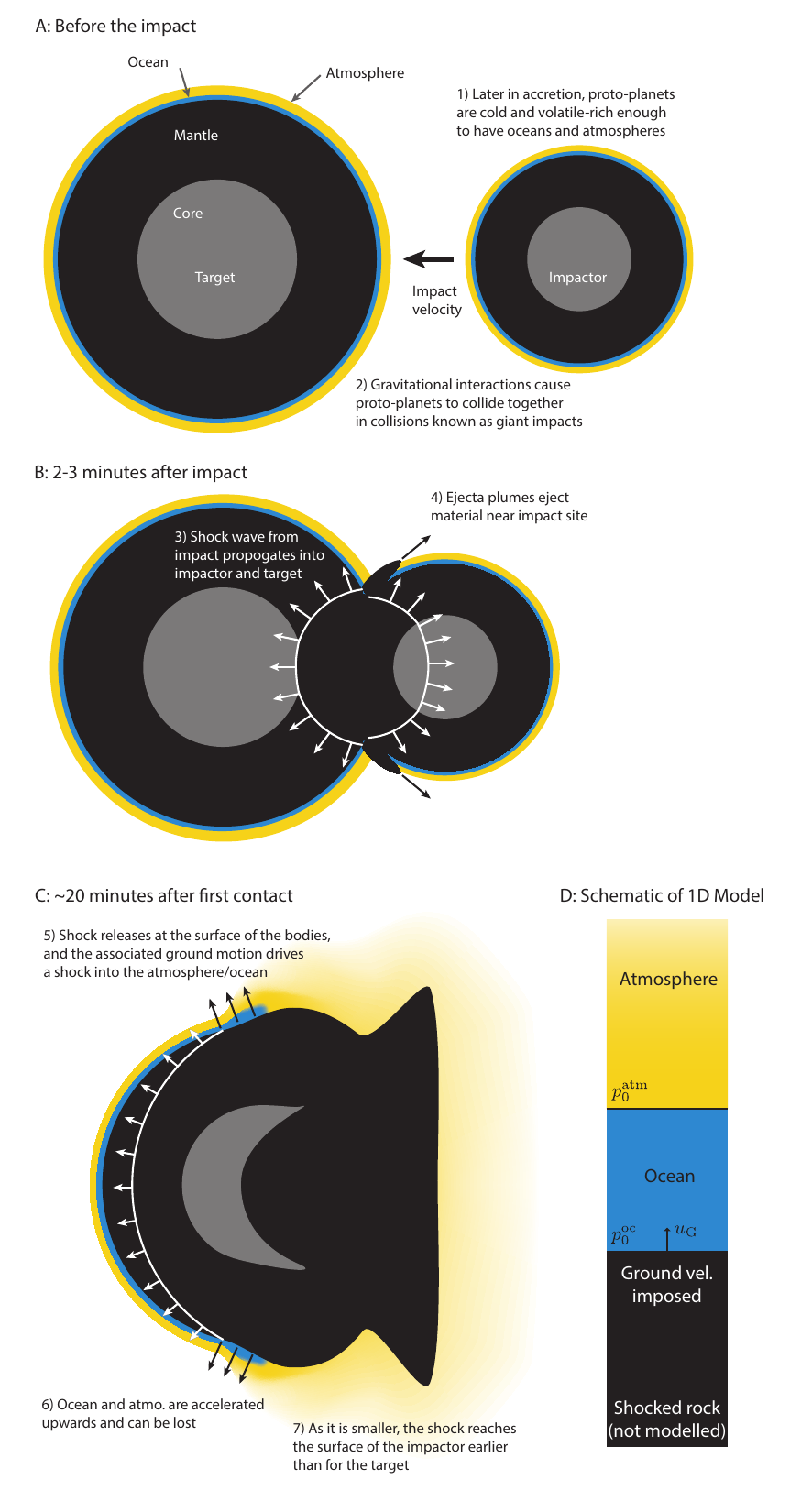}
        \caption{Caption on next page.}
\end{figure*}
\addtocounter{figure}{-1}
\begin{figure*}[t]
    \caption{Atmospheric loss from giant impacts occurs through ejection in impact plumes or from ground motion further from the impact site. A: A schematic of a giant impact a few minutes before first contact with different material layers indicated by different colors. B: The same collision a few minutes after initial contact. Melted and partially vaporized plumes are extruded from the impact site, carrying away a fraction of the atmosphere and ocean near the contact site. A shock wave (white line) propogates away from the impact site into the rest of the impactor and target. C: A schematic of a giant impact approximately 20 minutes after first contact. The shock wave travels through the planet and breaks out at the surface. The resulting acceleration of the ground drives a shock into the ocean and atmosphere, driving loss. D: A schematic of the 1D simulations performed for this study which is similar to that used in previous work \citep{Chen1997,Genda2003,Genda2005}. A hydrostatic ocean and atmosphere are initialized at a radius equal to the planetary radius in a spherical geometry. The mantle and core of the planet are not modelled directly and the breakout of the shock from the planet is mimicked by giving the lower boundary of the domain an initial velocity, $u_{\rm G}$.}

    \label{fig:Overview_cartoon}
    
\end{figure*}

How Earth acquired its unique atmosphere and ocean is a fundamental, unanswered question. Earth is thought to have gained a large fraction of its current budget of highly volatile elements (e.g., N, C, H, noble gases) during the main stages of accretion \citep[e.g.,][]{Halliday2013,Mukhopadhyay2019}. Accretion is a violent, stochastic process, and there are many mechanisms by which planets and their building blocks can gain and lose volatiles \citep[e.g.,][]{OBrien2014,Marty2016,Olson2019,Raymond2017a,Schlichting2015,Schlichting2018,Young2019,Odert2018_Mars_sized}. Determining how each potential mechanism works is vital for understanding the origin of Earth's volatile budget and that of other planets.

Giant impacts, collisions between planet-sized bodies, likely play a significant role in the chemical evolution of terrestrial planets \citep{Genda2005,Genda2003,Kegerreis2020a,Kegerreis2020,Denman2020,Denman2022_atmoloss_sE,Carter2018}. Most terrestrial planets experience several giant impacts during their formation \citep[e.g.,][]{Raymond2007,Quintana2016}. Such collisions are incredibly dramatic events with large fractions of the mantles of the colliding bodies being melted or vaporized, variable amounts of crust, mantle, and core being ejected, and the post-impact body often left rapidly rotating \citep{Canup2004anrev,Lock2017,Nakajima2015,Carter2020,Carter2018,Rufu2017,Lock2020}. Giant impacts have a particular significance for Earth as it is thought that the last giant impact \citep[or potentially the last few impacts:][]{Rufu2017,Asphaug2021} onto the proto-Earth injected material into orbit out of which our Moon formed \citep{Cameron1976,Hartmann1975}. The exact scenario for the so-called Moon-forming giant impact and the mechanisms by which the Moon formed in the aftermath are highly debated \citep{Canup2001,Reufer2012,Canup2012,Cuk2012,Lock2018moon,Rufu2017}, but the event marks the end of the main stage of Earth's accretion. 

In giant impacts, only a proportion of the volatiles on the colliding bodies is inherited by the final post-impact body, with the fraction of volatiles retained varying substantially between impacts. Volatiles are carried away dissolved or trapped in the ejected silicate and metal mass, and the atmospheres and oceans of the colliding bodies can also be directly ejected from the system. The latter process will be the focus of this paper.

There are two principal mechanisms by which atmosphere and ocean are ejected during giant impacts (Figure~\ref{fig:Overview_cartoon}). First, close to the initial contact point between the colliding bodies, the crust and upper mantle of both bodies is ejected as melted and vaporized plumes \citep[Figure~\ref{fig:Overview_cartoon}B, e.g.,][]{Carter2018}. Some of this material remains bound to the system, but a large fraction is typically ejected. What happens to any atmosphere or ocean close to the contact point during this process has not been studied in detail. At such high temperatures it is likely that the volatiles are fully soluble in the silicate \citep{Lock2018moon,Fegley2023_BSE_cond} and, if there is efficient mechanical mixing and chemical equilibration between the volatiles and silicate, the volatiles could share the same fate as the crust and upper mantle. Alternatively, the ocean and atmosphere may constitute a separate part of the escaping plumes and be lost at an efficiency dictated by the thermodynamics of shock and release of water and gas mixtures \citep[e.g.,][]{Kegerreis2018,Kegerreis2020}, which could be somewhat different from that of the silicates. The reality is likely somewhere between these two extremes but, in either case, most accretionary giant impacts would drive loss of atmosphere and ocean from near the contact point \citep[e.g.,][]{Carter2018,Kegerreis2018,Kegerreis2020}.

Away from the contact point, ocean and atmosphere can be lost through breakout of the impact shock wave from the surface of the planet \citep[Figure~\ref{fig:Overview_cartoon}C,][]{Chen1997,Genda2003,Genda2005,Schlichting2015}. The strong shock wave generated by the impact travels through each of the colliding bodies until it reaches the surface where the shock wave is transmitted to the atmosphere or ocean. The transmission of the shock wave from the planet to the atmosphere or ocean is known as the breakout of the shock wave, and leads to acceleration of the planets surface to velocities above that of the particle velocity of the shock within the planet (see Section~\ref{sec:bckgrnd}). The acceleration of the planet's surface upon breakout means that transmission of the shock into the atmosphere/ocean is sometimes described as the atmosphere/ocean being `kicked' by the silicate surface. The shock accelerates up the strong hydrostatic density gradient of the atmosphere and some fraction of the top of the atmosphere reaches escape velocity and is lost from the system (see Section~\ref{sec:bckgrnd} for a more extensive description of this process). The efficiency of loss due to the breakout of the shock wave from the impact has been quantified using 1D hydrodynamic simulations \citep{Chen1997,Genda2003,Genda2005} and semi-analytical calculations \citep{Schlichting2015} of the shock driven in the atmosphere for a given groud motion. 1D calculations have the advantage of being mathematically tractable or numerically inexpensive but, in order to determine the total loss from a given impact, knowledge of the ground velocity around the planet is required. \cite{Genda2003} used a single value for the average ground velocity from numerical simulations of the canonical Moon-forming giant impact \citep[a grazing collision by a Mars-mass impactor at near the escape velocity onto the proto-Earth,][]{Canup2001} and concluded that only about 20\% of the atmosphere would be lost from a planet with no ocean. \cite{Schlichting2015} used a simple 2D shock propagation model to calculate the ground velocity distribution across the surface and showed that, due to the highly non-linear relation between ground velocity and loss, the efficiency of loss was strongly sensitive to the distribution of ground motion and that using an average value for ground velocity underestimates the total loss by a factor of two. \cite{Yalinewich2018} took such calculations to their logical conclusion by using 3D hydrodynamic giant-impact simulations to determine the ground velocity distribution and hence the fraction of atmosphere that would be lost due to ground motion as a function of the impact velocity and the ratio of the sizes of the two colliding bodies. 

It is possible to simultaneously capture both near and far-field loss by explicitly including atmospheres in 3D numerical simulations of giant impacts \citep{Kegerreis2020, Kegerreis2020a,Denman2020,Denman2022_atmoloss_sE}. However, accurately resolving the thin atmospheres expected on many terrestrial planets, such as Earth, during the giant-impact phase requires extremely high resolution simulations. Advances in the scalability of hydrodynamic codes and the expansion in high performance computing resources have recently allowed direct simulation of atmospheres with surface pressures as low as 3.2~kbar \citep{Kegerreis2020, Kegerreis2020a}. Kegerreis and coworkers \citep{Kegerreis2020, Kegerreis2020a} used their simulations to develop a scaling law that relates the loss due to a given impact to the parameters of the impact (impact velocity, impactor mass, etc.). Encouragingly, \cite{Kegerreis2020} found broad agreement between their results and those calculated by convolving 1D models of atmospheric loss with the ground velocity distributions from their impact simulations \citep[as in][]{Yalinewich2018}. The efficiency of atmospheric loss from giant impacts varies widely, but most impacts only lead to the loss of a few tens of percent of the atmosphere and near-total loss is only achieved in high-velocity, near head-on impacts. 

So far, the studies we have discussed considered atmospheric loss from planets that do not have oceans. However, during the giant impact phase of accretion, the time between impacts is long enough that the atmosphere of proto-planets would cool sufficiently between impacts for condensation of a surface ocean \citep{Abe1988}. \cite{Genda2005} explored the effect of a surface ocean on atmospheric loss and concluded that the presence of an ocean can significantly increase the efficiency of loss (a full explanation of this phenomena is given in Section~\ref{sec:results:R}). So far, oceans have not been included in 3D simulations and so the quantitative effect of the presence of an ocean on total loss is not known. However, the results of \cite{Genda2005} suggest that the thermal state of a planet's surface could make the difference between a proto-planet losing or retaining its atmosphere during a giant impact. 
 
In this paper, we explore the effect that the surface conditions (e.g., atmospheric pressure, temperature, and composition, and the depth of an ocean) on the colliding bodies have on the efficiency of atmospheric and ocean loss from planets with small to modest atmospheric mass fractions. Previous studies have considered only a limited range of surface conditions and it is not well known how parameters such as planetary mass, ocean depth, surface pressure, surface temperature and atmospheric composition affect the efficiency of loss. Full 3D impact simulations are limited by their resolution to calculating atmospheric loss from bodies with thick atmospheres, on the order of several kbar for an Earth-mass body \citep{Kegerreis2020a}. Similarly, the highest resolution simulations currently would only be able to resolve oceans deeper than $\sim 40$~km. For the formation of planets, at least in our own solar system, it is important to understand the loss of much thinner atmospheres and shallower oceans. For example, it is typically thought that ancient Earth and Venus had atmospheres of a few hundred bar \citep{Kasting1988_runaway_gh,Marty2012,Halliday2013,Sossi2020}. To resolve such thin atmospheres, we take a hybrid approach, using 1D hydrodynamic simulations of loss due to a given ground motion to relate the surface properties to the efficiency of loss. These results can then be convolved with ground velocity distributions calculated from 3D giant-impact simulations to quantify the efficiency of loss from any given impact. In this paper we describe our 1D simulations which we will combine with 3D giant impact simulations in future work.

We begin by providing an overview of the processes controlling the breakout of the shock wave from the planet and impedance match calculations (Section~\ref{sec:bckgrnd}). We will then describe our methods (Section~\ref{sec:methods}) and report our results for the relationship between surface conditions, ground motion, and loss without (Section~\ref{sec:results:NO}) and with an ocean (Section~\ref{sec:results:R}). Section~\ref{sec:results:fit} outlines a parameterization that describes the efficiency of atmospheric and ocean loss as a function of ground velocity, planetary mass and the ratio of atmospheric to ocean mass. In Sections~\ref{sec:results:up_ug_relation_NO} and \ref{sec:discussion:issues} we explore the relationship between the strength of the shock in the planet and the ground motion and bound the effect of more complicated ground motions on the efficiency of loss. In Section~\ref{sec:discussion} we discuss the implications of our results and conclude in Section~\ref{sec:conclusions}. A full description of the numerical methods is contained in the appendix.

%xxxxxxxxxxxxxxxxxxxxxxxxxxxxxxxxxxxxxxxxxxxxxxxxxxxxxxxxxxxxxxxxxxxxxxxxxxxxxxxxxxxxxxxxxxxxxxxxxxxxxxxxxxxxxxxxxxxxxxxxxxxxxxxxxxxxxxxx
%xxxxxxxxxxxxxxxxxxxxxxxxxxxxxxxxxxxxxxxxxxxxxxxxxxxxxxxxxxxxxxxxxxxxxxxxxxxxxxxxxxxxxxxxxxxxxxxxxxxxxxxxxxxxxxxxxxxxxxxxxxxxxxxxxxxxxxxx
%xxxxxxxxxxxxxxxxxxxxxxxxxxxxxxxxxxxxxxxxxxxxxxxxxxxxxxxxxxxxxxxxxxxxxxxxxxxxxxxxxxxxxxxxxxxxxxxxxxxxxxxxxxxxxxxxxxxxxxxxxxxxxxxxxxxxxxxx
\section{The physics of loss due to breakout of the impact shock wave}
\label{sec:bckgrnd}

\begin{figure*}[t]
    \centering
    \includegraphics[height=0.96\textheight]{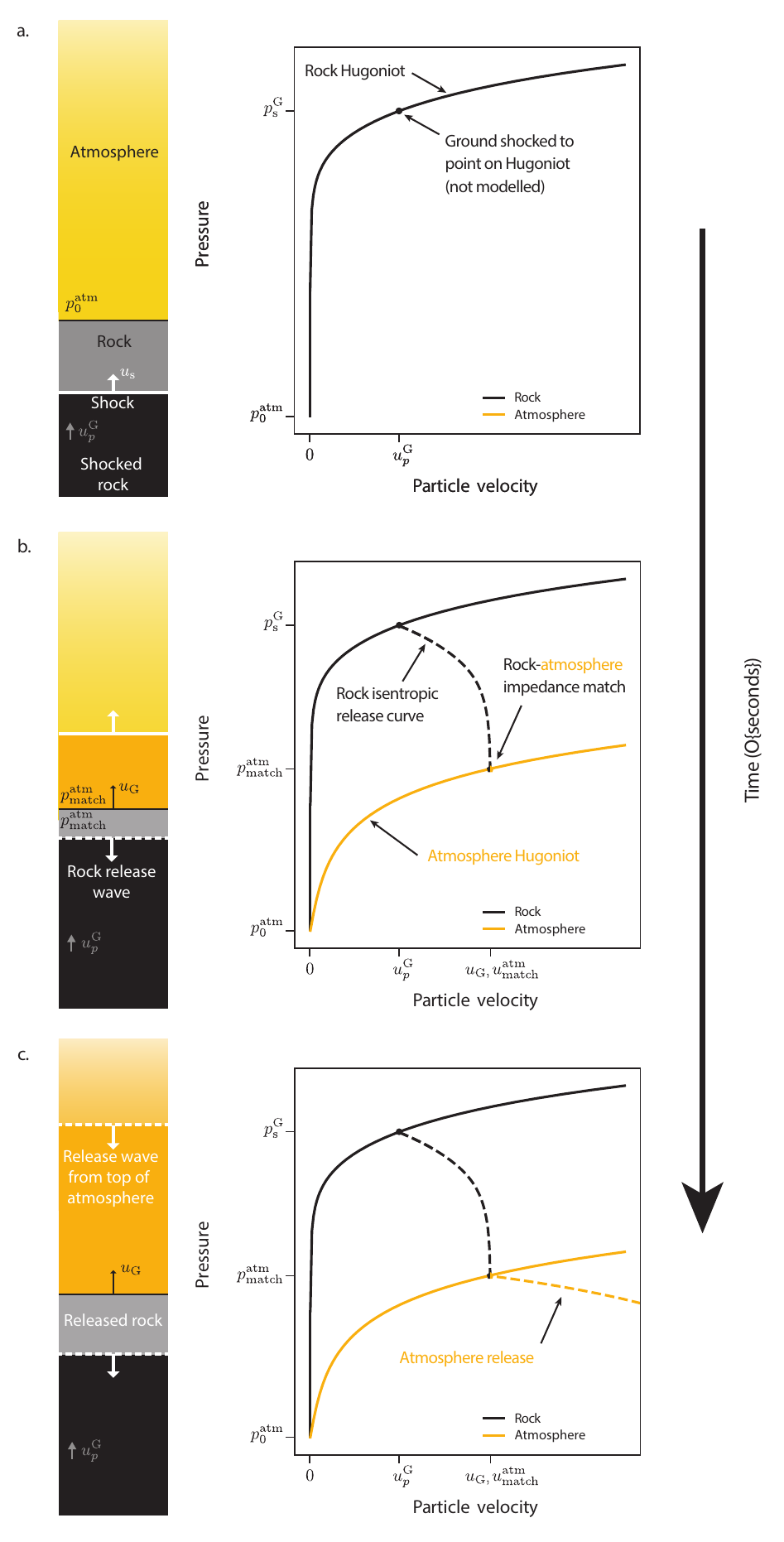}
    \caption{Caption on next page.}
\end{figure*}
\addtocounter{figure}{-1}
\begin{figure*}[t]
    \caption{The efficiency of loss is strongly influenced by the relative impedance of the atmosphere, ocean, and ground. Shown is a schematic that shows the relative position (left column) and the thermodynamic state (right) of the different material layers at different stages (rows) in the passage of the shock from the planet into the atmosphere. Left: Colors indicate different materials with rock in black, and atmosphere in orange. In the left column, darker shades of these colors indicate material under compression in the shock. Boundaries between materials are shown as thin black lines with their velocities shown as black arrows. The shock wave is indicated by a thick white line. Release waves are shown as white dashed lines and their velocities as white arrows. Where relevant, key dynamic and thermodynamic variables are noted. Right: Schematic particle velocity - pressure plots for the impedance match calculation between rock and atmosphere. Key pressures and particle velocities corresponding to different stages of the thermodynamic evolution of material are given on the axis, as labelled in the left column. Solid lines are shock Hugoniots, the locus of points that a shocked material can reach from an initial starting position. Hugoniots are not thermodynamic paths and the point reached material on each Hugoniot is indicated by a filled symbol. Dashed lines are release curves followed by material decompressing from a shocked states. Release curves are thermodynamic paths and the material moves along these lines. As in the left column, colors indicate different materials. A similar schematic for a planet with an ocean is shown in Figure~\ref{fig:impedance_match_cartoon}}
    \label{sup:fig:impedance_match_cartoon_NO}

\end{figure*}

In this section, we give an overview of the physical processes that occur when the impact shock wave reaches the surface of the planet (Figure~\ref{fig:Overview_cartoon}C) and how this results in loss of atmosphere/ocean. Here we will consider only the period immediately upon release of the shock and return to discuss the processes that complicate this simple picture later in evolution in Sections~\ref{sec:results} and \ref{sec:discussion}.

Figure~\ref{sup:fig:impedance_match_cartoon_NO} illustrates the stages in the breakout of an impact shock wave from the surface of a planet with an atmosphere but no ocean. The left hand column shows a schematic of the physical location and velocity of the material at different stages, and the right hand column shows the dynamics and thermodynamic states of material in pressure - particle velocity space. At the time illustrated in Figure~\ref{sup:fig:impedance_match_cartoon_NO}A the shock in the planet is approaching the surface. The shock accelerates and compresses the rock to a point along its Hugoniot, the locus of thermodynamic states and velocities that can be reached by shock compression (black solid line in the right column of Figure~\ref{sup:fig:impedance_match_cartoon_NO}). The point on the Hugoniot to which the rock is shocked gives the strength of the shock, which we quantify by the particle velocity of the shock in the planet, $u_p^{\rm G}$. 

When the shock wave reaches the surface of the planet (Figure~\ref{sup:fig:impedance_match_cartoon_NO}B), the pressure differential between the shocked rock and the atmosphere causes the surface to accelerate and the shock wave propagates into the atmosphere. However, the Hugoniot of a typical atmosphere (e.g., solid orange line in Figure~\ref{sup:fig:impedance_match_cartoon_NO}B, right) is shallower than that of rocks, due to the lower shock impedance (i.e., resistance to compression) of gases compared to rocks. As a result, the pressure in the shocked atmosphere is much lower than in the shocked rock for a given particle velocity. The ground must release to a lower pressure, following an isentrope (black dashed line), until it intersects the gas Hugoniot to achieve both pressure and particle velocity continuity with the atmosphere (black and orange pentagon). The particle velocity and thermodynamic properties at which the rock release curve and gas Hugoniot intersect is called the impedance match solution. The impedance-match velocity of the surface is greater than the particle velocity of the shock within the planet before release. The acceleration of the surface leads to a release wave that propagates back into the planet (white dashed line in Figure~\ref{sup:fig:impedance_match_cartoon_NO}B, left) and more and more of the rock accelerates to the impedance match velocity. Meanwhile, as the shock propagates up the atmosphere, the density of atmosphere in front of the shock falls, and so a higher particle velocity is required to achieve pressure continuity. The shock therefore accelerates as it moves upwards in the atmosphere. When the shock reaches the top of the atmosphere, a release wave, analogous to that in the rock, propagates downwards in the atmosphere (upper white dashed line in Figure~\ref{sup:fig:impedance_match_cartoon_NO}C, left), causing the atmosphere to accelerate to even higher velocity (orange dashed line in Figure~\ref{sup:fig:impedance_match_cartoon_NO}C, right). The process of acceleration of the shock wave upwards in the atmosphere and the subsequent release of the atmospheric shock drives a portion of the atmosphere to velocities above the escape velocity. Therefore, even when the initial ground-atmosphere impedance match velocity is much lower than escape, a portion of the atmosphere can be lost.

The presence of an ocean changes the efficiency of loss as the ocean has a different shock impedance than the rocky surface or atmosphere. Figure~\ref{fig:impedance_match_cartoon} shows the equivalent cartoon to that in Figure~\ref{sup:fig:impedance_match_cartoon_NO} for the same strength of impact shock, but when there is an ocean present. Before the shock wave breaks out from the surface of the planet (Figure~\ref{fig:impedance_match_cartoon}A) the situation is much the same as in the no-ocean case, except that the initial pressure of the rock is increased due to the mass of the ocean. The low compressibility of rocks means that the increased pre-shock pressure has little effect on the Hugoniot. When the shock wave from the impact reaches the surface of the planet (Figure~\ref{fig:impedance_match_cartoon}B), the shocked rock releases. In a similar manner to in the no-ocean case, the water is shocked to a point on its Hugoniot (blue solid line in Figure~\ref{fig:impedance_match_cartoon}B, right) that intersects the release curve for the rock. The impedance match between the rock and water (blue and black pentagon in Figure~\ref{fig:impedance_match_cartoon}B, right) is at a higher pressure and lower particle velocity than the impedance match between the rock and atmosphere as the compressibility of the ocean is lower than that of the gas. When the shock reaches the surface of the ocean, the water itself releases, accelerating the atmosphere to the ocean-atmosphere impedance match velocity. The release curve for water (blue dashed line in Figure~\ref{fig:impedance_match_cartoon}C, right) is shallower than that of rocks, and the impedance match velocity of the ocean surface with the atmosphere is higher than that of the ground in the no-ocean case. In the ocean case, the surface driving the loss of the atmosphere is effectively the surface of the ocean, not the ground, and the higher-velocity of the ocean surface has the potential to drive greater loss than in the no-ocean case. After release from the surface of the ocean, the shock propagates up the atmosphere in the same way as the no-ocean case (Figure~\ref{fig:impedance_match_cartoon}C).

In both the ocean and no-ocean cases, how a given strength of shock in the planet translates to the velocity of the ocean surface or ground depends on the slope of the atmospheric Hugoniot, and therefore on the properties of the atmosphere. We will discuss these effects in Section~\ref{sec:results:up_ug_relation_NO}.

\begin{figure*}[t]
    \centering
    \includegraphics[height=0.96\textheight]{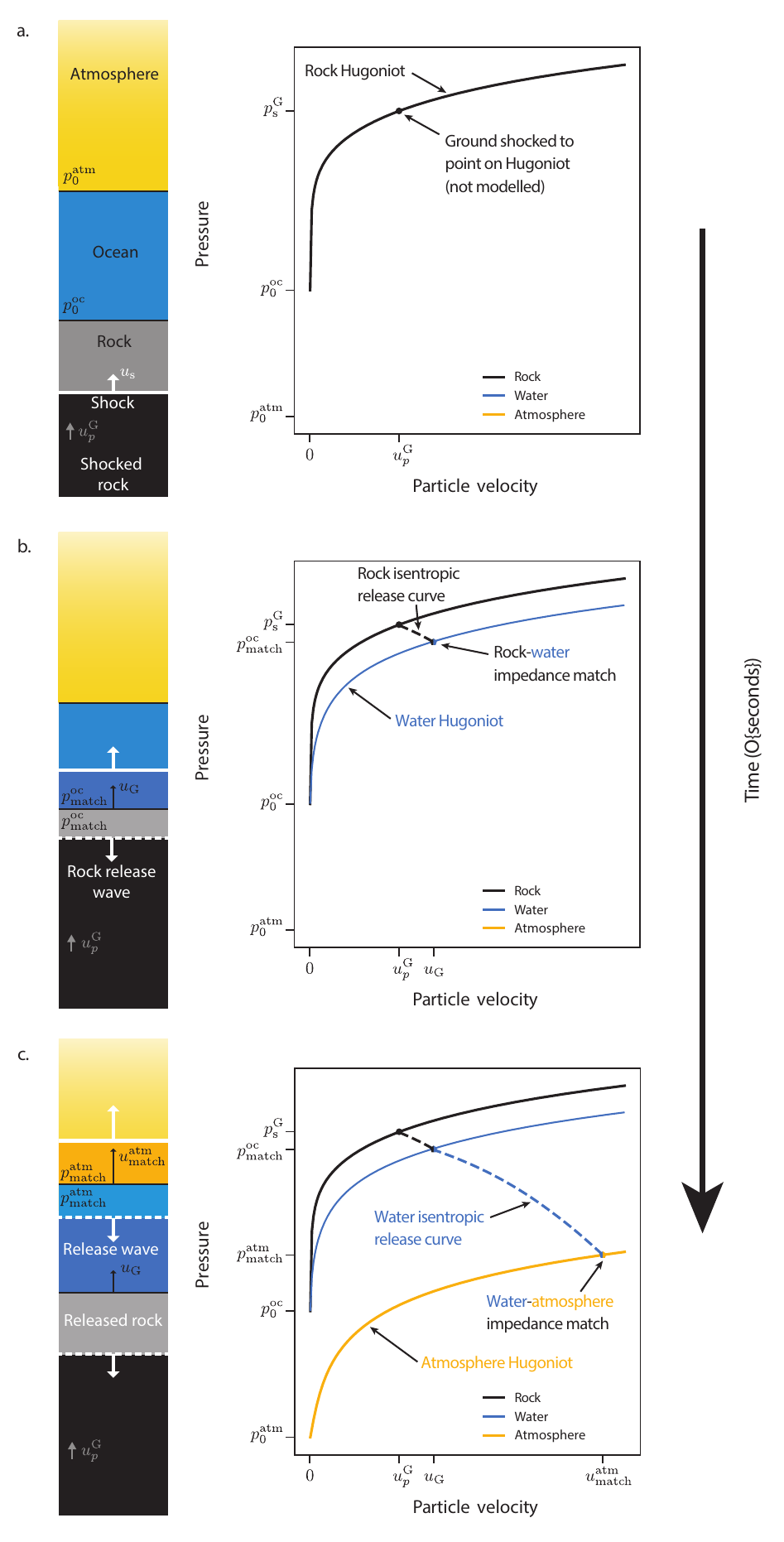}
    \caption{Caption on next page.}

\end{figure*}
\addtocounter{figure}{-1}
\begin{figure*}[t]
    \caption{The presence of an ocean can strongly influence the efficiency of loss. Shown is a schematic that shows the relative position (left column) and the thermodynamic state (right) of the different material layers at different stages (rows) in the passage of the shock from the planet, through the ocean, and into the atmosphere. Left: Different materials are indicated by different colors: rock - black, water - blue, and atmosphere - orange. In the left column, darker shades of these colors indicate material under maximum compression in the shock. Boundaries between materials are shown as thin black lines with their velocities shown as black arrows. The shock wave is indicated by a thick white line. Release waves are shown as white dashed lines and their velocities as white arrows. Where key dynamic and thermodynamic variables apply are noted. Right: Schematic particle velocity - pressure plots for the impedance match calculation between rock, water and atmosphere. Key pressures and particle velocities corresponding to different stages of the thermodynamic evolution of material are given on the axis, as labelled in the left column. Solid lines are shock Hugoniots, the locus of points that a shocked material can reach from an initial starting position. Hugoniots are not thermodynamic paths and the point reached material on each Hugoniot is indicated by a filled symbol. Dashed lines are release curves followed by material decompressing from a shocked states. Release curves are thermodynamic paths and the material moves along these lines. As in the left column, colors indicate different materials. A similar schematic for the case with no ocean is shown in Figure~\ref{sup:fig:impedance_match_cartoon_NO}}
    \label{fig:impedance_match_cartoon}

\end{figure*}

%xxxxxxxxxxxxxxxxxxxxxxxxxxxxxxxxxxxxxxxxxxxxxxxxxxxxxxxxxxxxxxxxxxxxxxxxxxxxxxxxxxxxxxxxxxxxxxxxxxxxxxxxxxxxxxxxxxxxxxxxxxxxxxxxxxxxxxxx
%xxxxxxxxxxxxxxxxxxxxxxxxxxxxxxxxxxxxxxxxxxxxxxxxxxxxxxxxxxxxxxxxxxxxxxxxxxxxxxxxxxxxxxxxxxxxxxxxxxxxxxxxxxxxxxxxxxxxxxxxxxxxxxxxxxxxxxxx
%xxxxxxxxxxxxxxxxxxxxxxxxxxxxxxxxxxxxxxxxxxxxxxxxxxxxxxxxxxxxxxxxxxxxxxxxxxxxxxxxxxxxxxxxxxxxxxxxxxxxxxxxxxxxxxxxxxxxxxxxxxxxxxxxxxxxxxxx
\section{Methods}
\label{sec:methods}

\subsection{1D hydrodynamic calculations}
\label{sec:methods:hydro}

\begin{figure*}[t]
    \centering
    \begin{interactive}{animation}{Figure4_animation.mp4}
    \includegraphics[width=0.91\textheight, angle=90,origin=c]{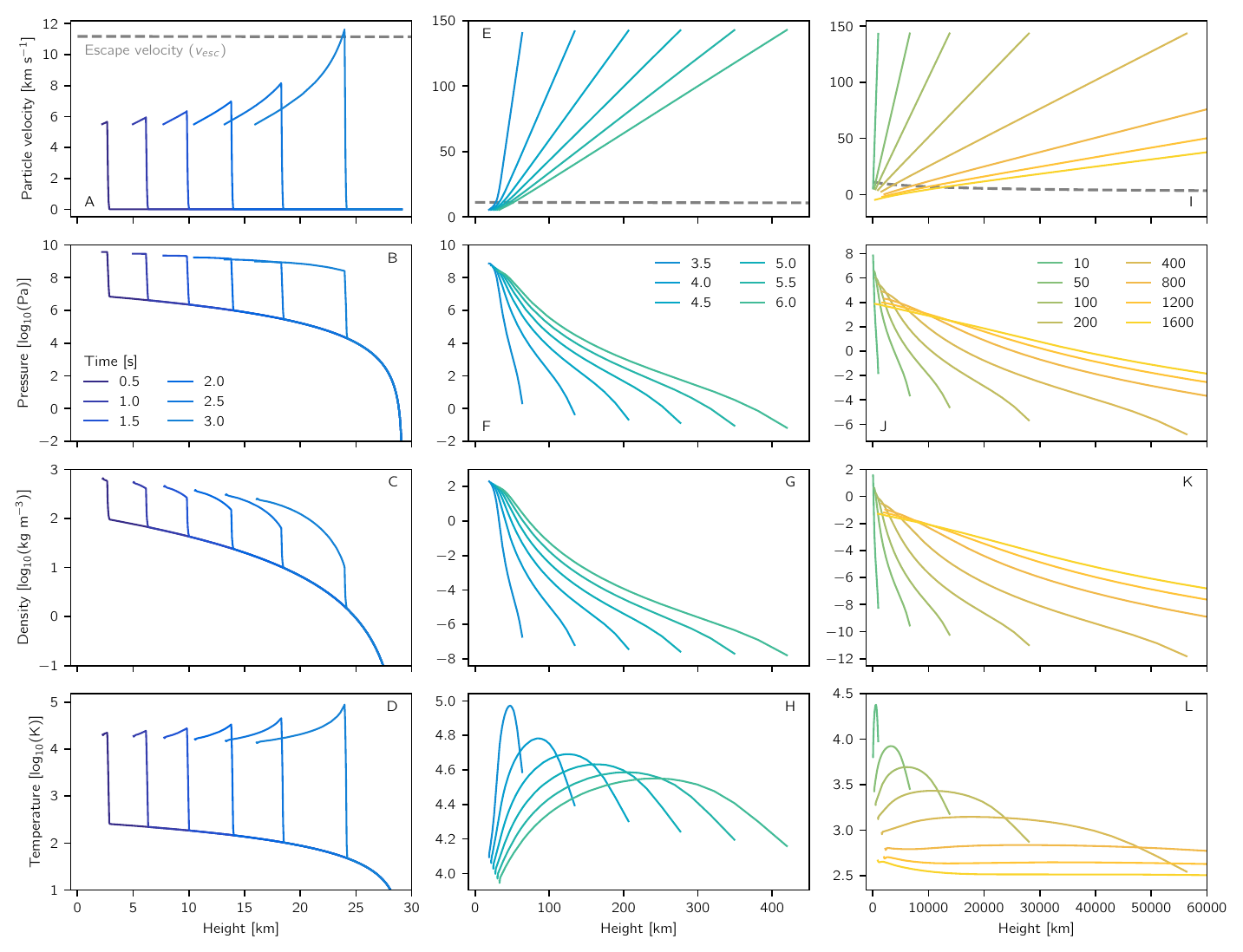}
    \end{interactive}
    \caption{Caption on next page.}
\end{figure*}

\addtocounter{figure}{-1}
\begin{figure*}[t]
    
    \caption{Atmospheric loss is driven by acceleration of the impact shock wave as it travels up the hydrostatic pressure profile. Shown are the velocity, pressure, density and temperature profiles of the atmosphere (with each row a different variable) at different times after break out of the shock wave from the planet in a 1D simulation. Lines of the same color in each panel show the atmospheric structure at the same time after initial breakout (see legends in the second row). Each profile is plotted as a function of position relative to the initial location of the ground (x-axis), and hence the location of the bottom of the atmosphere moves to higher values with time. Each column shows a different set of time steps with the axis scales altered to be appropriate for each set of time steps. The sharp increase in all parameters in the first column is the shock wave, which moves upwards in the atmosphere over time, i.e., as line colors become lighter. At about $\sim$~3~s, between the first and second columns, the shock reaches the top of the atmosphere and release to vacuum rapidly accelerates the top of the atmosphere to escape. For this simulation, there was no ocean and the atmosphere was Earth-like \citep[$m_{\rm a}=29$~g~mol$^{-1}$, $\gamma=1.4$:][]{Genda2003} with a surface pressure and temperature of $p_0=100$~bar and $T_0=283$~K, respectively. The initial ground velocity was $u_{\rm G}=5.5$~km~s$^{-1}$ and the final loss fraction was 0.25. Gray dashed lines give the escape velocity as a function of distance from the center of the planet. To avoid showing known numerical artifacts near the center of the calculation (see Section~\ref{sup:sec:boundary}), the density and temperature of the zones closest to the lower boundary are not plotted. This figure is comparable to Figure~3 of \cite{Genda2003}. An animated version of this figure is available lasting 11~s.
    }
    \label{fig:NO_evo}

\end{figure*} 

To calculate atmospheric and ocean loss due to ground motion we follow a similar approach to that of \cite{Genda2003,Genda2005}. A hydrostatic atmosphere, and in most cases a hydrostatic ocean, is initialized at the radius of the planet's surface in a 1D spherical geometry (Figure~\ref{fig:Overview_cartoon}D). The breakout of the shock wave is then simulated by giving the lower boundary a vertical velocity that generates a shock wave in the (ocean and) atmosphere  that accelerates a fraction of the (ocean and) atmosphere to escape.

We adapted the 1D WONDY hydrodynamic code \citep{Kipp1982} to calculate the evolution of the atmosphere (and ocean) in response to the motion of the ground. WONDY solves the Lagrangian 1D mass, momentum, and energy equations using a finite difference method. Artificial viscosity is used to resolve shocks by spreading the shock front over several Lagrangian cells. We have expanded the capabilities of the WONDY code by adding options for radial gravity, three additional equations of state (EOSs), and hydrostatic initialisation of an atmosphere and ocean. A full description of the adapted code and sensitivity tests are given in Appendix~\ref{sup:sec:WONDY}, but we provide a summary here.

The atmosphere and ocean were both modeled using 500 Lagrangian cells (or zones) each, and are initialized as stationary, hydrostatic, and adiabatic, isothermal or isoenergetic depending on the equation of state (EOS) used. By assuming that the atmosphere and ocean are stationary and hydrostatic we neglect the influence of the gravity of the other colliding body which could deform the surface and disturb the atmospheric structure. We will address the effect of pre-impact redistribution of the atmosphere and ocean in future work. The properties of the atmosphere were set by defining a surface temperature ($T_0$) and pressure ($p_0$) and the structure of the atmosphere was determined by integrating upwards from the surface. The top of the atmosphere is treated as a stress-free boundary, implemented by using an additional mass-less, zero-pressure cell at the top of the atmosphere. The atmosphere was modelled as an ideal gas with a constant molar mass ($m_{\rm a}$) and ratio of specific heat capacities ($\gamma$). When present, the ocean was initialized with a given depth ($H_{\rm oc}$) and the initial structure of the ocean was calculated by integrating downwards from the ocean surface, assuming thermal equilibrium with the atmosphere at the base of the atmosphere. In most simulations, we used the water EOS of \cite{Senft2008} to describe the thermodynamic properties of the ocean. In order to compare our results to those of \cite{Genda2005}, we also ran simulations using the International Association for the Properties of Water and Steam (IAPWS) EOS \citep{Wagner2002} and the Tilloston EOS \citep{Tillotson1962} using the parameters from \cite{O'Keefe1982}. We find good agreement between our results using the Senft \& Stewart tabulated EOS and the IAPWS EOS, which is not surprising as the Senft \& Stewart EOS was constructed partly using the IAPWS EOS, but find significant differences when using the Tillotson EOS which we discuss in Section~\ref{sec:results:R:previous_res}.

As in previous work \citep{Genda2003,Genda2005}, we do not directly model the shock in the planet. Instead, the propagation of the shock from the planet into the ocean/atmosphere, is simulated by imposing the velocity of the lower boundary of the domain, i.e., the ground motion. The boundary is given an initial velocity ($u_{\rm G}$) and then allowed to follow a ballistic trajectory, ignoring the influence of any forces other than gravity (see Section~\ref{sup:sec:boundary} for more details). Imposing a ballistic boundary condition assumes that the mass of the ground is much greater than the mass of any ocean and/or atmosphere and thus the ground is not slowed significantly by transferring momentum to the ocean and/or atmosphere. For the range of oceans and atmospheres we consider in this work, this is a good approximation. Even for the most massive atmosphere and ocean combination we simulated, the mass of the ocean and atmosphere combined is equivalent to a surface layer of only $\sim$~10~km, and is typically much lower. Using the ballistic boundary condition, if the ground velocity is below the escape velocity, the boundary eventually stops and then accelerates downwards towards its initial position. When the boundary approaches its initial position, it is gradually brought to a stop. The prescription of this later-stage evolution of the boundary rarely has an effect on the amount of loss. We discuss the effect of non-ballistic motion of the boundary in Section~\ref{sec:discussion:issues}. It is important to note that $u_{\rm G}$ is the velocity of the ground upon breakout of the shock into the atmosphere or ocean (i.e., the impedance-match velocity; see Section~\ref{sec:bckgrnd}), and is not the particle velocity of the shock in the planet. The relationship between the strength of the shock in the planet and $u_{\rm G}$ can vary depending on the properties of the atmosphere or ocean, and we discuss this in Section~\ref{sec:results:up_ug_relation_NO}. Furthermore, prescribing a ballistic trajectory ignores any further positive acceleration of the ground by decompression of the surface to pressures below the impedance match solution. We discuss the implications of this simplification in Section~\ref{sec:discussion:issues}.

Simulations were run for 5000~s to ensure that a plateau in atmospheric/ocean loss was achieved. A small number of runs failed before completion. Failed runs were typically for either particularly high or low ground velocities. Failure was generally due to either insufficient numerical viscosity in the first few time steps or due to complications with stopping of the boundary upon its ballistic descent late in time. If a plateau in loss had been reached prior to failure this value was taken as the final loss, otherwise the result was discounted and not considered in our results. In another subset of cases, the stopping of the boundary upon descent caused a secondary shock into the atmosphere leading to additional loss. This secondary shock is likely unrealistic as the surface would have spalled or vaporized shortly after release. Generally, the amount of additional loss is small as the initial hydrostatic structure of the atmosphere that allowed for the acceleration of the initial shock has been disrupted. To account for this effect, we identified the plateau in loss due to the original shock and took the loss just before the passage of the second wave as the final value for loss. We tested the sensitivity of our results to the intrinsic parameters of the code and found no variation within the range of reasonable values (Appendix~\ref{sup:sec:1Dsensitivity_tests}).

We ran simulations for a wide variety of surface pressures, surface temperatures, atmospheric compositions, ocean depths, planetary masses, ground velocities, and using the three different EOS for water to explore the dependence of each of these parameters on the efficiency of loss. The details of the surface conditions used in each set of calculations are described at the relevant point in Section~\ref{sec:results}.

%xxxxxxxxxxxxxxxxxxxxxxxxxxxxxxxxxxxxxxxxxxxxxxxxxxxxxxxxxxxxxxxx
%xxxxxxxxxxxxxxxxxxxxxxxxxxxxxxxxxxxxxxxxxxxxxxxxxxxxxxxxxxxxxxxx
\subsection{Impedance match calculations}

The impedance match velocities and pressures between different layers were found numerically, by solving for the intersection between the relevant release curve (of the ground/ocean) with the Hugoniot of the layer above. Hugoniots were calculated by iteratively finding the particle velocity, $u_{\rm p}$, that satisfied the Rankine-Hugoniot equations, or using the analytical expressions for shocks in an ideal gas \citep{ZelDovich&Raizer2002book,Melosh1989book}. Release curves were calculated as isentropes through the relevant EOS with the solutions found iteratively, if necessary. Jupyter notebooks, python scripts, and a widget that can calculate the impedance match between different materials will be made available on publication of this work.

The EOS used for water and atmospheres were the same as for the 1D hydrodynamics simulations (Section~\ref{sec:methods:hydro}), and the ground was modelled as forsterite \citep{Stewart2019forsteriteEOS}. To allow discussion of previous work \citep{Kegerreis2020a,Kegerreis2020}, we also calculated impedance matches using the EOS for \h2-He mixtures from \cite{Hubbard1980}. An early version of the \cite{Hubbard1980} EOS that was included in the SWIFT hydrodynamics code \citep{Schaller2018,Kegerreis2019} and used in previous work \citep{Kegerreis2020a,Kegerreis2020,Kegerreis2018} contained an error in the calculation of internal energy. The \cite{Hubbard1980} EOS is defined by expressions for pressure and heat capacity as functions of density and temperature. Previous work calculated the specific internal energy as
\begin{equation}
    \epsilon = c_V (\rho, T) \, T \;
\end{equation}
where $c_V$ is the specific heat capacity, $\rho$ is the density, and $T$ is temperature, but this expression neglects the change in heat capacity as a function of temperature and density. Here, we calculate internal energy as an integration first along the $T=0$~K isotherm, and then along an isochore:
\begin{equation}
\epsilon(\rho,T) = \int_{\rho_0}^{\rho}{\frac{\partial \epsilon}{\partial \rho'}\bigg|_{T=0}} d\rho'+ \int_{0}^{T}{\frac{\partial \epsilon}{\partial T'}\bigg|_{\rho'=\rho}} dT' \; ,
\end{equation}
where primes indicate integration variables. In the formulation of \cite{Hubbard1980} the first term, the integral along the isotherm, is zero and so
\begin{equation}
\epsilon(\rho,T) = \int_{0}^{T}{\frac{\partial \epsilon}{\partial T'}\bigg|_{\rho'=\rho}} dT' = \int_{0}^{T}{c_V(\rho,T')dT'} \; ,
\end{equation}
which can be solved analytically. This is now the method used for calculating internal energy in the \cite{Hubbard1980} EOS table included in the current and future releases of SWIFT. It is important to note that an EOS defined by expressions for heat capacity and pressure alone, such as the \cite{Hubbard1980} EOS, is non-conservative. In other words, integration of thermodynamics variable along different paths between points in phase space can give different values. There is thus no single definition of internal energy, and our choice of energy calculation only improves on previous work in that it gives a value that is consistent with the defined EOS.

%xxxxxxxxxxxxxxxxxxxxxxxxxxxxxxxxxxxxxxxxxxxxxxxxxxxxxxxxxxxxxxxxxxxxxxxxxxxxxxxxxxxxxxxxxxxxxxxxxxxxxxxxxxxxxxxxxxxxxxxxxxxxxxxxxxxxxxxx
%xxxxxxxxxxxxxxxxxxxxxxxxxxxxxxxxxxxxxxxxxxxxxxxxxxxxxxxxxxxxxxxxxxxxxxxxxxxxxxxxxxxxxxxxxxxxxxxxxxxxxxxxxxxxxxxxxxxxxxxxxxxxxxxxxxxxxxxx
%xxxxxxxxxxxxxxxxxxxxxxxxxxxxxxxxxxxxxxxxxxxxxxxxxxxxxxxxxxxxxxxxxxxxxxxxxxxxxxxxxxxxxxxxxxxxxxxxxxxxxxxxxxxxxxxxxxxxxxxxxxxxxxxxxxxxxxxx
\section{Results of 1D hydrocode simulations}
\label{sec:results}

In this section we discuss the results of our numerical calculations. First, we will consider the efficiency of loss as a function of ground velocity without (Section~\ref{sec:results:NO}) and with an ocean (Section~\ref{sec:results:R}), including comparisons to previous results. We present a parameterization of the relationship between ground velocity and loss in both cases in Section~\ref{sec:results:fit}. 

%xxxxxxxxxxxxxxxxxxxxxxxxxxxxxxxxxxxxxxxxxxxxxxxxxxxxxxxxxxxxxxxx
%xxxxxxxxxxxxxxxxxxxxxxxxxxxxxxxxxxxxxxxxxxxxxxxxxxxxxxxxxxxxxxxx
\subsection{The dependence of loss on ground velocity in the absence of an ocean}
\label{sec:results:NO}

Atmospheric loss in the no-ocean case has been considered in a number of previous studies \citep[e.g.,][]{Chen1997,Genda2003,Schlichting2015}. \cite{Genda2003} explored seven cases of ideal, adiabatic atmospheres with different atmospheric compositions, surface temperatures and pressures, and atmospheric compositions, and conducted simulations using different prescriptions for $\gamma$. They found that the degree of loss is relatively minimal unless the ground velocity exceeds $\sim0.5v_{\rm esc}$ and observed relatively little variation in the efficiency of loss between different atmospheric properties and prescriptions for $\gamma$. \cite{Schlichting2015} calculated loss for both isothermal and adiabatic atmospheres with $\gamma=4/3$ and $\gamma=5/3$. In agreement with \cite{Genda2003} they found little difference in the efficiency of loss between atmospheres with different $\gamma$, but found that loss from isothermal atmospheres was somewhat less efficient than for adiabatic atmospheres at the same ground velocity (a difference in loss of up to $\sim10$\%). Here we revisit atmospheric loss in the absence of an ocean to ground truth our numerical model and to further explore the effect of atmospheric properties and planetary mass on the efficiency of loss.

Figure~\ref{fig:NO_evo} shows the evolution of an atmosphere upon breakout of a shock a planet's surface as calculated using our 1D hydrocode. The evolution follows that expected based on the physics described in Section~\ref{sec:bckgrnd}. The pressure of the shock at the base of the atmosphere is set by the impedance-match solution at the particle velocity of the ground imposed in the simulation. The shock wave accelerates up the strong adiabatic density gradient of the atmosphere, heating and compressing the gas, until the shockfront reaches the top of the atmosphere (at $\sim$3.2s in Figure~\ref{fig:NO_evo}). When the shock reaches the low-density edge of the atmosphere the compressed gas expands rapidly, reaching speeds far in excess of the escape velocity, and the top of the atmosphere is lost from the gravitational well of the planet (Figure~\ref{fig:NO_evo}E, I). Momentum transfer to the portion of the atmosphere that is lost occurs in the first few seconds to tens of seconds after the release of the shock into the atmosphere. After this point the lost portion of the atmosphere behaves almost ballistically and its velocity begins to fall as it moves further from the planet. In our simulations, the ground eventually stops (at 830~s in Figure~\ref{fig:NO_evo}) and the remaining bound atmosphere begins to fall back down to the planet. In reality, before this point the release wave from other parts of the surface could have slowed the ground motion and the rock surface could have spalled, melted, or vaporized. However, as the momentum is transferred to the lost portion of the atmosphere very early, these complications likely do not affect the efficiency of loss from the initial shock wave \citep[see][and Section~\ref{sec:discussion:issues}]{Kegerreis2019}.

\begin{figure}
        \centering
    \includegraphics[width=\columnwidth]{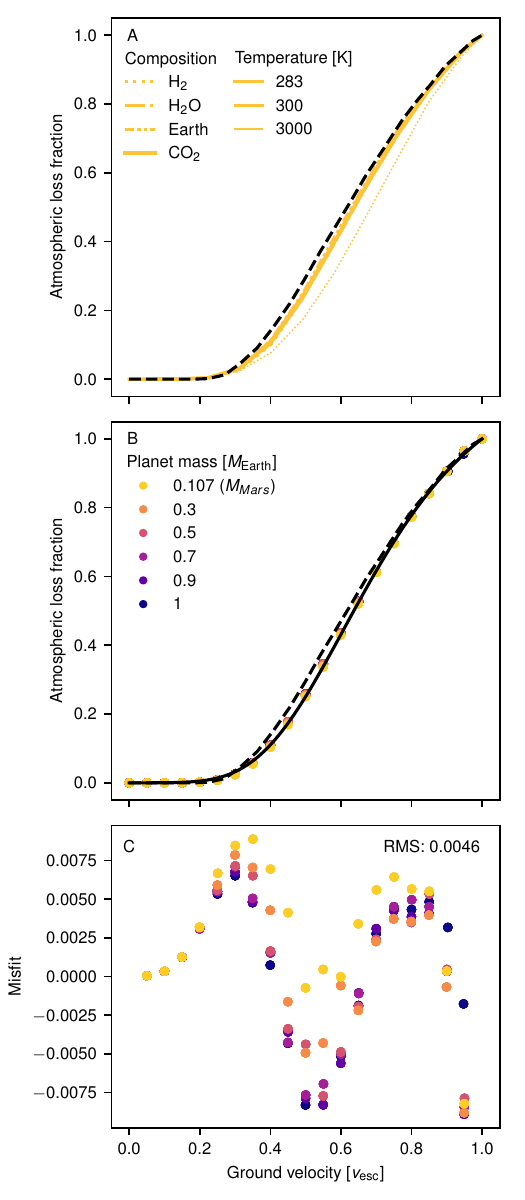}
    \caption{Caption opposite.}
    \label{fig:NO}
\end{figure}

We find good agreement between our results and those of previous studies, and demonstrate more completely that the relationship between ground velocity and atmospheric loss in the absence of an ocean is relatively insensitive to atmospheric composition, surface temperature and pressure, and planetary mass. Figure~\ref{fig:NO}A shows the efficiency of atmospheric loss for H$_2$ ($m_{\rm a}=2$~g~mol$^{-1}$, $\gamma=1.4$), H$_2$O ($m_{\rm a}=18$~g~mol$^{-1}$, $\gamma=1.25$), CO$_2$ ($m_{\rm a}=44$~g~mol$^{-1}$, $\gamma=1.29$) and an approximation to an Earth-like N$_2$ and O$_2$-dominated atmosphere \citep[$m_{\rm a}=29$~g~mol$^{-1}$, $\gamma=1.4$;][]{Genda2003} with surface temperatures of 283 and 3000~K (H$_2$, CO$_2$ and Earth-like) or 300 (H$_2$O) and surface pressures of 100~bar. The black dashed line is the result from \citet{Genda2003} for a 1~bar atmosphere of Earth-like composition and a surface temperature of 288~K. With the exception of the 3000~K, H$_2$ atmosphere, all of the simulations gave very similar results and were in good agreement with those of \cite{Genda2003} and \cite{Schlichting2015}. Loss of the high-temperature H$_2$ atmosphere is slightly less efficient for a given ground velocity (at most a few percent), which may be surprising given that the high-$T$ H$_2$ atmosphere is much more extended, and so more loosely bound, than the other example atmospheres. The high-$T$ H$_2$ atmosphere has a height of 2.3~$R_{\rm Earth}$ (Earth radii), compared to a maximum height of 0.07~$R_{\rm Earth}$ for the other examples. The pressure and density gradients are much lower, affecting how the shock wave accelerates through the atmosphere, and more of the mass of the atmosphere is at lower pressure. In addition, the hot H$_2$ atmosphere is so extended that for ground velocities less than $\sim0.6~v_{\rm esc}$ the ground has stopped and begun to fall back to its original postion even before the initial shock has reached the top of the atmosphere. The release wave from the reversal of the ground velocity propagates upwards in the atmosphere and may play a role in reducing the efficiency of loss compared to other cases where the shock wave is supported as the atmosphere is accelerated to escape.

% The structure of the atmosphere is determined by $\gamma$ and the ratio $m_{\rm a}/T_0$ with the height of the atmosphere given by  
% %
% \begin{equation}
%   H_{\rm atm} = R_{\rm p} \left( \frac{ \lambda_0}{\lambda_0 -\frac{\gamma}{\gamma-1} } -1 \right ) \; ,
% \end{equation}
% %
% where
% %
% \begin{equation}
%     \lambda_0=\frac{G M_{\rm p} m_{\rm a}}{N_{\rm A}k T_0 R_p} \;, 
% \end{equation}
% %
% where....

\addtocounter{figure}{-1}
\begin{figure}
    \caption{The relationship between ground velocity and loss in the no-ocean case is insensitive to atmospheric composition, surface temperature and pressure, and planetary mass. A: Fraction of atmosphere lost from an Earth-mass ($M_{\rm Earth}$) planet as a function of ground velocity for 100~bar atmospheres of different compositions (line styles) and surface temperatures (line thicknesses). The black dashed line is the result from \citet{Genda2003} for a 1~bar atmosphere of Earth-like composition and a surface temperature of 288~K. B: The fraction of atmosphere lost from bodies of different masses with surface pressures of 0.1, 0.5, 1, 5, 10, 50, 100, and 500~bar (colored symbols). Ground velocity is normalized to the escape velocity, $v_{\rm esc}$, of each body. Atmospheres were CO$_2$ with a surface temperature of 300~K. The solid black line was calculated using the parameterization described in Section~\ref{sec:results:fit}. C: The misfit of the results in B from the parameterization described in Section~\ref{sec:results:fit}.}
\end{figure}

Figure~\ref{fig:NO}B shows the efficiency of atmospheric loss for atmospheres of varying surface pressures on planets of between Mars and Earth mass. Points are for all combinations of atmospheric pressures of 0.1, 0.5, 1, 5, 10, 50, 100, and 500~bar and planetary masses of of 0.107 ($M_{\rm Mars}$), 0.3, 0.5, 0.7, 0.9, and 1~$M_{\rm Earth}$ with mass indicated by color. The black dashed line is the result from \citet{Genda2003} for an Earth-mass planet and the solid black line is a fit to our simulation results (see Section~\ref{sec:results:fit}). There is very little variation in the efficiency of loss as a function of ground velocity with atmospheric pressure and planetary mass, when ground velocity is normalized to the escape velocity. This confirms what has been assumed in other studies \citep{Genda2003,Genda2005,Schlichting2015} that the effect of planetary mass in the absence of an ocean is almost entirely accounted for by normalization to the escape velocity.

%xxxxxxxxxxxxxxxxxxxxxxxxxxxxxxxxxxxxxxxxxxxxxxxxxxxxxxxxxxxxxxxx
%xxxxxxxxxxxxxxxxxxxxxxxxxxxxxxxxxxxxxxxxxxxxxxxxxxxxxxxxxxxxxxxx
\subsection{Dependence of loss on ground velocity in the presence of an ocean}
\label{sec:results:R}

\begin{figure}
    \centering
    \includegraphics[width=\columnwidth]{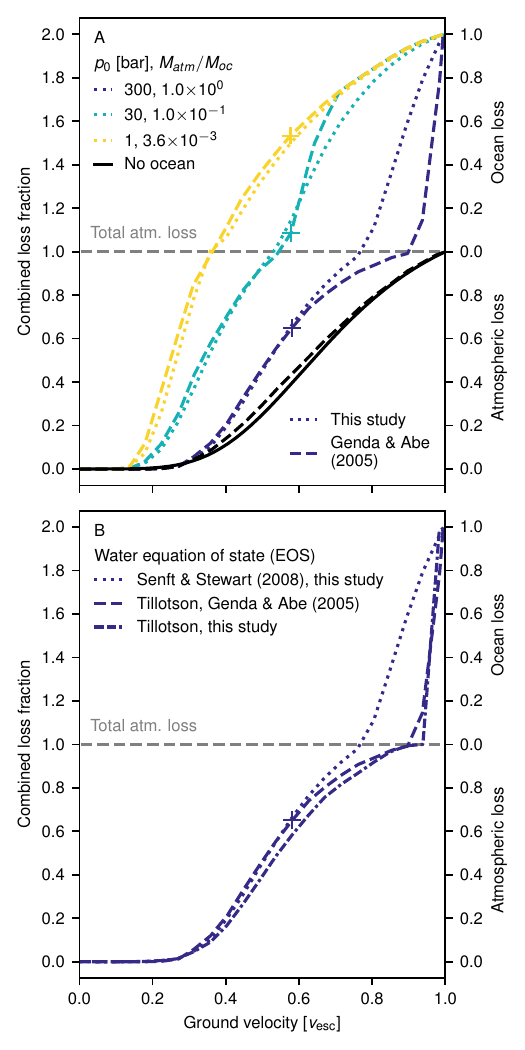}
     \caption{Caption opposite.}
\end{figure}
\addtocounter{figure}{-1}
\begin{figure}
    \caption{Our results agree well with those of \citet{Genda2005} at low ground velocities but deviate at high ground velocities due to using an improved equation of state (EOS) for water. A: Atmosphere and ocean loss from an Earth mass, $M_{\rm Earth}$, body as a function of ground velocity for H$_2$ atmospheres of different surface pressures (colored lines) above 3~km oceans. Dotted lines are the results of our calculations and dashed lines are the results of \citet{Genda2005}. The ocean surface temperature in each case was 300~K. Black lines are for the case of no ocean from \cite{Genda2003} (dashed line) or our parameterization described in Section~\ref{sec:results:fit} (solid line). Crosses indicate the maximum velocity at which \citet{Genda2005} calculated loss using the IAPWS water EOS, due to reaching the maximum pressure of validity for that EOS. Cumulative loss fraction is the sum of atmospheric and ocean loss. Grey dashed lines indicates total atmospheric loss but zero ocean loss. B: Fraction of atmosphere and ocean lost for a 300~bar, \h2 atmosphere over a 3~km ocean, calculated using different water EOS in our study and in \citet{Genda2005}.
    }
    \label{fig:ug_loss}
    
\end{figure}

The efficiency of atmospheric loss for a given ground velocity can be significantly enhanced if the colliding bodies have part or all of their surfaces covered by water \citep{Genda2005}. 
In the following sections, we compare our results to those of \cite{Genda2005}, and explore how the the efficiency of loss is dependent on the initial surface conditions (e.g., atmospheric pressure, ocean depth, etc.), and the mass of the planet.

%%%%%%%%%%%%%%%%%%%%%%
\subsubsection{Comparison to previous results}
\label{sec:results:R:previous_res}
 
Figure~\ref{fig:ug_loss}A shows the efficiency of loss as a function of ground velocity for H$_2$ atmospheres of 300, 30, and 1~bar (dotted lines) above 3~km deep oceans on Earth-mass planets with surface temperatures of 300~K. The pressure at the base of ocean were approximately 600, 330, and 300~bar, respectively. For reference, loss in the no-ocean case is shown in black. The examples shown in Figure~\ref{fig:ug_loss}A were also explored by \cite{Genda2005} and their results are shown as dashed lines. In their study, \cite{Genda2005} considered H$_2$ atmospheres with six different atmospheric pressures but we have chosen to only show three here to allow clear comparison with our results. At ground velocities less than $\sim6$~km~s$^{-1}$ (0.54~$v_{\rm esc}$) we find relatively good agreement between our results and those of \cite{Genda2005}, but at higher velocities our results diverge, particularly for higher pressure atmospheres. This difference is due to the EOS for water used at high ground velocities in each study. \cite{Genda2005} ran simulations using both the IAPWS EOS \citep{Wagner2002} and the Tilloston EOS \citep{Tillotson1962}. At lower ground velocities the two EOS gave relatively similar results, but the maximum pressure limit of the IAPWS EOS precluded its use for ground velocities above 6~km~s$^{-1}$ where the pressure in the shocked stated exceeded the range of the EOS (marked by crosses on each loss curve in Figure~\ref{fig:ug_loss}A). It is therefore in the regime in which \cite{Genda2005} only performed calculations using the Tillotson EOS in which our results significantly differ from theirs. 
 
 To confirm that the only difference between our results and those of \cite{Genda2005} is the water EOS used, we ran additional simulations using the Tillotson EOS for the ocean and found good agreement between our results when using the same EOS. Figure~\ref{fig:ug_loss} shows an example set of simulations for a Earth-mass planet with a 300~bar, H$_2$ atmosphere over a 3~km ocean with our simulations using the \cite{Senft2008} EOS shown by the dotted line, our results using the Tillotson EOS as a dash-dash-dot line, and the results of \cite{Genda2005} as a dashed line. The treatment of expanded states in the Tillotson EOS often leads to unphysical solutions at low densities, causing simulations to fail. As a result, very few of our calculations using the Tillotson EOS reached the prescribed run-time of 5000~s which likely accounts for the slightly lower loss we calculate in some cases. 
 
The Tillotson EOS is designed to model the behaviour of material in the shocked state but does not provide a good description of material properties in lower-density, expanded states. This is particularly an issue in the multi-phase liquid and vapor region where a minimum density cutoff is imposed which is typically a sizeable fraction of the reference density. In contrast, the water EOS from \cite{Senft2008} is designed for use in planetary collisions and includes a liquid-vapor phase region to more accurately describe expanded states. The efficiency of loss in the ocean case is dictated by the complex combination of multiple waves (see discussion below) and so the use of high-quality EOS for water is critical to accurately determine loss. At high ground velocities, more of the water reaches expanded states and hits the minimum density cutoff, likely accounting for the lower loss at high ground velocities when using the Tilltoson EOS. Given that the \cite{Senft2008} EOS provides a more accurate description of expanded states, our results are likely more realistic than those of \cite{Genda2005} using the Tillotson EOS. For a discussion of the comparison between the Tillotson and more advanced EOS see \citet{Stewart2020_key_req_EOS}.

%%%%%%%%%%%%%%%%%%%%%%
\subsubsection{Dependence on ocean depth and atmospheric pressure}
\label{sec:results:R:R}

To explore the effect of surface pressure and ocean depth, as well as planetary mass (Section~\ref{sec:results:R:Mp}), on the efficiency of loss as a function of ground velocity, we conducted simulations with every combination of six different planetary masses (0.107, 0.3, 0.5, 0.7, 0.9, and 1~$M_{\rm Earth}$), seven surface pressures ($p_0=1$, 5, 10, 50, 100, 300, and 500~bar), and nine ocean depths (0.1, 0.5, 1, 2, 3, 5, 10, 20, and 30~km). We also conducted additional simulations for each planetary mass with 900~bar atmospheres and oceans of 0.1~km depth. Atmospheres were CO$_2$ ($m_{\rm a}=44$~g~mol$^{-1}$, $\gamma=1.29$) with surface temperatures of 300~K. Simulations were performed for ground velocities at 0.05~$v_{\rm esc}$ intervals between 0.05 and 0.95~$v_{\rm esc}$.

\begin{figure*}[t]
    \centering
    \includegraphics[width=\textwidth]{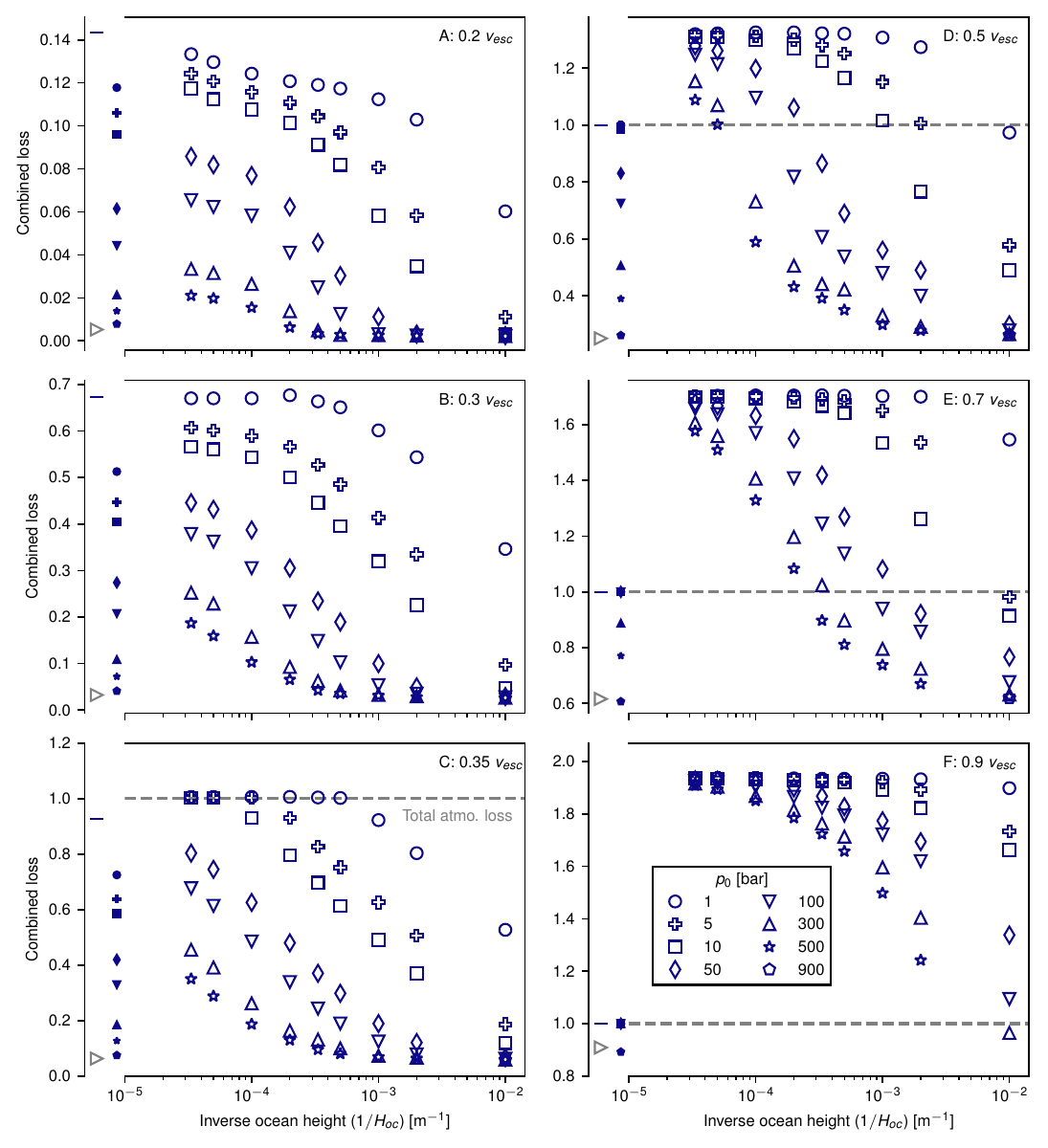}
     \caption{Caption on next page.}
\end{figure*}
\addtocounter{figure}{-1}
\begin{figure*}[t]
    \caption{The efficiency of atmospheric loss as a function of ground velocity depends on the depth of the ocean and the atmospheric pressure. Each panel shows the combined loss (sum of atmosphere and ocean fraction lost) as a function of inverse ocean depth for different ground velocities. Loss was calculated for different initial atmospheric pressures (open symbols) on an Earth-mass planet. Grey dashed lines indicate total atmospheric loss (a combined loss of one). The open grey triangle to the left of each panel indicates the loss expected in the absence of an ocean. Filled symbols to the left of each panel show the loss determined by convolving the velocity of the ocean surface determined from an impedance-match solution with a parameterization for atmospheric loss in the case of no ocean (see Section~\ref{sec:results:fit}). The  blue dash mark to the left of each panel shows a similar calculation except using the velocity of the ocean surface expected upon release of the ocean to 1~Pa.}
    \label{fig:Hloss_Earth}
\end{figure*}

\begin{figure*}[t]
    \centering
    \begin{interactive}{animation}{Figure8_animation.mp4}
    \includegraphics[width=0.91\textheight,angle=90,origin=c]{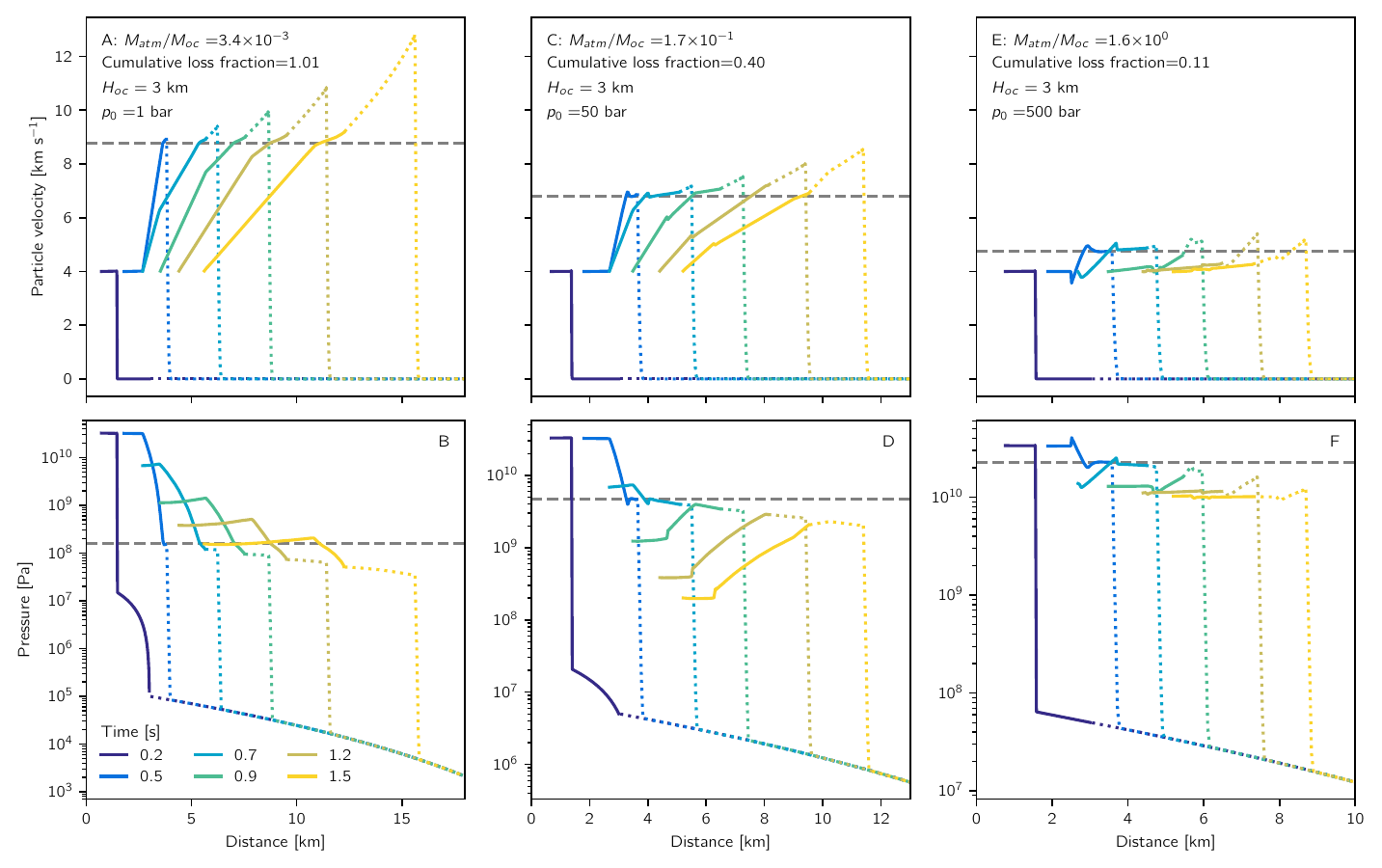}
    \end{interactive}
    \caption{Caption on next page.}
\end{figure*}
\addtocounter{figure}{-1}
\begin{figure*}[t]
    \caption{Acceleration of the ocean surface upon release of the shock can lead to a significant increase in the efficiency of atmospheric loss for a given ground velocity. Shown are velocity (top row) and pressure (bottom row) profiles of the ocean (solid lines) and atmosphere (dotted lines) at different times (colors) for simulations with a ground velocity of 4~km~s$^{-1}$. Columns show the evolution for planets with the same ocean depth (3~km) but different atmospheric pressures, resulting in different ratios of the mass of the atmosphere to the mass of the ocean. The same time steps are shown in each column. The atmosphere in all cases was CO$_2$ ($m_{\rm a}=44$~g~mol$^{-1}$, $\gamma=1.29$) with a surface temperature of 300~K, and the planet was Earth-mass with an escape velocity of 11.2~km~s$^{-1}$. Gray dashed lines show the impedance match solution for the release of the ocean to the corresponding atmosphere. An animated version of this figure accompanies this manuscript.
    }
    \label{fig:evo}

\end{figure*}

Figure~\ref{fig:Hloss_Earth} shows the efficiency of loss from an Earth-mass planet for six example surface velocities for different atmospheric pressures (symbols) and ocean depths (x-axis). The amount of atmosphere and ocean lost is presented as the sum of the atmospheric and ocean loss (which we refer to as combined loss) with one being total loss of atmosphere and two being the total loss of both ocean and atmosphere. For reference, the lower open grey triangle to the left of each panel shows the loss expected in the absence of an ocean. Note that the x-axis is the inverse of ocean height ($1/H_{\rm oc}$).

Variation in the atmospheric pressure and ocean depth can make the difference between almost zero and total loss of an atmosphere, and between zero and almost total loss of the ocean.
Loss is more efficient from planets that initially have deeper oceans and/or lower-pressure atmospheres. For shallow oceans and high-pressure atmospheres, the efficiency of loss tends towards a low-loss limit where the efficiency of loss in the no-ocean case (open grey triangle to the left of each panel in Figure~\ref{fig:Hloss_Earth}). Increasing the ocean depth while keeping the atmospheric pressure constant leads to an increase in the efficiency of loss until loss plateaus at a high-loss limit for very deep oceans. Over the range of conditions we have considered only the lowest pressure atmospheres plateau in the high-loss limit. At lower ground velocities, where only the atmosphere is being lost, the value of the plateau is dependent on the initial atmospheric pressure. As we will describe below, this phenomena is due to the fact that the lower the initial atmospheric pressure, the higher the velocity of the ocean surface upon release to the atmosphere and so the stronger the driver for atmospheric loss. When the atmosphere is totally lost ocean loss in the high-loss limit is almost invariant of initial atmospheric pressure 

It is evident from Figure~\ref{fig:Hloss_Earth} that the physics of atmospheric loss in the presence of an ocean is more complicated than a simple impedance match calculation (Section~\ref{sec:bckgrnd}). Filled symbols to the left of each panel in Figure~\ref{fig:Hloss_Earth} show the loss determined by convolving the velocity of the ocean surface determined from an impedance match solution (see Figure~\ref{fig:imp_match_vdiff}) with the parameterization for atmospheric loss in the case of no ocean (see Section~\ref{sec:results:fit}), i.e., the loss that would be expected if loss of the atmosphere was only controlled by the ocean surface driving a shock at the impedance-match velocity. The dark blue line on the left of each panel in Figure~\ref{fig:Hloss_Earth} show a similar calculation except using the velocity of the ocean surface expected upon release of the ocean to very low pressure (1~Pa). The plateau in loss seen in our simulations is typically higher than that calculated assuming the impedance match velocity as the driving velocity for loss, and additional processes must be at play.

To explain the dependence of loss on atmospheric pressure and the depth of the ocean, it is necessary to understand the dynamics of the system beyond the initial breakout of the shock (Section~\ref{sec:bckgrnd}). Figure~\ref{fig:evo} shows examples of the early evolution of the ocean and atmosphere after breakout of the impact shock from the ground. Each column shows the evolution for planets with the same depth of ocean (3~km) and the same ground velocity (4~km~s$^{-1}$), but with increasing initial atmospheric pressures going from left to right (1, 50, and 500~bar).
 Upon breakout from the planet, the system evolves as dictated by the impedance-match between the different layers, as described in Section~\ref{sec:bckgrnd}. The shock propagates through the ocean, compressing the water and accelerating it to the ground velocity. When the shock front reaches the surface of the ocean the water releases to the impedance match pressure between the water and the atmosphere (grey dashed lines in Figure~\ref{fig:evo}), driving a shock wave into the atmosphere. The shock accelerates as it travels up the atmosphere and the initial evolution is similar to that seen in Figure~\ref{fig:NO_evo} for the no-ocean case, but with the ocean surface taking the role of the ground. 
 
 Lamentably, further evolution of the system after the initial transmission of the shock into the atmosphere complications the simple impedance-match picture. As discussed in Section~\ref{sec:bckgrnd}, the release wave propagating downwards into the ocean causes more of the ocean to accelerate. Furthermore, as the pressure at the base of the atmosphere falls (Figure~\ref{fig:evo}), the ocean continues to decompress and accelerate and can reach velocities significantly above the impedance match velocity at later times. The ocean is stretched between the rapidly moving ocean surface and the ground, and the pressure in the ocean decreases rapidly. In cases with low pressure atmospheres and deep oceans (e.g., Figure~\ref{fig:evo}A and B), the pressure in the ocean remains above that of the atmosphere for most, if not all, of the subsequent evolution and the atmospheric shock is well supported. The ocean surface continues to decompress, accelerating to velocities approaching those expected upon release of the ocean to a vacuum. This effect is responsible for the plateauing of loss in the high-loss limit.
 
\begin{figure*}[t]
    \centering
    \includegraphics[width=\textwidth]{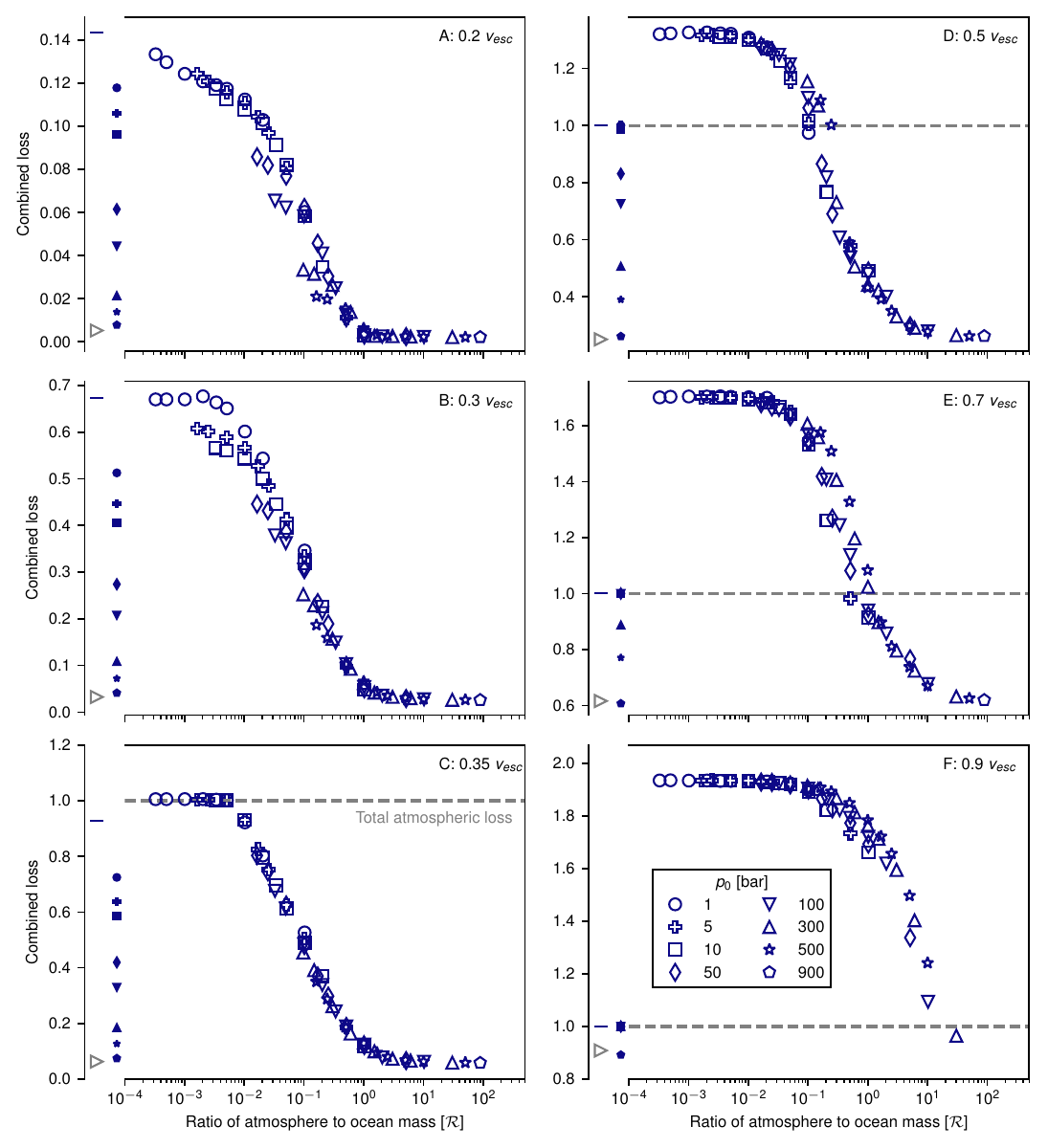}
    \caption{The transition between the high and low loss regimes scales well with the ratio of atmospheric to ocean mass ($\mathcal{R}$). Each panel shows the combined loss (sum of atmosphere and ocean fraction lost) as a function of $\mathcal{R}$. Symbols and lines are the same as in Figure~\ref{fig:Hloss_Earth}.}
    \label{fig:Rloss_Earth}
\end{figure*}

 However, in most cases the ballistic expansion of the ocean causes the ocean pressure to drop below that of the lower atmosphere after a time dependent on the initial surface conditions and strength of the shock (Figure~\ref{fig:evo}C-F). The difference in pressure exerts a force to slow the ocean and the ocean velocity is decreased by a series of pressure waves traversing the ocean. The shock is no longer fully supported in the atmosphere and a release wave from the bottom of the atmosphere can retard the acceleration of the upper fraction of the atmosphere, leading to decreased loss. In cases with either very thin oceans or very high-pressure atmospheres (e.g., Figure~\ref{fig:evo}E and F), pressure waves can equalize the pressure throughout the ocean before the shock wave has propagated far into the atmosphere. The velocity of the ocean surface slows to that of the ground, and the evolution of the atmosphere is very similar to that in the no-ocean case. This explains why the atmospheric loss tends to that in the no-ocean case in the low-loss limit (Figure~\ref{fig:Rloss_Earth}).
 
 In all cases, for sufficiently high ground velocities continued expansion of the ocean, contributed to by a release wave from the top of the atmosphere, leads to a slow acceleration of the ocean surface and loss of the top of the ocean (e.g., Figure~\ref{fig:evo}A, top of the ocean on the right side of the yellow solid line). When the whole atmosphere is lost the ocean effectively releases to zero pressure and the initial pressure of the atmosphere becomes almost irrelevant. The high-loss limit for ocean loss is therefore relatively insensitive to atmospheric pressure.

The time at which the pressure in the ocean becomes less than that at the base of the atmosphere is critical in governing the efficiency of loss. The earlier in time this transition occurs, the earlier the ocean surface and atmosphere are slowed, and the lower the degree of atmospheric/ocean loss. The timing of this transition depends on two factors: the depth of the ocean; and the initial atmospheric pressure. For deeper oceans, the increase in the depth of the ocean layer as the ocean surface expands is a smaller fraction of the total depth of the ocean. The density, and hence pressure, of the ocean therefore falls more slowly. The dependence of loss on atmospheric pressure is more complicated as there are two competing effects. The atmospheric pressure dictates the impedance match pressure and velocity of the ocean surface and base of the atmosphere. The lower the initial atmospheric pressure, the higher the impedance-match velocity of the ocean surface and the greater the driver for atmospheric loss. In addition, the lower the initial atmospheric pressure, the lower the pressure in the atmospheric shock. All else being the same, as the ocean expands it takes longer for the pressure in the ocean to fall below the pressure in the shocked lower atmosphere, leading to slowing of the ocean surface later in time. However, working against this effect is that the higher the surface velocity of the ocean, the more rapidly the ocean expands. The pressure in the ocean decreases more rapidly and so falls below the pressure in the lower atmosphere earlier in time, leading to an earlier slowing of the ocean surface by pressure waves. Determining the balance between these two effects of atmospheric pressure is non-trivial.

The efficiency of atmospheric loss is therefore controlled by the initial depth of the ocean and atmospheric pressure in two principle ways: the atmospheric pressure controls the impedance-match velocity between the ocean and atmosphere that drives enhanced loss; and a combination of the ocean depth and initial atmospheric pressure determine when the ocean surface begins to slow. \cite{Genda2005} previously suggested that loss was dependent on the ratio of the mass of the atmosphere to the mass of the ocean ($\mathcal{R}=M_{\rm atm}/M_{\rm oc}$). For a planet of a given mass, this mass ratio is proportional to the ratio of initial atmospheric pressure to ocean depth ($p_0/H_{\rm oc}$):
\begin{equation}
\mathcal{R} \sim \frac{p_0}{H_{\rm oc}} \frac{1}{g} \; ,
\end{equation}
where $g$ is the gravitational acceleration at the base of the atmosphere. We find that loss correlates well with both $\mathcal{R}$ (Figure~\ref{fig:Rloss_Earth}) and $p_0/H_{\rm oc}$ over a wide range of atmospheric pressures and ocean depths, including across the transition between the low and high loss regimes. This demonstrates the key role that atmospheric pressure and ocean depth play in loss. 

The correlation of loss with $\mathcal{R}$ provides an alternative way of understanding the limits on loss. In the high-$\mathcal{R}$ limit, the ocean is much less massive than the atmosphere and so the release of the ocean cannot provide sufficient momentum to drive any enhancement in atmospheric loss. The efficiency of loss is then the same as if it was just driven by the ground motion alone, and the whole atmosphere, or any amount of ocean, is not lost until the ground velocity is very close to the escape velocity. In the low-$\mathcal{R}$ limit, the ocean is much more massive than the atmosphere and so the atmosphere offers little impediment to the release of the ocean to very low pressures. The transition between the high and low loss regimes occurs when the mass of the atmosphere is comparable to that of the ocean (i.e.,  $\mathcal{R}\sim1$) and takes place over one to two orders of magnitude in $\mathcal{R}$, dependent on the ground velocity. The transition occurs at higher $\mathcal{R}$ as ground velocity increases as the ocean itself begins to be lost and the influence of the atmosphere diminishes. 

We find that considering loss as a function of $\mathcal{R}$ is more useful than $p_0/H_{\rm oc}$ when considering planets with different masses (Section~\ref{sec:results:R:Mp}). We therefore use $\mathcal{R}$ as the principal control on loss from here on out. However, it is important to bear in mind the close relationship between the two ratios.

Scaling with $\mathcal{R}$ does not capture all the factors that influence the relationship between ground velocity and loss. In Figure~\ref{fig:Rloss_Earth}A and B, the dependence of the upper limit of loss on atmospheric pressure is noticeable, but the scaling with $M_{\rm atm}\sim p_0$ of the x-axis means that the offset in loss at a given $\mathcal{R}$ is smaller than at the same $H_{\rm oc}$. In addition, loss in transition between the high and low $\mathcal{R}$ regimes does not scale perfectly with $\mathcal{R}$, particularly in the regime where ocean is being lost, with loss being higher in cases with initially higher pressure atmospheres at the same $\mathcal{R}$. In such cases, the scaling with $M_{\rm atm}\sim p_0$ is over correcting for the effect of atmospheric pressure as the ocean is able to decompress to lower pressures with limited restriction from the atmosphere. Over the range of parameters we have considered, variation in atmospheric pressure leads to differences in loss that are much smaller than the overall $\mathcal{R}$ effect, with the exception of in the ocean loss transition region when the variation due to atmospheric pressure can be $\sim1/3$ of the total variation in loss. If we were to consider cases with high pressure atmospheres but much larger ocean depths the simple scaling with $\mathcal{R}$ would not well describe the value to which loss plateaus in the high-loss regime. The lower impedance-match velocity of higher pressure atmospheres would result in loss plateauing at lower values than for lower pressure atmospheres (Figure~\ref{fig:Hloss_Earth}). This effect would cause deviation on the order of the total variation in loss for cases with a similar $\mathcal{R}$. Considering lower pressure atmospheres than those we examine here could compound this effect as they could plateau at higher loss fractions than the 1~bar minimum pressure we simulated. A wider range of parameters will be considered in future work but, for now, we advise caution when using the results of this work beyond the parameter regime simulated. 

\newpage
%%%%%%%%%%%%%%%%%%%%%%
\subsubsection{Dependence on planetary mass}
\label{sec:results:R:Mp}

\begin{figure*}[t]
    \centering
    \includegraphics[width=0.9\textwidth]{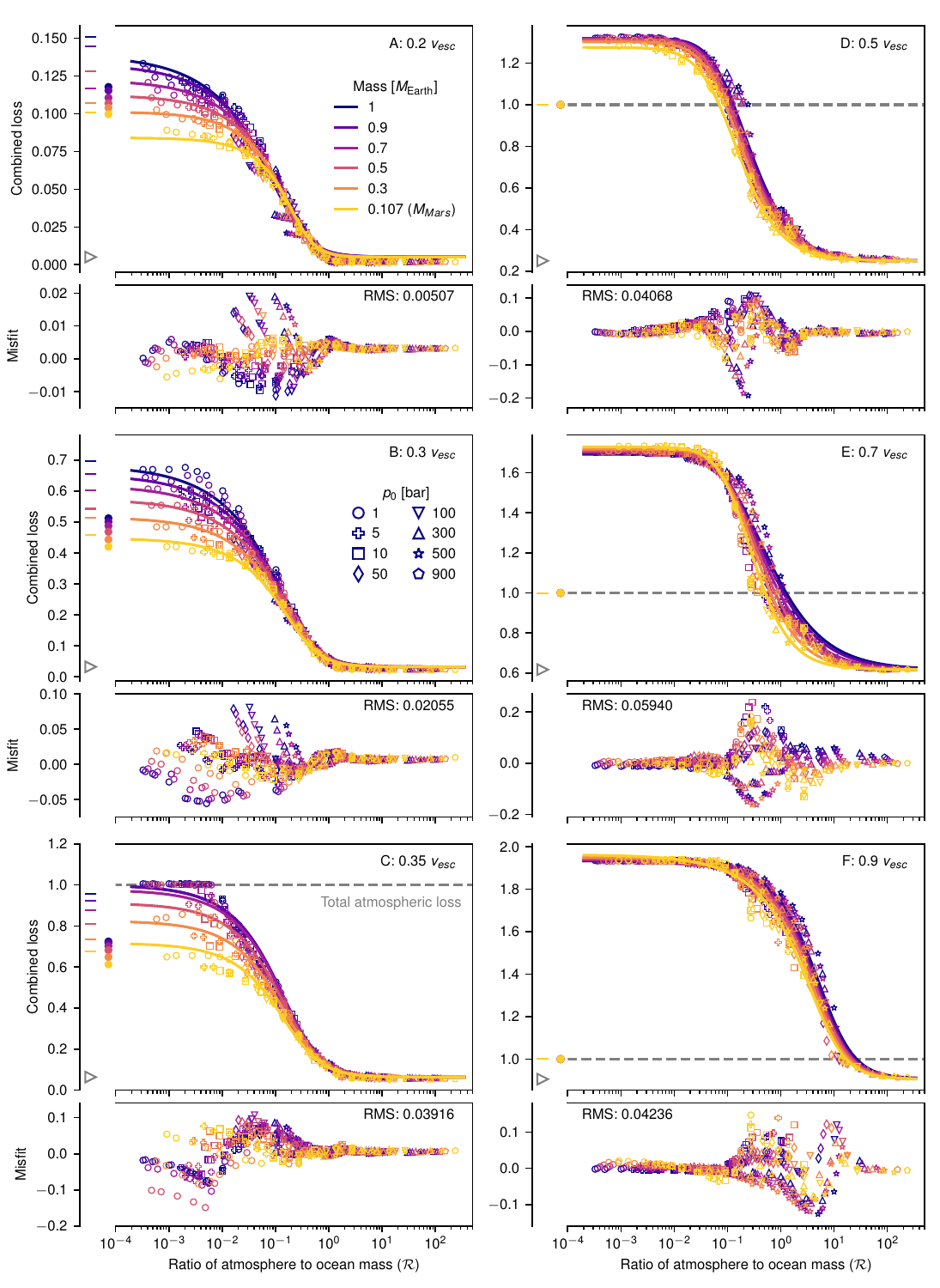}
    \caption{Caption on next page.}
\end{figure*}
\addtocounter{figure}{-1}
\begin{figure*}[t]
    \caption{The efficiency of atmospheric loss depends strongly on the ratio of atmospheric to ocean mass ($\mathcal{R}$). Each panel shows the combined (atmosphere and ocean) loss for a different ground velocity, normalized to the escape velocity of each planet, as a function of $\mathcal{R}$. Points are the results of simulations for a range of planetary masses (colors), initial atmospheric pressures (symbols) and initial ocean depths (decreasing left to right). Colored solid lines show a parameterized fit of the simulation results (Section~\ref{sec:results:fit}) for each planetary mass. Grey dashed lines indicate total atmospheric loss (a combined loss of one). The open grey triangle to the left of each panel shows the loss expected in the absence of an ocean. Filled colored circles to the left of each panel show the loss determined by convolving the velocity of the ocean surface determined from an impedance match solution for an atmosphere of 1~bar (the lowest pressure considered in this work) with a parameterization for atmospheric loss in the case of no ocean (Section~\ref{sec:results:fit}). Colored dash marks to the left of each panel show a similar calculation except using the velocity of the ocean surface expected upon release of the ocean to 1~Pa. In both cases, the color of the symbols indicates the planetary mass. The mass of the planet affects the degree of loss in as it determines the absolute velocity that corresponds to the given fraction of the escape velocity which dictates the absolute impedance match velocity. The lower axis in each panel shows the misfit between the simulations and the parameterized fit. Indicated is the root mean squared (RMS) misfit at each velocity.  }
    \label{fig:Rloss}
\end{figure*}

The scaling of loss with $\mathcal{R}$ holds well, with some caveats, when considering planets of different masses. Figure~\ref{fig:Rloss} shows the combined loss as a function of $\mathcal{R}$ for six example velocities (normalized to $v_{\rm esc}$) and six different mass planets (colors) between the mass of Mars ($M_{\rm Mars}$) and the mass of Earth ($M_{\rm Earth}$). The symbols are the same as in Figures~\ref{fig:Hloss_Earth} and \ref{fig:Rloss_Earth}, and colored lines show the results of our paramterization for atmospheric loss for each mass planet (Section~\ref{sec:results:fit}). The dependence on loss is generally well captured by scaling with $\mathcal{R}$, with the exception that in the low-$\mathcal{R}$ regime when there is only partial atmospheric loss the plateau in loss is lower for lower mass planets. For a ground velocity which is a given fraction of $v_{\rm esc}$, the absolute ground velocity, and hence the strength of the shock, for a lower mass planet is lower. The resulting impedance-match velocity for the ocean-atmosphere interface is a lower fraction of the escape velocity of the smaller planet, leading to less efficient loss (filled symbols to the left of each panel in Figure~\ref{fig:Rloss}). The influence on loss is compounded by the fact that the atmospheric loss function is highly non-linear at low velocities (Figure~\ref{fig:NO}) while the impedance match velocity is relatively linear with respect to absolute velocity. In the regime in which the entire atmosphere is lost, the loss of ocean is controlled by expansion of the ocean to low pressure and the maximum loss is relatively insensitive to planetary mass. 

There is also substantial deviation from a simple $\mathcal{R}$ scaling in the transition region between the low and high loss regimes, with loss from more massive planets being more efficient. This variation is likely largely due to the sensitivity of the dynamics of loss to the shock and release path of water which itself is a function of the absolute strength of the shock. The variation due to planetary mass is compounded by the variation due to initial atmospheric pressure/ocean depth (Section~\ref{sec:results:R:R}) and the difference in combined loss can be as much as 50\% at the same $\mathcal{R}$ in regions where the loss fraction is varying rapidly with $\mathcal{R}$. These variations are the main cause of error in our parameterization of loss (Section~\ref{sec:results:fit}).

%%%%%%%%%%%%%%%%%%%%%%
\subsubsection{Effect of atmospheric composition}
\label{sec:results:R:comp}

\begin{figure}
    \centering
    \includegraphics[width=\columnwidth]{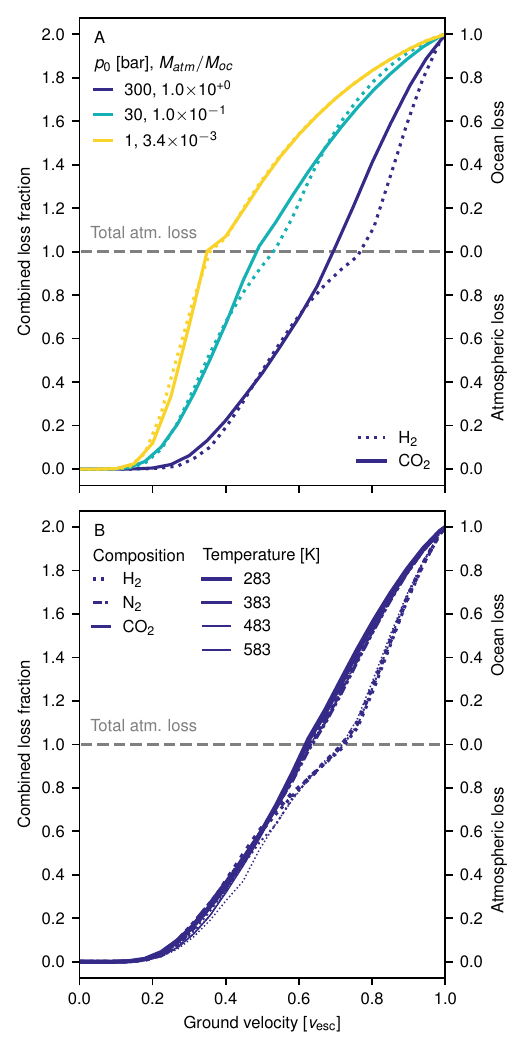}
    \caption{Loss in the ocean case is somewhat sensitive to the composition of the atmosphere. A: Atmosphere and ocean loss from an Earth-mass body as a function of ground velocity for H$_2$ (dotted lines) and CO$_2$ (solid lines) atmospheres of different surface pressures (colors) above 3~km oceans.  B: Loss for atmospheres of different compositions (line styles) and ocean surface temperatures (line thicknesses) for a 100~bar atmosphere over a 3~km ocean.
    }
    \label{fig:atmo_params}
\end{figure}

In the presence of an ocean, the effect of atmospheric composition on the relationship between ground velocity and loss is complex as the composition of the atmosphere can affect the efficiency of loss in a number of different ways. First, the composition of the atmosphere controls the compressibility of the shocked gas, and  the impedance match velocity and pressure of the ocean surface. For an ideal gas in the strong-shock limit
\begin{equation}
    \frac{\rho_{\rm s}}{\rho_0} \approx \frac{\gamma +1}{\gamma - 1} \; ,
\end{equation}
where $\rho_{\rm s}$ and $\rho_0$ are the density of the shocked and unshocked material, respectively, and $\gamma$ is the ratio of specific heat capacities of the gas. The compression of material due to the shock is entirely dependent on $\gamma$ which varies from 1.25 to 1.4 for the gases we consider, resulting in a variation of density in the shocked gas of 6 to 9. The particle velocity, $u_{\rm p}$, at a given shock pressure, $p_{\rm s}$, in the strong shock limit is 
\begin{equation}
    u_{\rm p} \approx \sqrt{\frac{p_{\rm s}}{\rho_0} \left ( \frac{2}{\gamma +1} \right )} \; .
\end{equation}
The lower the initial density of the gas, the higher the particle velocity required to reach a given shock pressure. As a result, the impedance match velocity at the ocean surface is higher for lighter gases while the impedance match pressure is lower. In particular, H$_2$ is much less dense than gases such as N$_2$ and CO$_2$ and the initial velocity of the ocean surface can be tens of percent larger for H$_2$ than for the heavier gases. The lower pressure of the impedance match solution in such cases means that the pressure at the base of the ocean can remain above that at the base of the atmosphere for longer, helping sustain the velocity of the ocean surface. Finally, the height of the atmosphere, and hence the time it takes for the shock to reach the top of the atmosphere, can vary significantly depending on its composition. For example, it can take the shock more than ten times longer to reach the top of a H$_2$ atmosphere than a CO$_2$ atmosphere. As a result, the surface of the ocean, and the ground, slow earlier in the evolution relative to the progress of the shock through the atmosphere. So, although the velocity of the ocean surface is sustained for longer for H$_2$ atmospheres in absolute time, the ocean surface is typically slowed earlier relative to the evolution of the atmosphere, acting to reduce the efficiency of loss. Finally, the release wave from the top of the atmosphere reaches the ocean surface earlier during the loss of \co2 atmospheres compared to \h2 atmospheres and so the pressure in the ocean drops more rapidly and the ocean reaches higher velocities earlier in the evolution. 

The net effect of these competing factors depends on the initial conditions and the regime of loss, and can be difficult to isolate. For example, Figure~\ref{fig:atmo_params}A shows the same cases as in Figure~\ref{fig:ug_loss}A for both H$_2$ and CO$_2$ atmospheres which show some of the common tradeoffs. The loss for different heavy atmospheres (e.g., N$_2$ and CO$_2$) is very similar, but the loss of H$_2$ atmospheres can be significantly different. In the high loss limit (e.g., yellow curves in Figure~\ref{fig:atmo_params}A), the shock in the atmosphere is supported late into the evolution in both the CO$_2$ and H$_2$ cases and the efficiency of loss is very similar. In the regime where only part of the atmosphere is lost, loss in the H$_2$ case is slightly greater due to the larger impedance match velocity. However, this effect is muted by the continued decompression of the ocean surface and the ocean surface in the \co2 case reaches greater velocities than in the H$_2$ cases within a few seconds. In the high-loss regime when the whole atmosphere is lost, the loss of ocean in the CO$_2$ and H$_2$ cases are nearly identical as the low-mass atmosphere provides little impediment to the loss of the ocean, regardless of atmospheric composition.

In the transition between the high and low loss limits (e.g., teal and blue lines in Figure~\ref{fig:atmo_params}A) the differences between CO$_2$ and H$_2$ atmospheres are more complex. At low ground velocities, loss of CO$_2$ atmospheres is more efficient largely because the ocean surface in the H$_2$ cases slows significantly before the shock reaches the top of the atmosphere. There is then an intermediate ground velocity regime where the situation is similar to that in the high loss limit and loss of H$_2$ atmospheres is more efficient. Close to near total loss of the atmosphere, and continuing into the ocean-loss regime, loss of \CO2 atmospheres and their oceans once again becomes more efficient. This effect is likely a result of acceleration caused by the more rapid drop in the pressure of the ocean in the \co2 cases allowing the ocean to reach higher velocities. In cases where the ocean is almost entirely lost, the situation can reverse (e.g., teal line in Figure~\ref{fig:atmo_params}A). This may indicate that the higher impedance match velocity in the \h2 case controls the loss of the deep ocean.

The difference in loss between \co2 and \h2 atmospheres can be tens of percent, particularly in the ocean-loss regime. However, for the range of parameters we have explored, the loss of \h2 atmospheres is within, if sometimes at the extreme, of the variation we see between different $H_{\rm oc}$, $p_0$ and $M_{\rm p}$ combinations for CO$_2$ atmospheres with similar \r ~(Figure~\ref{fig:Rloss}). Therefore, although the effect of atmospheric composition can be significant, it is a second order effect compared to the dominant $\mathcal{R}$ scaling.

%%%%%%%%%%%%%%%%%%%%%%
\subsubsection{Effect of surface temperature}
\label{sec:results:R:T}

The range of possible surface temperatures in the ocean case is limited to the range over which liquid water is stable on the surface (we do not consider an ice-covered surface in this work). The exact temperature range depends on the initial atmospheric pressure but, over the range considered here, the effect of atmospheric temperature on loss is minor. Figure~\ref{fig:atmo_params}B shows loss for an example case with H$_2$, N$_2$, and CO$_2$ atmospheres and surface temperatures between 283 and 583~K. Higher temperature atmospheres typically lead to slightly less efficient loss, likely because the greater atmospheric height means that the atmospheric shock becomes unsupported earlier in the evolution (see Section~\ref{sec:results:R:comp}). This effect is only really noticeable for \h2 atmospheres in the atmospheric loss regime (e.g., see thin dotted lines in Figure~\ref{fig:atmo_params}B)  where the efficiency of loss can be a few percent lower than in colder cases. 

%xxxxxxxxxxxxxxxxxxxxxxxxxxxxxxxxxxxxxxxxxxxxxxxxxxxxxxxxxxxxxxxxxxxxxxxxxxxxxxxxxxxxxxxxxxxxxxxxxxxxxxxxxxxxxxxxxxxxxxxxxxxxxxxxxxxxxxxx
%xxxxxxxxxxxxxxxxxxxxxxxxxxxxxxxxxxxxxxxxxxxxxxxxxxxxxxxxxxxxxxxxxxxxxxxxxxxxxxxxxxxxxxxxxxxxxxxxxxxxxxxxxxxxxxxxxxxxxxxxxxxxxxxxxxxxxxxx
%xxxxxxxxxxxxxxxxxxxxxxxxxxxxxxxxxxxxxxxxxxxxxxxxxxxxxxxxxxxxxxxxxxxxxxxxxxxxxxxxxxxxxxxxxxxxxxxxxxxxxxxxxxxxxxxxxxxxxxxxxxxxxxxxxxxxxxxx
\section{The relationship between shock strength and ground velocity and the implications for the efficiency of loss}
\label{sec:results:up_ug_relation_ocean}
\label{sec:results:up_ug_relation_NO}

So far we have considered the effect of the surface conditions on the efficiency of atmospheric and ocean loss for a specified ground velocity. However, as discussed in Section~\ref{sec:bckgrnd}, the impedance-match ground velocity that results from a shock wave of a given strength within a planet depends strongly on parameters such as the initial surface pressure and temperature. We now turn to consider how the surface conditions influence the ground velocity due to a given impact and the effect on atmospheric and ocean loss. In this section we will only consider the magnitude of the initial ground velocity and return to discuss non-ballistic effects on the ground velocity in Section~\ref{sec:discussion:issues}.

For a given impact, the strength of the shock inside the planet before breakout is insensitive to the properties of a thin atmosphere and ocean. In this section, we will therefore use the particle velocity in the shock in the planet, prior to breakout into the atmosphere/ocean, as the independent parameter when comparing loss from planets with different surface conditions. In other words, we will consider how much of the atmosphere/ocean would be lost from a given impact if the only parameters that changed were the initial surface conditions. 

%xxxxxxxxxxxxxxxxxxxxxxxxxxxxxxxxxxxxxxxxxxxxxxxxxxxxxxxxxxxxxxxx
%xxxxxxxxxxxxxxxxxxxxxxxxxxxxxxxxxxxxxxxxxxxxxxxxxxxxxxxxxxxxxxxx
\subsection{The no-ocean case}
\label{sec:results:up_ug_relation:NO}

\begin{figure*}[t]
    \includegraphics[width=0.97\textheight,angle=90,origin=c]{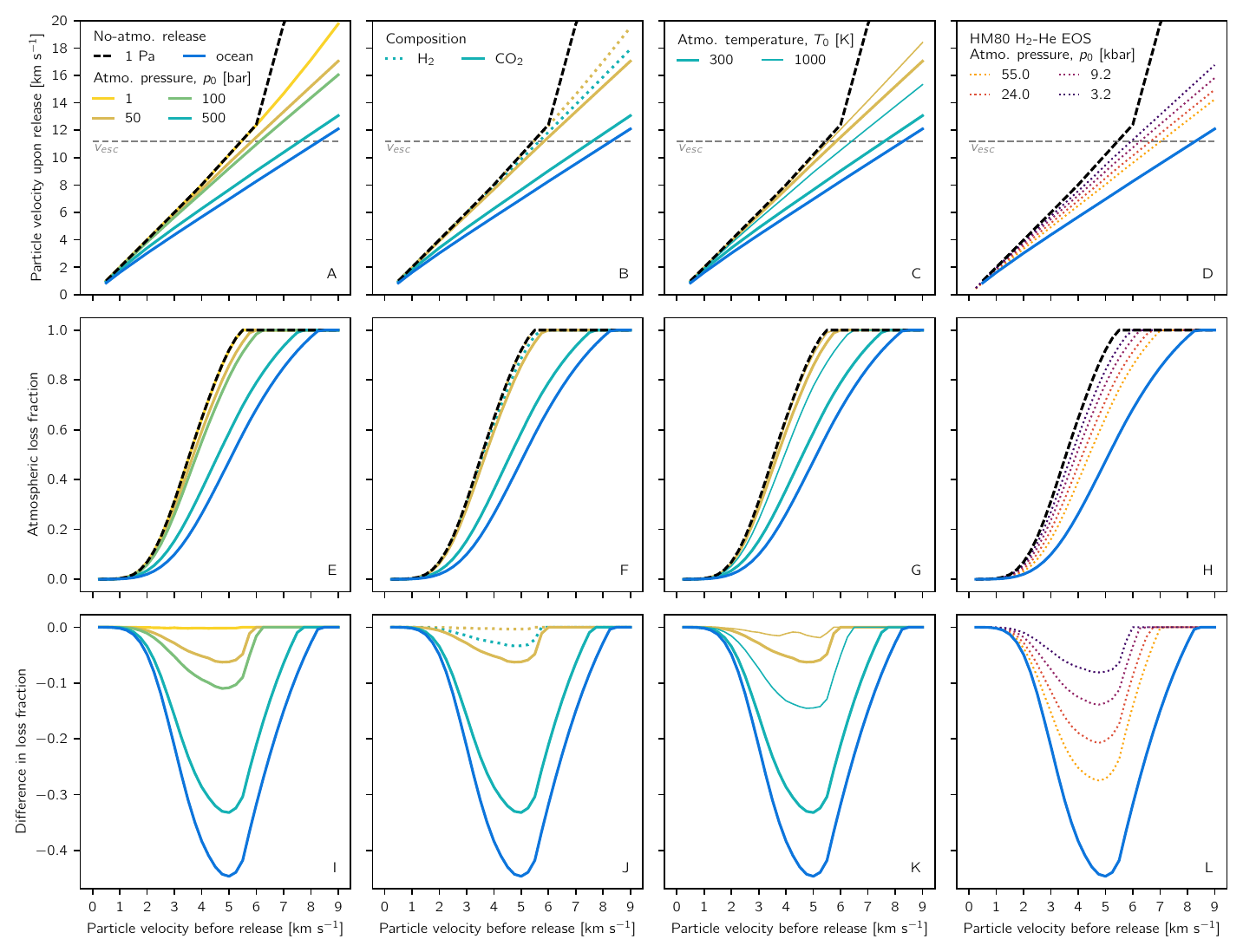}
  \caption{Caption on next page.}

\end{figure*}
\addtocounter{figure}{-1}
\begin{figure*}[t]

    \caption{ The initial velocity of the ground or ocean surface depends on the properties of the atmosphere. A-C: Release velocity of the ground, as determined by an impedance match calculation, as a function of the particle velocity of the shock in the planet before release for atmospheres with different pressures (A, colors), compositions (B, line styles), and temperatures (C, line thicknesses). The particle velocity of the ground on release to very low pressures (1~Pa: black dashed line), and the impedance match velocity for release of the ground into a water ocean (deep blue line) are also shown. The surface was modelled as forsterite \citep{Stewart2019forsteriteEOS}, the atmospheres as ideal gases, and the ocean using the \cite{Senft2008} EOS. The sudden increase in the slope of the low-pressure release line is due to the onset of vaporization. D: As A-C but for atmospheres modelled using the H$_2$-He EOS of \citet{Hubbard1980} with the initial pressures and temperatures (500~K) as used in the 3D loss simulations of \citet{Kegerreis2019}. Note that the implementation of the \citet{Hubbard1980} in \citet{Kegerreis2019} had an inconsistency in the calculation of internal energy which we have corrected here (see Section~\ref{sec:methods}). E-H: The fraction of atmosphere lost due to the breakout of a shock as a function of particle velocity in the planet before release to the atmosphere for the atmospheres shown in A-D. Calculations are for loss from an Earth-mass planet. Loss was calculated by inputting the calculated impedance match velocity for a each atmosphere and shock strength into our parameterization for loss in the no-ocean case (Section~\ref{sec:results:fit}). The loss that could be driven by release of the ground to very low pressures (1~Pa) is shown as a black dashed line. I-L: The fractional difference between the loss calculated for the atmospheres shown in E-H and the loss calculated assuming that the ground released to very low pressures (1~Pa, a theoretical upper limit on loss, black dashed line in A-D). For low-pressure, high temperature, and lighter atmospheres the loss is close to maximal, but loss of heavier and higher pressures atmospheres can be tens of percent less.}
            \label{fig:imp_match_vdiff}
            \label{fig:imp_match_loss_diff_NO}

\end{figure*}

% B: The release velocity of the ocean surface upon breakout of the shock into the atmosphere as a function of the particle velocity of the shock in the planet before release. Labelling is the same as in A, with the release of the ocean surface to very low pressure (1~Pa) shown as a darker blue dashed line. The variation in atmospheric temperature for which it is possible to have a liquid water ocean is small and so the difference in the particle velocity upon release is small. We have therefore not shown different temperature atmospheres for clarity. The resulting surface velocity is much larger than in the no-ocean case, and can vary depending on atmospheric pressure and composition. However, continued decompression of the ocean after the initial release of the shock from the ocean-atmosphere interface can result in the surface velocity tending to that in the no-atmosphere case regardless of the properties of the atmosphere, reducing the impact of the variation due to atmospheric properties (Section~\ref{sec:results:R}).

\begin{figure}
        \centering
    \includegraphics[width=\columnwidth]{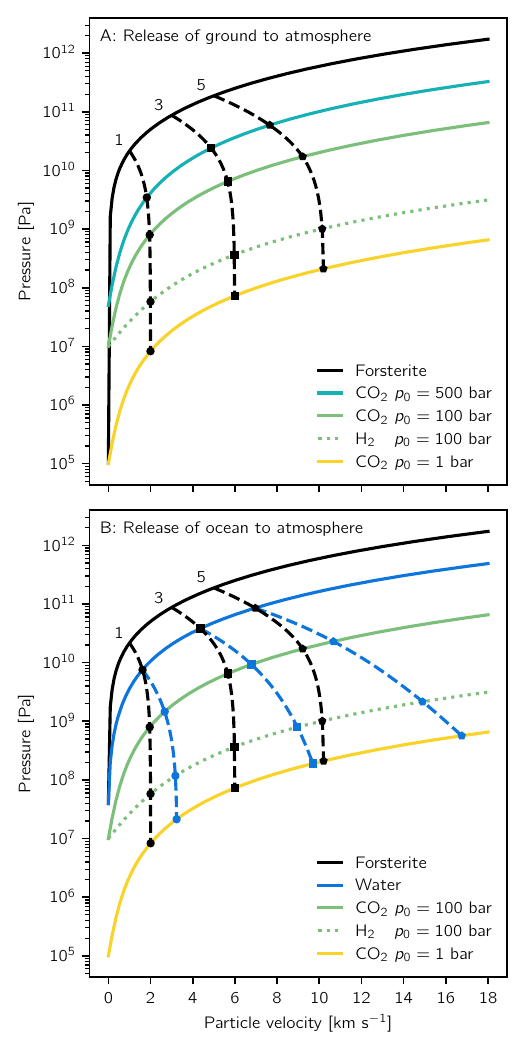}
    \caption{Caption opposite.}

\end{figure}
\addtocounter{figure}{-1}
\begin{figure}
    \caption{The initial velocity of the surface that is driving atmospheric loss depends on the properties of the atmosphere. A: Impedance match solution in the no-ocean case. Solid or dotted lines show the Hugoniots (the locus of physical states produced by shock waves in a material) for forsterite (black, a proxy for the planet's surface) and atmospheres of varying composition and initial pressure. Black dashed lines show the path for isentropic release of the forsterite shocked to particle velocities of 1, 3, and 5~km~s$^{-1}$. The pressure and velocity of both the planet's surface and the atmosphere upon breakout of the shock is set by the intersection of the forsterite release curve and the Hugoniot of the relevant atmosphere (black symbols). The impedance match ground velocity is higher for lower pressure, lower molecular weight, and hotter atmospheres (Figure~\ref{fig:imp_match_vdiff}A-C). B: Impedance match solution in the presence of an ocean. Lines and symbols are the same as in A with the addition of the shock Hugoniot of water (blue solid line), isentropic release curves for water (blue dashed lines), and blue symbols marking the impedance match solutions for the ocean with different atmospheres. The release curve of water is shallower that that of forsterite and intersects the atmosphere Hugoniots at higher particle velocities than the rock release curve. Similarly to the no-ocean case, lower pressure and lighter atmospheres lead to larger initial velocities of the ocean surface (Figure~\ref{fig:imp_match_loss_diff}A).
    }
        \label{fig:imp_match_up_p}
\end{figure}

As discussed in Section~\ref{sec:bckgrnd}, in the no-ocean case, the ground velocity is set by an impedance match between the rock surface and the atmosphere. Figure~\ref{fig:imp_match_vdiff}A-C shows the ground velocity as a function of the particle velocity in the shock in the planet before breakout for atmospheres with different pressures, compositions, and temperature (colored lines). For reference, the particle velocity of the ground on release to very low pressures (1~Pa: black dashed line), and the impedance match velocity for release of the ground into an ocean (deeper blue line) are also shown. For high-temperature,  low-pressure atmospheres the ground velocity is close to the low-pressure release velocity, as the impedance match pressure is low enough that the forsterite release curve is very steep in $u_p$-$p$ space by the time it intersects the shock Hugoniot of the gas. This can be seen in the example impedance-match calculations show in $u_p$-$p$ space in Figure~\ref{fig:imp_match_up_p}A (similar to the schematics in Figure~\ref{sup:fig:impedance_match_cartoon_NO} and \ref{fig:impedance_match_cartoon}) where the black dashed line show rock release curves and black symbols show the impedance-match solutions. For the same strength of shock in the planet, the impedance match velocity decreases with increasing atmospheric pressure, and decreasing temperature and is also lower for heavier atmospheres (Figure~\ref{fig:imp_match_vdiff}B). Therefore, even though the dependence of loss on the ground velocity is largely insensitive to the atmospheric properties (Section~\ref{sec:results:NO} and Figure~\ref{fig:NO}), it is easier to lose hotter, lighter, and lower pressure atmospheres from terrestrial planets due to the higher ground velocity resulting from any given impact.

The effect of the varying impedance-match velocities on the magnitude of loss is shown in Figure~\ref{fig:imp_match_loss_diff_NO}. Panels E-G show the loss calculated by inputting the impedance-match velocities shown in  Figure~\ref{fig:imp_match_vdiff}A-C into our parameterization of the relationship between ground velocity and atmospheric loss (Section~\ref{sec:results:fit}). Panel H shows similar results but for the four atmospheres considered in the 3D loss simulations of \citet{Kegerreis2019} utilizing the EOS of \citet{Hubbard1980}. Panels I-L show the difference between the calculated loss from each of the atmospheres and the loss that would be driven by release of the surface to very low pressures (1~Pa, black dashed line in panels A-H). Figure~\ref{fig:imp_match_loss_diff_NO} also shows an ocean line which we will discuss in Section~\ref{sec:results:up_ug_relation:ocean}.  The efficiency of loss for a given strength of impact shock can vary by 10s~\% between different atmospheres (Figure~\ref{fig:imp_match_loss_diff}I-L), purely as a function of the difference in the impedance match velocity between the ground and the atmosphere. The ability of giant impacts to remove volatiles from planets is therefore dictated, in part, by the surface conditions on the colliding bodies before the impact.

As a result of the dependence of ground velocity on atmospheric properties, care needs to be taken when combining $u_{\rm G}$-loss scalings (Section~\ref{sec:results:fit}) with the results of 3D impact simulations and when applying the results of 3D impact loss simulations to accretion models. The peak ground motion in a 3D impact simulation will be dictated by the pressure of the atmosphere, or the lack of atmosphere, used in that simulation.  When combining 1D simulations with the ground velocity from 3D impact simulations \cite[see e.g.,][]{Kegerreis2019} to calculate the loss expected from a planet with a different atmosphere than that used in the 3D simulation, the peak ground motion must be corrected to that dictated by the impedance match with the alternative atmosphere. Furthermore, results of directly calculated 3D loss simulations from a planet with an atmosphere of one composition, pressure, and temperature cannot necessarily be applied to loss from a planet with a different atmosphere without significant error. For example, the difference in the loss efficiency curves shown in Figure~\ref{fig:imp_match_loss_diff_NO}~H and L may be at least partly responsible for the difference in atmospheric loss between different mass atmospheres observed by \citet{Kegerreis2019} (e.g., their Figure~7). The dependence of ground velocity on atmospheric properties complicates efforts to develop universal scaling laws for atmospheric loss.

%xxxxxxxxxxxxxxxxxxxxxxxxxxxxxxxxxxxxxxxxxxxxxxxxxxxxxxxxxxxxxxxx
%xxxxxxxxxxxxxxxxxxxxxxxxxxxxxxxxxxxxxxxxxxxxxxxxxxxxxxxxxxxxxxxx
\subsection{The ocean case}
\label{sec:results:up_ug_relation:ocean}

Figure~\ref{fig:imp_match_loss_diff}A shows the impedance match velocity between the ocean and the atmosphere for a given strength of shock in the planet. The impedance match velocities are much larger than that between the ground and atmosphere in the no-ocean case (Figure~\ref{fig:imp_match_vdiff}A-D), explaining the significant increase of loss achieved in the high-loss regime (Figure~\ref{fig:Rloss}). However, as we discussed in Section~\ref{sec:results:R}, the loss efficiency can be significantly perturbed from that expected from an impedance-match calculation. 

When there is an ocean on the pre-impact body, the ground velocity is set, not by release of the ground to the atmosphere, but by an impedance match between the rock surface and the bottom of the ocean. The darker blue line in Figure~\ref{fig:imp_match_vdiff}A-D shows the ground velocity upon release to the ocean as a function of the shock particle velocity in the planet before breakout. Note that, although the Hugoniot depends on the initial pressure/density of the material, over the range of ocean depths and atmospheric pressures considered in this paper, the forsterite and water Hugoniots are similar and there is little variation in the ground velocity upon release. The ground velocity in the ocean case can be tens of percent lower than the ground velocity in the no-ocean case for the same impact (Figure~\ref{fig:imp_match_vdiff}I-L). 

The reduced ground velocity in the ocean case can lead to a previously unrecognized phenomenon: the presence of an ocean can actually reduce the efficiency of loss \citep[c.f.,][]{Genda2005}. In the large atmosphere/ocean mass ratio, low-loss regime the loss efficiency is the same as the loss in the no-ocean case for the same ground velocity. However, the ground velocity in the ocean case, for a given impact, is lower than in the no-ocean case. Therefore, the amount of atmospheric loss in a given impact will be lower if an ocean is present than if it were absent. Figure~\ref{fig:imp_match_loss_diff}B shows the fraction of atmosphere lost due to a given particle velocity depending on the ocean to atmosphere mass ratio. The black dashed line shows the equivalent relationship for low-pressure atmospheres in the absence of an ocean. The atmosphere to ocean mass ratio needs to be sufficiently small in order for the loss efficiency in the ocean case to exceed that in the no-ocean case. Figure~\ref{fig:imp_match_loss_diff}C shows the critical mass ratio required for loss in the ocean case to equal that in the no-ocean case, assuming release of the ground to very low pressure ($\sim1$~Pa). The mass of the ocean must be at least comparable to the mass of the atmosphere in order for the presence of an ocean to enhance loss, over that in the no-ocean case. The critical mass ratio is lower for higher pressure atmospheres as the release velocity is comparatively lower in the no-ocean case.

\begin{figure}
\centering
    \includegraphics[width=\columnwidth]{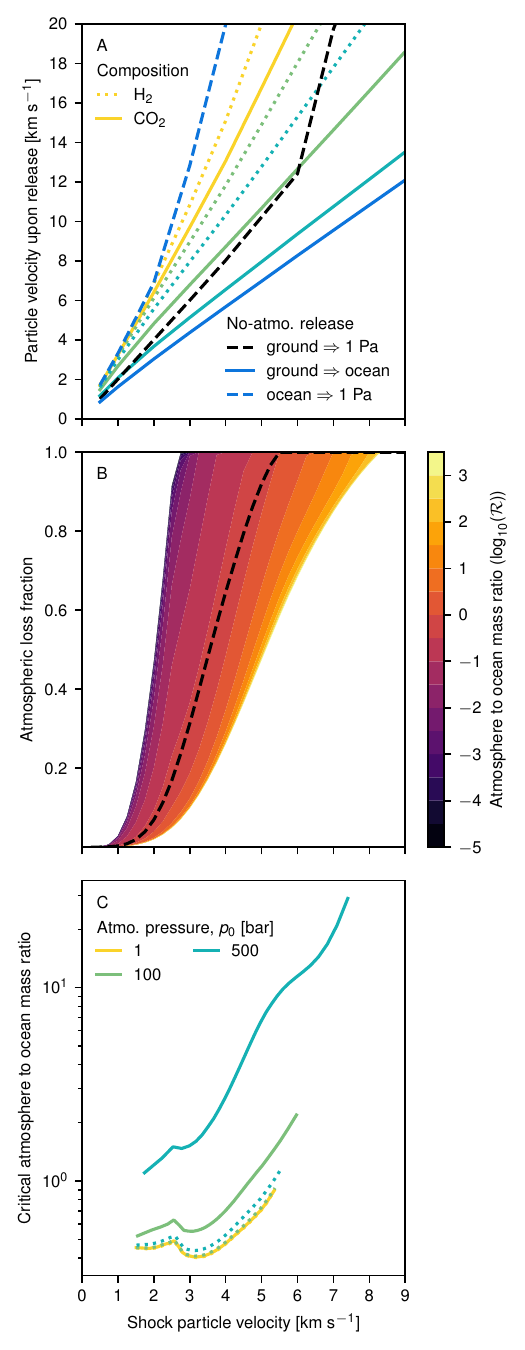}
    \caption{Caption opposite.}

\end{figure}
\addtocounter{figure}{-1}
\begin{figure}
    \caption{The presence of an ocean can both increase and decrease the efficiency of atmospheric loss. A: Release velocity of the ocean surface, as determined by an impedance match calculation, as a function of the particle velocity of the shock in the planet before release to the base of the ocean for atmospheres with different pressures (colors), and compositions (line styles). The particle velocity of the ground on release to very low pressures (1~Pa: black dashed line), the impedance match velocity for release of the ground into a water ocean (deep blue solid line), and the impedance match velocity of the ocean when released to low pressure (1~Pa: blue dashed line) are also shown. B: The fraction of atmosphere lost due to the breakout of a shock wave of a given particle velocity before release to the ocean, for different ratios of atmospheric to ocean mass ($\mathcal{R}$). Calculations are for an Earth-mass planet.  Loss was calculated by convolving our parameterization for loss in the ocean case (Section~\ref{sec:results:fit}) with the impedance match velocity on release of the shock from the planet to the base of the ocean. The loss that could be driven by release of the ground to very low pressures (1~Pa) is shown as a black dashed line (as in Figure~\ref{fig:imp_match_loss_diff_NO}). C: The atmosphere to ocean mass ratio required for a shock of the same strength to remove the same amount of atmosphere as in the no-ocean case. Bodies with a higher atmosphere to ocean mass ratios than this critical ratio will experience less efficient loss than in the no-ocean case.}
            \label{fig:imp_match_loss_diff}

\end{figure}

%xxxxxxxxxxxxxxxxxxxxxxxxxxxxxxxxxxxxxxxxxxxxxxxxxxxxxxxxxxxxxxxxxxxxxxxxxxxxxxxxxxxxxxxxxxxxxxxxxxxxxxxxxxxxxxxxxxxxxxxxxxxxxxxxxxxxxxxx
%xxxxxxxxxxxxxxxxxxxxxxxxxxxxxxxxxxxxxxxxxxxxxxxxxxxxxxxxxxxxxxxxxxxxxxxxxxxxxxxxxxxxxxxxxxxxxxxxxxxxxxxxxxxxxxxxxxxxxxxxxxxxxxxxxxxxxxxx
%xxxxxxxxxxxxxxxxxxxxxxxxxxxxxxxxxxxxxxxxxxxxxxxxxxxxxxxxxxxxxxxxxxxxxxxxxxxxxxxxxxxxxxxxxxxxxxxxxxxxxxxxxxxxxxxxxxxxxxxxxxxxxxxxxxxxxxxx
\section{Non-ballistic boundary conditions}
\label{sec:discussion:issues}
\label{sup:sec:bound_change}

As in previous work \citep{Genda2003,Genda2005}, we have not directly simulated the shock in the planet. Instead, in most of our simulations, we have modeled the breakout of the shock from the planet to the atmosphere/ocean by prescribing the motion of the ground, assuming that the ground reaches its peak velocity instantaneously and then evolving ballistically. However, there are multiple processes that could perturb the motion of the ground from this ideal that we must consider.

\begin{figure}
    \centering
    \includegraphics[width=\columnwidth]{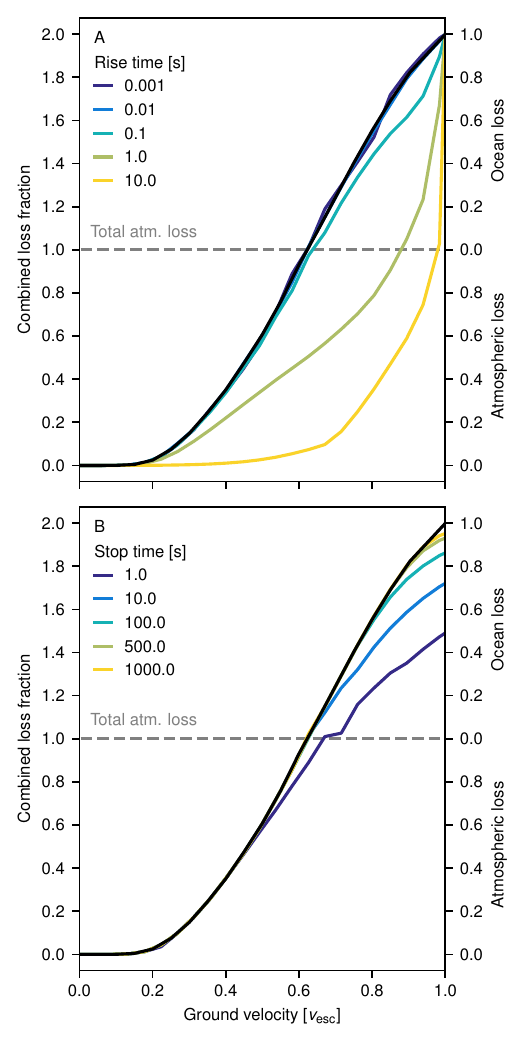}
    \caption{Non-ballistic motion of the ground can affect the efficiency of loss. Shown is loss as a function of maximum ground velocity with different rise times (A) and stopping times (B) for the ground motion. Loss was calculated for an Earth-mass planet with a 3~km ocean and an atmospheric pressure of 100~bar. For reference, the instantaneous rise time solution is shown in black. Axis labels are the same as in Figure~\ref{fig:ug_loss}.
    }
    \label{fig:rise_stop_time}
\end{figure}

Firstly, the acceleration of the ground due to the release of the shock to the atmosphere/ocean will occur over some finite rise time. That is, there is a finite rise time over which any point on the surface of the planet accelerates from stationary to the impedance-match velocity. Experimental studies have found that rise times for shocks in silicates are on the order of $10^{-7}$~s \citep[e.g.,][]{Grady1987}. However, rises times in the shocks from nuclear explosions have been found to be much longer, up to several $10^{-2}$~s \citep{Melosh2003}. \cite{Genda2003} investigated the effect of rise time on loss in the no-ocean case and found that rise times less than $\sim1$~s had only minimal effect on loss for a ground velocity of 5.6~km~s$^{-1}$ on an Earth-mass planet with an atmosphere similar to the present-day Earth ($m_{\rm a}=29$~g~mol$^{-1}$, $\gamma=1.4$). To further examine the effect of rise time on loss, we ran simulations in which the boundary accelerated linearly over a given rise time, $t_{\rm rise}$ (Appendix~\ref{sup:sec:boundary}). We confirm the result of \cite{Genda2003} in the no-ocean case, but find there can still be a substantial decrease in loss at higher ground velocities for rise times $>\sim 0.1$~s. The rise times required to effect loss in H$_2$ atmospheres are much longer ($>\sim$10~s), likely due to the larger scale height of the atmosphere and hence longer timescale for evolution. Regardless, we concur with \cite{Genda2003} that the rise times typical of shocks are too short to significantly impact the efficiency of loss. Similarly, we find that in cases with an ocean rise times typical of the rise time of shocks have minimal effect on loss (Figure~\ref{fig:rise_stop_time}A). 

The key limitation to our approach is that in prescribing that the maximum velocity of the ground is reached instantaneously and that any reduction in pressure at the base of the atmosphere/ocean below the impedance match pressure does not lead to any additional acceleration of the ground. In effect, we are assuming that the release curve of the rock is vertical in $u_{\rm p}$-$p$ space when the impedance match is reached (similar to the $u_{\rm p}=1$~km~s$^{-1}$ curve in Figure~\ref{fig:imp_match_up_p}). This is a very good assumption for low pressure atmospheres and relatively weak shocks in the absence of an ocean, but for higher pressure atmospheres and stronger shocks, or for release into an ocean, the release curve of the rock intersects the air or water Hugoniot while it still has a significant slope in $u_{\rm p}$-$p$ space (see e.g., 5~km~s$^{-1}$ example in Figure~\ref{fig:imp_match_up_p}A, B). Therefore, when the base of the atmosphere/ocean decompresses as it expands outwards, the decrease in pressure at the base of the atmosphere/ocean will cause the ground to accelerate to velocities beyond the initial impedance match. The increase in ground velocity is particularly large for very strong shocks when the ground can vaporize upon release. This later acceleration has the potential increase the efficiency of loss. 

Direct simulation of the planet's surface, or parameterization of the ground motion, is beyond the scope of this paper. However, we can examine the potential effect on atmospheric loss of continued acceleration of the ground by calculating the anticipated velocity of the ground based on the forsterite release curve and the base ocean/atmosphere pressure from our simulations, and comparing this to the results of our simulations with finite rise times. It is important to note that our finite-rise time simulations are not capturing the full conditions of continued acceleration of the boundary, as in the finite-rise time simulations the boundary is accelerating at a constant rate into an initially hydrostatic atmosphere/ocean rather than an atmosphere/ocean that have already been significantly perturbed, and accelerated, by the initial shock. The finite rise-time simulations are therefore likely to overestimate the loss for the same timing of late acceleration of the ground. 

We find that the effect of rise time is intrinsically linked to the decompression of the base of the ocean/atmosphere, and hence the anticipated acceleration of the ground. Accelerations  of the ground coincident with or after the release of the base of the ocean/atmosphere typically lead to much less additional loss than accelerations that occur earlier. This is likely due to the fact that any additional pressure/shock waves driven by the accelerating boundary do not travel as quickly in the decompressed ocean/atmosphere and so do not accelerate as quickly up the much shallower post-release pressure gradient in the atmosphere. It is therefore likely that additional acceleration of the boundary upon release would have only a modest effect on the efficiency of atmospheric loss. 

The pressure evolution of the bottom of the ocean/atmosphere can vary significantly between different surface conditions and shock strengths, and including the effects of continued acceleration of the boundary will likely add significant complexity to the parameterization of atmospheric and ocean loss. We intend to explore these effects in future work. In the meantime, loss from a given impact can be bounded by convolving our parameterization of loss with both the impedance match velocity and the ground velocity upon release to low pressure.

Finally, the ballistic assumption could be violated by slowing of the ground earlier than under the ballistic assumption. This is expected due to release of the shock in the planet by release waves from elsewhere on the surface. Figure~\ref{fig:rise_stop_time}B  shows loss as a function of ground velocity for an example case, where the motion of the ground has been stopped at different times after the start of the simulation (colors). In the regime of atmospheric loss, stopping the boundary after the initial release of the shock (10$^{-2}$-10$^2$ for the range of ocean depths considered here) causes very little change in loss efficiency, even though it can take tens of seconds for the fraction of the atmosphere that is lost to be accelerated to escape. This is because the acceleration of the atmosphere is supported by the expansion of the ocean. The passage time of the shock across the ocean in the example case shown in Figure~\ref{fig:rise_stop_time}B is $<1$~s. If the ocean begins to expand before the boundary is stopped, and before the shock in the ocean is no longer supported by the ground, the expanding ocean continues to support the shock in the atmosphere until much later in evolution. In the ocean loss regime, stopping the boundary later in time (up to a few hundred seconds) can still have a significant affect on loss, with the effect greatest at the highest ground velocities (Figure~\ref{fig:rise_stop_time}B), despite the lost ocean fraction reaching escape within a few seconds. When the boundary slows, release waves propagating upwards, slowing the escaping ocean and reducing the fraction that is lost. In giant impacts, the time between the breakout of the shock and release of the impact shock varies across the planet. When determining the loss from 3D impact simulations using the results of 1D simulations, it is important to account for the fact that loss of ocean will be less efficient from areas where the shock in the planet is released in less than a few hundred seconds. Such points are, however, likely near the impact site where a lot of additional processes are occurring, and the 1D approximation is potentially invalid regardless.

%xxxxxxxxxxxxxxxxxxxxxxxxxxxxxxxxxxxxxxxxxxxxxxxxxxxxxxxxxxxxxxxxxxxxxxxxxxxxxxxxxxxxxxxxxxxxxxxxxxxxxxxxxxxxxxxxxxxxxxxxxxxxxxxxxxxxxxxx
%xxxxxxxxxxxxxxxxxxxxxxxxxxxxxxxxxxxxxxxxxxxxxxxxxxxxxxxxxxxxxxxxxxxxxxxxxxxxxxxxxxxxxxxxxxxxxxxxxxxxxxxxxxxxxxxxxxxxxxxxxxxxxxxxxxxxxxxx
%xxxxxxxxxxxxxxxxxxxxxxxxxxxxxxxxxxxxxxxxxxxxxxxxxxxxxxxxxxxxxxxxxxxxxxxxxxxxxxxxxxxxxxxxxxxxxxxxxxxxxxxxxxxxxxxxxxxxxxxxxxxxxxxxxxxxxxxx
\section{Discussion}
\label{sec:discussion}

We now discuss the implications of our results for our understanding of planetary volatile and surface evolution (Section~\ref{sec:discussion:R}), and the potential for reproducing the observed fractionation of different volatile elements due to preferential loss of atmosphere from giant impacts (Section~\ref{sec:discussion:frac}). We then discuss the potential impact of Rayleigh-Taylor instabilities on the efficiency of loss (Section~\ref{sec:discussion:vap}), and considerations for when combining our 1D loss results with 3D impact simulations (Section~\ref{sec:discussion:coupling}).

%xxxxxxxxxxxxxxxxxxxxxxxxxxxxxxxxxxxxxxxxxxxxxxxxxxxxxxxxxxxxxxxx
%xxxxxxxxxxxxxxxxxxxxxxxxxxxxxxxxxxxxxxxxxxxxxxxxxxxxxxxxxxxxxxxx
\subsection{Sensitivity of loss to volatile budgets and pre-impact surface conditions}
\label{sec:discussion:R}

We have shown that the surface conditions on colliding bodies before giant impacts can have a strong influence on the efficiency of atmospheric and ocean loss. In the no-ocean case, hotter, lower pressure, and lower molecular weight atmospheres are more easily lost due to the effect of atmospheric properties on the impedance-match velocity between the ground and atmosphere. If terrestrial planets grow to a large enough size while the nebula is still present, they can accrete a primary H$_2$-He atmosphere \citep[e.g.,][]{Ikoma2006} as inferred in the case of many exoplanets \citep[e.g.,][]{Jontof-Hutter2019_exo_comp}. After the nebular dissipates, such a light atmosphere would be comparatively susceptible to loss by giant impacts, in addition to more generally considered thermal or stellar-radiation driven loss \citep[e.g.,][]{Lammer2014}. Over time, the atmospheres of planets can become dominated by heavier molecular weight species produced by degassing of accreted solids and from the planet's interior. These heavier molecular weight atmospheres, often called secondary atmospheres, would be less susceptible to loss by giant impacts compared to H$_2$-He atmospheres. For atmospheres of all compositions, depending on the efficiency of volatile accretion, repeat impact events could reduce the atmospheric pressure, providing a positive feedback for increasingly efficient loss through accretion. The ability of giant impacts to remove atmospheric volatiles thus can evolve substantially during planet formation as the atmospheres of planets change composition, mass, and oxidation state.

Towards the end of planet formation, the cadence of giants impacts and the flux of planetesimal accretion in our solar system are both low enough to allow significant cooling between giant impacts. Thus, water to condense on proto-planets, given a suitable oxidation state, in the time between giant impacts \citep{Abe1988}. This may also be the case in many exosystems, but will depend strongly on the timescale of accretion. Therefore, giant impacts that occur later in accretion are likely to be between planets with pre-impact oceans. We have shown that, over the range of atmospheres we considered, loss from planets with oceans is only weakly dependent on the atmospheric composition, temperature and pressure, and is largely dominated by the atmospheric to ocean mass ratio \citep[as proposed by][]{Genda2005}. We find a relatively rapid transition between a high and a low loss regime, over one to two orders of magnitude in the atmosphere to ocean mass ratio. Contrary to previous thinking, the presence of an ocean can reduce the efficiency of loss in the low-loss regime. Condensation of an ocean could then actually protect the surface volatile inventory from loss by giant impacts. However, if the ocean is sufficiently massive, with the critical mass being typically within a factor of a few of the atmospheric mass, the presence of an ocean can significantly increase the efficiency of atmospheric loss, in agreement with \citet{Genda2005}. Therefore, a critical factor in understanding the evolution of planetary atmospheres during accretion is determining whether they are in the high or low loss regimes. 

 In our solar system, the compositions of chondritic meteorites are often used as proxies for the compositions of the building blocks of Earth. For most chondrites, if their full budget of H, C and N were converted into a CO$_2$ and N$_2$ atmosphere over a pure H$_2$O ocean, the resulting atmosphere to ocean mass ratio would be of order $10^0$-$10^1$ (see Table~\ref{tab:R_values}). For all but the highest ground velocities, such an atmosphere and ocean would be at the edge of the low-loss efficiency regime and only high energy impacts would drive substantial atmospheric loss \citep{Kegerreis2020,Kegerreis2020a}. However, since loss of ocean is less efficient than loss of atmosphere (Section~\ref{sec:discussion:frac}), there is a positive feedback where loss events push the atmosphere to ocean mass ratio lower, increasingly the susceptibility of the remaining atmosphere to loss in subsequent giant impacts. The rapid transition in the efficiency of loss with decreasing atmosphere to ocean mass ratio could result in planets reaching a tipping point in volatile evolution once their atmosphere to ocean mass ratio becomes sufficiently small to be in the high-loss regime. Such an effect relies on a planet having a sufficient number of giant impacts with a relative timing that allows an ocean to condense between impacts. The strong sensitivity to the order and timing of giant impacts could result in planets with similar accretionary histories obtaining substantially different final volatile budgets.

\begin{table}
    \centering
    \begin{tabular}{l c c  }
        Reservoir &  $\mathcal{R}$ (oxidised) & $\mathcal{R}$ (reduced) \\
        \hline 
         Earth's surface today & 3.6$\times 10^{-3}$ & -   \\ 
Earth's exosphere & 0.21 & 0.13 \\ 
BSE \citep{Marty2012} & 0.7$\pm$0.5 & 0.4$\pm$0.5\\ 
BSE \citep{Halliday2013} & 0.3$\pm$0.1 & 0.2$\pm$0.1\\ 
CC \citep{Marty2012} & 1.1$\pm$0.3 & 0.7$\pm$0.3\\ 
% CI & 0.73 & 0.47 \\ 
CI & 8.85 & 5.71 \\ 
% CV & 594. & 380. \\ 
H & 12.8 & 8.24 \\ 
L & 17.2 & 11.0 \\ 
LL & 13.6 & 8.75 \\  
EH & 15.9 & 10.4 \\ 
% Eucrite & 1.04 & 0.66 \\ 

    \end{tabular}
    \caption{The likely volatile budgets of accreting protoplanets span from the high to the low loss regimes. Presented are the atmosphere to ocean mass ratios for the Earth today, and different reservoirs or meteorite groups assuming that all the carbon and nitrogen were partitioned into the atmosphere and the hydrogen was in the ocean. Estimates assuming an N$_2$ and \co2 (oxidized) or N$_2$ and CO (reduced) atmospheres are given in different columns. Estimate of Earth's exosphere volatile budget from \citet{Marty2012}. Bulk silicate Earth (BSE) estimates from \citet{Marty2012} and \citet{Halliday2013}. Average carbonaceous chondrite (CC) estimated by \citet{Marty2012} with data from \cite{Kerridge1985} and \cite{Robert2003_DH}. CI values based on Orgueil \cite{Lodders2003}. Average composition of ordinary chondrites (H, L, LL) taken from compilation by \cite{Schaefer2007_ord_outgas}. Enstatite (EH) chondrite values based on Indarch with data from \citet{Wiik1956_stony}, \cite{Moore1966_E_C}, \cite{Moore1969_N_chondrites}, and \cite{Grady1986_enstatite} as compiled by \cite{Schaefer2017}.}
    \label{tab:R_values}
\end{table}

Beyond just the total volatile budget, critical to the loss or retention of volatile elements is the partitioning of volatiles between reservoirs in forming planets. As an example, the present-day Earth has an atmosphere to ocean mass ratio of $\sim 3.6\times10^{-3}$, firmly in the high loss efficiency regime. However, if all the H, C and N in the present-day atmosphere, ocean and sedimentary rocks (exosphere) were present as a CO$_2$ and N$_2$ atmosphere and a H$_2$O ocean, the atmosphere to ocean mass ratio would be $\sim$0.21 (Table~\ref{tab:R_values}), which is in the transition between the high and low loss regimes. Similar calculations using two different estimates for the volatile content of the entire BSE give atmosphere to ocean mass ratios of $0.7\pm0.6$ \citep{Marty2012} and $0.3\pm0.15$ \citep{Halliday2013}. The oxidation state of a planet's surface also plays a role in dictating the efficiency of atmospheric loss. For example, if CO was the dominate carbon species in the atmospheres calculated above the atmospheric mass would be $\sim1/3$ lower (right column of Table~\ref{tab:R_values}).

How volatiles partition between reservoirs, and the oxidation state of the surface, in the period between giant impacts depends on a number of factors including: the separation of volatiles from the silicate during condensation of the silicate vapor produced in the giant impact \citep{Stewart2018LPSC, Lock2020,Caracas2023}; degassing of the mantle during magma ocean solidification \citep[see e.g.;][]{Elkins-Tanton2012,Bower2019}; dissolution of carbon in the ocean and potentially precipitation and storage of abiogenic (or even potentially biogenic) carbonates; weathering of crust transferring volatiles into the crust or sediments; and the thermal state of the atmosphere and surface \citep[e.g.,][]{Zahnle2010}.  Given the difficulty in understanding each of these processes, and the interactions between them, it is likely not possible to precisely determine the mass and composition of ocean and atmosphere before each impact. We therefore advocate taking a statistical approach to exploring the effect of giant impacts on the volatile budgets of terrestrial planets, making use of the high and low-loss efficiency regimes described here to bound the evolution of planetary volatile budgets. 

%xxxxxxxxxxxxxxxxxxxxxxxxxxxxxxxxxxxxxxxxxxxxxxxxxxxxxxxxxxxxxxxx
%xxxxxxxxxxxxxxxxxxxxxxxxxxxxxxxxxxxxxxxxxxxxxxxxxxxxxxxxxxxxxxxx
\subsection{Fractionation of hydrophile and atmophile species}
\label{sec:discussion:frac}

One of the main outstanding questions in regard to Earth's chemistry is that the C/N and H/N ratios of the BSE are significantly fractionated from those of known chondrites \citep{Halliday2013}. It has been proposed that preferential loss of atmosphere to ocean in impacts could fractionate N and C from H \citep{Tucker2014}. If C was dissolved in the ocean or stored in the crust or sediments, more C could be retained, leading to N/C fractionation. 

Our results confirm that giant impacts drive significantly more loss of atmospheric species than loss of water. In the high-loss efficiency regime, total atmospheric loss can be achieved from sections of the colliding bodies where ground motions reach only $\sim0.35$~$v_{\rm esc}$, for an Earth-mass planet, whereas substantial ocean loss requires ground velocities that are a large fraction of $v_{\rm esc}$. The amount of ocean loss can also be reduced by slowing of the ground by release waves within the planet, a process which atmospheric loss is relatively insensitive to (Section~\ref{sec:discussion:issues}). If the proto-Earth, or the planetary embryos that accreted to form it, underwent a giant impact when an ocean was present, there is significant potential for H/N and C/N fractionation. Such an effect would be particularly strong if the ratio of atmospheric to ocean mass was low enough to be in the high-loss efficiency regime. Smaller impacts can also lead to significant atmospheric loss \citep{Schlichting2015}, but the degree of ocean/atmosphere fractionation in such events has not been determined. If loss by giant impacts is required to explain the Earth's H/N and C/N ratios, this would imply that Earth experienced, potentially multiple, giant impacts late in formation when it had an ocean and potentially a low enough atmospheric to ocean mass ratio to make atmospheric loss efficient.

%xxxxxxxxxxxxxxxxxxxxxxxxxxxxxxxxxxxxxxxxxxxxxxxxxxxxxxxxxxxxxxxx
%xxxxxxxxxxxxxxxxxxxxxxxxxxxxxxxxxxxxxxxxxxxxxxxxxxxxxxxxxxxxxxxx
\subsection{Potential for Rayleigh-Taylor instabilities}
\label{sec:discussion:vap}

Rayleigh-Taylor instabilities develop when a fluid attempts to accelerate (i.e., push) another fluid of greater density than itself. The growth of the instability disrupts the material interface and significantly reduces the acceleration of the denser fluid due to the forcing of the lighter fluid. If such instabilities were to develop at the ground/ocean, ground/atmosphere, or ocean/atmosphere boundaries during the process of atmospheric loss the efficiency of loss could be significantly reduced. Our 1D simulations cannot capture instabilities and so it is necessary for us to compare the density of the different layers during our simulations to determine the likelihood for Rayleigh-Taylor instabilities.

The ground/ocean boundary is not at risk of experiencing Rayleigh-Taylor instabilities over the range of parameters considered in this work. As discussed in Section~\ref{sec:discussion:issues}, we do not directly model the planet in our simulations, but we can calculate the expected density of the ground by using the pressure at the base of the ocean/atmosphere from our calculations combined with the calculated release curve of forsterite (Figure~\ref{fig:imp_match_loss_diff}). We find that the density of the ground is never lower than the base of the ocean, even in cases where the ground begins to vaporize, and so we would expect the ground/ocean boundary to remain stable.

Determining the stability of the ground/atmosphere and ocean/atmosphere boundaries is more difficult due to the limitations of the available EOS. High quality EOS for heavy gases such as N$_2$, CO$_2$, and air mixtures that cover the range of conditions considered here are not publicly available or easily attainable. In this work we therefore decided to use an ideal gas EOS for atmospheres. For lower pressure atmospheres ($<\sim 150$~bar for CO$_2$, $\sim 4$~kbar for H$_2$) the density of the ideal gas Hugoniot is lower than the density of forsterite and water on the release curve at the impedance match point, even in cases where the ground/atmosphere begins to vaporize. However, for higher pressures the density of the ideal gas EOS exceeds the density of the water and even forsterite. This is likely due to the fact that the ideal gas EOS significantly overestimates the density of gases at high pressure as it neglects the volumes and interactions of particles. More sophisticated EOS for gases \citep{Span1996,Lemmon2000,Span2000} predict a much lower density for the Hugoniot of atmospheric gases, below that of the forsterite and water at the impedance match pressure. However, the maximum pressure covered by these EOS preclude using them to calculate shocks over much of the range considered here. It is therefore likely that the ocean/atmosphere and ground/atmosphere boundaries are indeed stable to Rayleigh-Taylor instabilities but wider-ranging, high-quality gas EOS are required to confirm this.

%xxxxxxxxxxxxxxxxxxxxxxxxxxxxxxxxxxxxxxxxxxxxxxxxxxxxxxxxxxxxxxxx
%xxxxxxxxxxxxxxxxxxxxxxxxxxxxxxxxxxxxxxxxxxxxxxxxxxxxxxxxxxxxxxxx
\subsection{Combination of 1D and 3D simulations}
\label{sec:discussion:coupling}

We have produced scaling laws for loss as a function of ground velocity (Section~\ref{sec:results:fit}) and scripts to calculate impedance match velocities (available through GitHub and Zenodo), which allow for linking of our results with ground velocity distributions calculated from 3D impact simulations \citep[e.g.,][]{Kegerreis2019,Yalinewich2018}. We hope that this will allow for investigations of the effect of surface conditions on global atmospheric loss which are currently not possible due to computational limitations or expense.

When combining our results with 3D simulations there are two particularly important factors to consider. First, it is important to correct the ground motion from the 3D simulation to the impedance match velocity for the atmosphere/ocean under consideration from that used in the 3D simulations. This can be done with the scripts and functions made available with this manuscript.

Second, by considering a 1D system in our simulations, we have inherently assumed that the shock in the planet breaks out perpendicular to the ground. That is, that the ground velocity is always vertical relative to the local surface. In reality, the shock wave will be travelling at an angle relative to the surface. On breakout of the shock, the wave will be refracted depending on the impedance of the materials either side of the boundary \cite{Henderson1989_refraction_shock}. It is not clear what the best approach is to account for a non-perpendicular incidence angle, but we suggest that a correction to the local escape velocity due to the component of the shock particle velocity parallel to the surface may be adequate. However, different methods will need to be tested using full 3D simulations as ground truth before proceeding with hybrid 1D-3D models.

%xxxxxxxxxxxxxxxxxxxxxxxxxxxxxxxxxxxxxxxxxxxxxxxxxxxxxxxxxxxxxxxxxxxxxxxxxxxxxxxxxxxxxxxxxxxxxxxxxxxxxxxxxxxxxxxxxxxxxxxxxxxxxxxxxxxxxxxx
%xxxxxxxxxxxxxxxxxxxxxxxxxxxxxxxxxxxxxxxxxxxxxxxxxxxxxxxxxxxxxxxxxxxxxxxxxxxxxxxxxxxxxxxxxxxxxxxxxxxxxxxxxxxxxxxxxxxxxxxxxxxxxxxxxxxxxxxx
%xxxxxxxxxxxxxxxxxxxxxxxxxxxxxxxxxxxxxxxxxxxxxxxxxxxxxxxxxxxxxxxxxxxxxxxxxxxxxxxxxxxxxxxxxxxxxxxxxxxxxxxxxxxxxxxxxxxxxxxxxxxxxxxxxxxxxxxx
\section{Conclusions}
\label{sec:conclusions}

We have conducted a large number of hydrodynamic simulations and impedance-match calculations to explore the effect of surface conditions on the efficiency of atmospheric loss from terrestrial planets due to giant impacts. We have found that pre-impact surface conditions can have a significant effect on the efficiency of loss driven by the impact shock wave, in the far-field away from the impact site. The higher ground release velocities, as given by impedance-match solutions, for lower molecular weight, hotter, and lower pressure atmospheres mean that such atmospheres are more efficiently lost. 

The presence of an ocean can also substantially influence atmospheric loss. In the ocean case the loss efficiency is largely dictated by the atmospheric to ocean mass ratio, with a relatively rapid transition between a low loss and a high loss regime as the mass ratio of atmosphere to ocean decreases. If the ocean is above a critical mass, typically within a factor of a few of the atmospheric mass, the presence of an ocean can significantly increase the efficiency of loss. However, contrary to previous thinking, the presence of an ocean can reduce the efficiency of loss if the ocean is not sufficiently massive.

The efficiency of loss due to giant impacts is therefore highly dependent on the surface conditions on the colliding bodies before the impact. As the surface conditions on planets evolve during accretion so will the potential for substantial atmospheric loss as a result of giant impacts. In particular,  later in accretion there is sufficient time for oceans to condense between most impact events \citep{Abe1988} potentially allowing for much more efficient atmospheric loss. Preferential loss of atmosphere over ocean, coupled with the fact that loss efficiency increases with decreasing atmosphere to ocean mass ratio, could generate a positive feedback that accelerates atmospheric loss from planets that experience multiple late giant impact events. 

To allow our 1D results to be combined with 3D giant impact simulations \citep[e.g.,][]{Kegerreis2020} to calculate the total loss from specific impacts, we have developed a scaling law that relates ground velocity due to an impact, the escape velocity of the planet, and the ratio of atmosphere to ocean mass, to the efficiency of loss. Future work will use this approach to develop scaling laws that approximate loss as a function of both impact parameters and surface conditions. Such laws will allow atmospheric loss from giant impacts to be included directly in combined dynamical and chemical models of planet formation.

\vspace{20pt}
%xxxxxxxxxxxxxxxxxxxxxxxxxxxxxxxxxxxxxxxxxxxxxxxxxxxxxxxxxxxxxxxxxxxxxxxxxxxxxxxxxxxxxxxxxxxxxxxxxxxxxxxxxxxxxxxxxxxxxxxxxxxxxxxxxxxxxxxx
%xxxxxxxxxxxxxxxxxxxxxxxxxxxxxxxxxxxxxxxxxxxxxxxxxxxxxxxxxxxxxxxxxxxxxxxxxxxxxxxxxxxxxxxxxxxxxxxxxxxxxxxxxxxxxxxxxxxxxxxxxxxxxxxxxxxxxxxx
%xxxxxxxxxxxxxxxxxxxxxxxxxxxxxxxxxxxxxxxxxxxxxxxxxxxxxxxxxxxxxxxxxxxxxxxxxxxxxxxxxxxxxxxxxxxxxxxxxxxxxxxxxxxxxxxxxxxxxxxxxxxxxxxxxxxxxxxx
\begin{acknowledgements}
    
The authors acknowledge the late, great, Jay Melosh for useful discussions and stimulating questions on the topic of atmospheric loss. SJL thanks Erik Asphaug and Hidenori Genda for help in replicating the results of \cite{Genda2005} using the Tillotson EOS, and Matthew Roche for his feedback on earlier versions of the manuscript. We also thank Hidenori Genda and an anonymous reviewer for their encouraging and constructive comments which helped improve the manuscript. SJL acknowledges the support of an NASA Earth and Space Science Fellowship (grant NNX13AO67H), NSF (awards EAR-1947614 and EAR-1725349), the Earth and Planetary Science Department of Harvard University, the Division of Geological and Planetary Sciences of the California Institute of Technology, and the UK Natural Environment Research Council (grant NE/V014129/1). STS acknowledges support from NASA through awards 80NSSC18K0828 and NNX15AH54G. This work was carried out using the FASRC Odyssey cluster, supported by the FAS Division of Science Research Computing Group at Harvard University, and the computational facilities of the Advanced Computing Research Centre, University of Bristol.

Our processed results, scripts for reproducing the figures in this manuscript, and the modifications made to the WONDY hydrodynamics code are available on GitHub (\href{https://github.com/sjl499/Lock_Stewart_2023_PSJ_atmospheric_loss}{sjl499/Lock\_Stewart\_2023\_PSJ\_atmospheric\_loss}) and archived on Zenodo \citep{LockStewart2023_zenodo_v1_forced_general}. A Python package, {\it SIMPLES} (Shock Impedance Match Package and Loss Event Simulator) that can calculate impedance match velocities for real materials and the resulting atmospheric loss is also available on GitHub (\href{https://github.com/sjl499/simples}{sjl499/simples}) and archived on Zenodo \citep{simples_Zenodo_v1_general}.

\end{acknowledgements}

% \section*{References}

% %%Supplemental

% %make all the sections labelled with an S for supp
% % \renewcommand{\thepage}{\arabic{page}}  
% \renewcommand{\thesection}{A\arabic{section}}   
% \renewcommand{\thetable}{A\arabic{table}}   
% \renewcommand{\thefigure}{A\arabic{figure}}
% \renewcommand{\theequation}{A\arabic{equation}}

% %reset page count and section count
% % \setcounter{page}{1}
% \setcounter{section}{0}
% \setcounter{figure}{0}
% \setcounter{table}{0}
% \setcounter{equation}{0}

\appendix

%xxxxxxxxxxxxxxxxxxxxxxxxxxxxxxxxxxxxxxxxxxxxxxxxxxxxxxxxxxxxxxxxxxxxxxxxxxxxxxxxxxxxxxxxxxxxxxxxxxxxxxxxxxxxxxxxxxxxxxxxxxxxxxxxxxxxxxxx
%xxxxxxxxxxxxxxxxxxxxxxxxxxxxxxxxxxxxxxxxxxxxxxxxxxxxxxxxxxxxxxxxxxxxxxxxxxxxxxxxxxxxxxxxxxxxxxxxxxxxxxxxxxxxxxxxxxxxxxxxxxxxxxxxxxxxxxxx
%xxxxxxxxxxxxxxxxxxxxxxxxxxxxxxxxxxxxxxxxxxxxxxxxxxxxxxxxxxxxxxxxxxxxxxxxxxxxxxxxxxxxxxxxxxxxxxxxxxxxxxxxxxxxxxxxxxxxxxxxxxxxxxxxxxxxxxxx
\section{Description of 1D Hydrocode}
\label{sup:sec:WONDY}

For calculating atmospheric and ocean loss in 1D we follow an approach similar to that of \cite{Genda2003, Genda2005}.
We used a modified version of the WONDY hydrocode \citep{Kipp1982} which solves the Lagrangian 1D mass, momentum and energy equations using a finite difference method with artificial viscosity to allow accurate modelling of shock waves by spreading the shock front over several zones. 
We modified the code by adding radial gravity and several different equation of state (EOS) options as well as allowing for hydrostatic initialisation of an atmosphere and ocean. 
The details of our adapted code are described in the following sections. 

%
%XXXXXXXXXX \\
%FIGURE: Schematic of 1D Model \\
%XXXXXXXXXX \\

%xxxxxxxxxxxxxxxxxxxxxxxxxxxxxxxxxxxxxxxxxxxxxxxxxxxxxxxxxxxxxxxx
%xxxxxxxxxxxxxxxxxxxxxxxxxxxxxxxxxxxxxxxxxxxxxxxxxxxxxxxxxxxxxxxx
\subsection{Finite Difference Implementation of the Fundamental Fluid Dynamics Equations}

The core of the code is the solution of the fundamental fluid dynamics equations in 1D. 
Conservation of momentum can be written as
\begin{equation}  
\label{sup:eqn:mom_conc}
\rho a = - \frac{\partial p}{\partial x} -  \frac{\partial q}{\partial x} + \rho g \; ,
\end{equation}
where $\rho$ is density, $a$ is acceleration, $p$ is pressure, $x$ is the Lagrangian spatial coordinate, $g$ is gravitational acceleration and $q$ is the viscous stress. For our purposes $q$ is simply the artificial viscosity.
The acceleration in a Lagrangian reference frame is given by
\begin{equation}  
\label{sup:eqn:acc}
a= \frac{\partial u}{\partial t} \; ,
\end{equation}
where $u$ is the velocity,
\begin{equation}  
\label{sup:eqn:vel}
u =\frac{\partial x}{\partial t} \; .
\end{equation}
Conservation of energy is the balance between the rate of change of internal energy and the rate at which work is being done,
\begin{equation}  
\label{sup:eqn:E_conc}
\rho  \frac{\partial \epsilon }{\partial t} = (p + q) \frac{1}{\rho} \frac{\partial \rho}{\partial t} \; ,
\end{equation}
where $\epsilon$ is the specific internal energy.

In WONDY, the conservation equations are solved using a finite difference method. 
The fluid is divided into Lagrangian zones (or cells) and the parameters of each cell are advanced in incremental time steps using the conservative equations. The bulk fluid properties ($m$, $\rho$, $q$, $p$, $\epsilon$, $c$, $T$) are defined at the center of each zone and the kinematic variables ($a$, $u$, $x$) are defined at the boundaries of the zones. $c$ is adiabatic sound speed, $T$ is the temperature, $m$ is the mass of a given Lagrangian zone and the other variables are described above. For our purposes all kinematic variables are defined in reference to the center of the planet. Linear interpolation is used to find values between these points, e.g., to find the bulk fluid properties at the edge of a zone. Zones are labelled in order with an index, $j$, starting at 1 with the lower boundary referred to with the index 0. The notation we employ here, after \cite{Kipp1982}, for the fluid properties is $\psi ^{n}_{j}$ which donates an arbitrary quantity,  $\psi$, at the $j$\textsuperscript{th} zone and the $n$\textsuperscript{th} time step. 
Integer $j$ denote values of the quantity at the upper boundary of a zone whereas half integer values indicate quantities defined at the center of the zone. 
Similarly integer $n$ indicate values known at that time step and half integer values indicate quantities defined halfway between time steps. 
Bulk fluid properties are defined at $\psi ^{n}_{j-\frac{1}{2}}$ and the kinematic variables at $\psi ^{n}_{j}$ with the exception of $u$ which is defined at $u^{n+\frac{1}{2}}_{j}$. These choices of when and where to define variables are for numerical convenience which will become evident below. The second order centred difference method is used to calculate both time and spatial derivatives of all quantities.

Using this notation the fundamental equations can be defined in finite difference form.
The finite difference momentum equation is
\begin{align}
\begin{aligned}
\label{sup:eqn:mom_conc_fd} 
a^n_j = 2 & \left [ \frac{(p^n_{j-\frac{1}{2}} +  q^n_{j-\frac{1}{2}} ) -  (p^n_{j+\frac{1}{2}} + q^n_{j+\frac{1}{2}} )}{\rho^n_{j+\frac{1}{2}} (x^n_{j+1} -x^n_j) + \rho^n_{j-\frac{1}{2}} (x^n_{j} -x^n_{j-1})}\right ] \\
& - \frac{G M_p}{\left (x_j^n \right ) ^2} \; ,
\end{aligned}
\end{align}
where $g_j^n$ has been substituted for a radial gravity profile for a planet of mass $M_p$.
$G$ is the universal gravitational constant. 
Equation~\ref{sup:eqn:mom_conc_fd} can be used to calculate the acceleration at the $n$\textsuperscript{th} time step from quantities that are already known from that time step.
The velocity at the next half time step can then determined from the finite difference form of Equation~\ref{sup:eqn:acc}:
\begin{equation} 
\label{sup:eqn:vel_fd}
u^{n+\frac{1}{2}}_{j}=u^{n-\frac{1}{2}}_{j}+\frac{1}{2} \left (t^{n+1} - t^{n-1} \right) a_j^n \; .
\end{equation}
Similarly from Equation~\ref{sup:eqn:vel} the position of zone $j$ at the next full time step is
\begin{equation} 
\label{sup:eqn:x_fd}
x^{n+1}_j=x^{n}_j + \left ( t^{n+1} - t^n \right ) u^{n+\frac{1}{2}}_{j} \; .
\end{equation}

Having used Equations~\ref{sup:eqn:mom_conc_fd}, \ref{sup:eqn:vel_fd} and \ref{sup:eqn:x_fd} to determine the kinematic variables for the next time step the code recalculates the thermodynamic properties for the zone.
Conservation of mass in spherical coordinates provides the updated density,
\begin{equation} 
\rho^{n+1}_{j-\frac{1}{2}} = \frac{ 3 m_{j-\frac{1}{2}} }{4 \pi \left [  \left ( x^{n+1}_{j} \right ) ^3 -  \left ( x^{n+1}_{j-1} \right ) ^3 \right]} \; .
\end{equation}
To reduce round off error the code actually implements the following, equivalent, expression,
\begin{align}
\begin{aligned}
\rho^{n+1}_{j-\frac{1}{2}} =  & \left \{  \frac{1}{\rho^{n}_{j-\frac{1}{2}}}+ \right. \\
 & \; \; \left. \frac{ 4 \pi  \Delta t^{n+\frac{1}{2}} }{3 m_{j-\frac{1}{2}}} \left [ \xi^{n+\frac{1}{2}}_{j} u^{n+\frac{1}{2}}_{j} - \xi^{n+\frac{1}{2}}_{j-1} u^{n+\frac{1}{2}}_{j-1} \right ] \right \}^{-1} \; ,
\end{aligned}
\end{align}
where
\begin{equation} 
\xi^{n+\frac{1}{2}}_{j}  = \left( x^{n+1}_{j} \right )^2 + x^{n+1}_{j} x^{n}_{j} + \left ( x^{n}_{j} \right )^2 \; ,
\end{equation}
and
\begin{equation} 
\Delta t^{n+\frac{1}{2}} = t^{n+1} - t^n \; .
\end{equation}
The internal energy and pressure for the next time step are found by a combination of the energy equation and the EOS of the material. 
The finite difference form of the energy equation is
\begin{equation}
\begin{multlined}
\label{sup:eqn:E_conc_fd}
 \epsilon^{n+1}_{j-\frac{1}{2}} = \epsilon^{n}_{j-\frac{1}{2}} + \\
 2 \left ( p^{n+1}_{j-\frac{1}{2}} + p^{n}_{j-\frac{1}{2}} +2 q^{n+\frac{1}{2}}_{j-\frac{1}{2}}\right ) 
\frac{\left ( \rho^{n+1}_{j-\frac{1}{2}} - \rho^{n}_{j-\frac{1}{2}}\right )}{ \left( \rho^{n+1}_{j-\frac{1}{2}} - \rho^{n}_{j-\frac{1}{2}}\right ) ^2} \; .
\end{multlined}
\end{equation}
We will examine the solution of this equation for each of the EOS used in this study in Appendix~\ref{sup:sec:EOS}. 

Resolving shocks in numerical codes is challenging due to the very rapid change in material properties and velocity across the shock front. The natural viscosity of the fluids we consider is low, leading to very thin shock fronts that cannot be feasibly resolved at planetary scales. Artificial viscosity is used in WONDY to smear the shock front over several zones and avoid discontinuities.
WONDY includes both a quadratic viscosity,
\begin{equation} 
q_1=\rho b_1^2 \left ( \frac{1}{\rho} \frac{\partial \rho }{\partial t} \right )^2 \;,
\end{equation}
and a linear viscosity,
\begin{equation} 
q_2= b_2 c \left (  \frac{\partial \rho }{\partial t} \right ) \;,
\end{equation}
with the total viscosity given as a simple sum of $q_1$ and $q_2$ (i.e., $q=q_1+q_2$). $b_1$ and $b_2$ are parameters that control the strength of the artificial viscosity. To make the artificial viscosity insensitive to the absolute spatial scales, $b_i$ are scaled to the size of the zone,
\begin{equation} 
b_i=B_i \Delta x \; ,
\end{equation}
so the true constants, $B_i$, are approximately equal to the number of zones that any disturbance will be smoothed over. The values for all constants used in the code are given in Table~\ref{sup:tab:code_constants} and the sensitivity of our results to each parameter is examined in Appendix~\ref{sup:sec:1Dsensitivity_tests}.

$q_1$ is large only when the rate of change is large (i.e. in the shock front) and small elsewhere so it is used mainly to control gradients in the shock front and avoid discontinuities. 
$q_2$ is more effective at controlling spurious oscillations elsewhere where $q_1$ is negligible. 
Both forms of the viscosity are only used when the material is in compression, i.e. where
\begin{equation} 
\frac{\partial \rho }{\partial t} < 0 \;  .
\end{equation}
The finite difference form of the combined viscosity is
\begin{equation}
\label{sup:eqn:q_fd}
\begin{multlined}
q_{j-\frac{1}{2}}^{n+\frac{1}{2}} = \frac{\rho^{n+1}_{j-\frac{1}{2}} + \rho^{n}_{j-\frac{1}{2}}}{2} \\
\left [ B_2 \left ( \frac{x_j^{n+1}- x_{j-1}^{n+1}+x_j^{n}-x_{j-1}^{n}}{2} \right ) c^n_{j-\frac{1}{2}} 
\left ( \frac{1}{\rho} \frac{\partial \rho }{\partial t} \right )   \right.  \\
\left.  +B_1^2 \left (  \frac{x_j^{n+1}- x_{j-1}^{n+1}+x_j^{n}-x_{j-1}^{n}}{2} \right ) ^2  \right. \\
\times \left. \left ( \frac{1}{\rho} \frac{\partial \rho }{\partial t} \right )
\left | \frac{1}{\rho} \frac{\partial \rho }{\partial t} \right |
\right ] \;  ,
\end{multlined}
\end{equation}
where
\begin{equation}
\frac{1}{\rho} \frac{\partial \rho }{\partial t} = \frac{2 \left ( \rho_{j-\frac{1}{2}}^{n+1} - \rho_{j-\frac{1}{2}}^{n} \right)}
{\Delta t^{n+\frac{1}{2}} \left ( \rho_{j-\frac{1}{2}}^{n+1} + \rho_{j-\frac{1}{2}}^{n} \right) } \;  ,
\end{equation}
and
\begin{equation} 
\Delta t^{n+\frac{1}{2}} = t^{n+1} - t^n \; .
\end{equation}

The time step is chosen to ensure numerical stability. 
The criterion for linear stability of the difference equations used in the WONDY code \citep{Hicks1978} is
\begin{equation}
\begin{multlined}
 \Delta t < \Delta x \left [ B_2c  + 2B_1^2 \left | \frac{\dot{\rho}}{\rho} \right | \Delta x \right. \\
\left. + \sqrt{\left (B_2 c + + 2B_1^2 \left | \frac{\dot{\rho}}{\rho} \right | \Delta x \right )^2 + c^2} \, \right ]^{-1} \; .
\end{multlined}
\label{sup:eqn:tcriteria}
\end{equation}
At the end of each cycle, the value of the right hand side of Equation~\ref{sup:eqn:tcriteria} is evaluated for each cell to determine the maximum time step that can be used to ensure stability.
Since the criteria in Equation~\ref{sup:eqn:tcriteria} is based on a linear stability analysis, and then assumed to apply generally, the critical value for each cell is scaled by a constant, $K_{\rm t_1}$, to account for any non-linear effects and ensure stability, such that 
\begin{equation}
\begin{multlined}
\Delta t^{n+\frac{3}{2}}_{j-\frac{1}{2}} = K_{\rm t_1} \Delta x^{n+1}_{j-\frac{1}{2}}  
\left [ B_2 c_{j-\frac{1}{2}}^{n+1}+2 B_1^2 \left | \frac{\dot{\rho}}{\rho} \right | \Delta x^{n+1}_{j-\frac{1}{2}}  \right. \\
\left. + \sqrt{\left ( B_2 c_{j-\frac{1}{2}}^{n+1}+ 2 B_1^2 \left | \frac{\dot{\rho}}{\rho} \right | \Delta x^{n+1}_{j-\frac{1}{2}} \right )^2 + \left(c_{j-\frac{1}{2}}^{n+1} \right )^2} \, \right ]^{-1} \; .
\end{multlined}
\end{equation}
The next time step is then set as 
\begin{equation}
\Delta t^{n+\frac{3}{2}} = \min{\left ( \Delta t^{n+\frac{3}{2}}_{j-\frac{1}{2}}, K_{\rm t_2} \Delta t^{n+\frac{1}{2}} \right )} \;  ,
\end{equation}
to ensure stability and to limit the increase in the time step to a factor $K_{\rm t_2}$.

Although we will discuss the code in dimensional parameters here, these equations are implemented in the code in a dimensionless form to reduce numerical errors.
All the parameters, with the exception of temperature, are non-dimensionalised using combinations of: the ground velocity, $u_{\rm G}$; the radius of the planet, $R_{\rm p}$; and $p_0$.

Note that the gravity field in the 1D calculation is assumed to be fixed and radial.
Since the time for loss is small ($\sim100$~s) compared to the timescale of the impact, any change to the field over this period is minimal, but the field could still be distorted by the deformation of the target and the presence of the impactor.
As long as the gravity field does not vary significantly from a $1/r^2$ dependence, a local escape velocity in the direction of the ground motion should be sufficient to scale our results to calculate the loss in 3D impact simulations \citep{Kegerreis2020,Kegerreis2020a}.

%xxxxxxxxxxxxxxxxxxxxxxxxxxxxxxxxxxxxxxxxxxxxxxxxxxxxxxxxxxxxxxxx
%xxxxxxxxxxxxxxxxxxxxxxxxxxxxxxxxxxxxxxxxxxxxxxxxxxxxxxxxxxxxxxxx
\subsection{Implementation of Equations of State}
\label{sup:sec:EOS}

In this work we have used a variety of equations of state, in particular for water. 
The EOS dictates the method used to solve the finite difference form of the energy equation (Equation~\ref{sup:eqn:E_conc_fd}). 
We will now outline the equations of state used and their implementation.

%xxxxxxxxxxxxxxxxxxxxxxxxxxx
\subsubsection{Ideal Gas Equation of State}

The ideal gas EOS is the simplest EOS for gases and its implementation is straightforward. 
The EOS can be written in the form
\begin{equation}
p= ( \gamma - 1 ) \rho \epsilon \;  ,
\end{equation}
where $\gamma$ is the ratio $c_{\rm p}/c_{\rm V}$, and $c_{\rm p}$ and $c_{\rm V}$ are the specific heat capacities at constant pressure and volume respectively.
This linear relation between pressure and energy allows for an analytical solution to the energy equation,
\begin{equation}
\begin{multlined}
\label{sup:eqn:E_conc_idealG_fd}
 \epsilon^{n+1}_{j-\frac{1}{2}} = \left [ \epsilon^{n}_{j-\frac{1}{2}} + \right. \\
 \left. 2 \left ( p^{n}_{j-\frac{1}{2}} +2 q^{n+\frac{1}{2}}_{j-\frac{1}{2}}\right ) 
\frac{\left ( \rho^{n+1}_{j-\frac{1}{2}} - \rho^{n}_{j-\frac{1}{2}}\right )}{ \left( \rho^{n+1}_{j-\frac{1}{2}} - \rho^{n}_{j-\frac{1}{2}}\right ) ^2}  \right ] \\
\left [ 1- (\gamma -1 ) \rho^{n+1}_{j-\frac{1}{2}} \right ]^{-1} \;  ,
\end{multlined}
\end{equation}
which exclusively uses quantities that are known from the previous time step. 
The pressure at the next time step is given by
\begin{equation}
p^{n+1}_{j-\frac{1}{2}}= ( \gamma - 1 ) \rho^{n+1}_{j-\frac{1}{2}} \epsilon^{n+1}_{j-\frac{1}{2}} \; .
\end{equation}
The only parameter that needs to be specified for an ideal gas EOS is $\gamma$, although other parameters are needed to initialise the atmosphere (see Appendix~\ref{sup:sec:init}).

%xxxxxxxxxxxxxxxxxxxxxxxxxxx
\subsubsection{Tabulated Equations of State}
\label{sup:sec:tab_EOS}

We have also added to WONDY the ability to use a tabulated EOS in the form of a $T$-$\rho$ table. In this work we have used the water table form \cite{Senft2008}, a high quality water EOS based partially on the IAPWS water EOS (see Appendix~\ref{sup:sec:IAPWS_EOS}) but extended to the high pressures relevant for planetary impacts.

Use of a tabulated EOS requires a numerical solution to the energy equation. 
We solve for the temperature at each time step that solves the energy equation using a Newton-Raphson method.
The solution is found iteratively using
\begin{equation}
T_{i+1}=T_{i} - \frac{ f (T_i)}{f'( T_i)} \;  ,
\end{equation}
where the subscripts mark the number of the iteration,
\begin{equation}
\begin{multlined}
f(T_i) = \epsilon(\rho^{n+1}, T_i) -\epsilon^{n} \\
-\frac{\left ( \rho^{n+1} - \rho^{n} \right )}{ \left( \rho^{n+1} - \rho^{n} \right ) ^2} \left ( p(\rho^{n+1}, T_i) -  p^{n} -2 q^{n+\frac{1}{2}} \right ) \;  ,
\end{multlined}
\end{equation}
and
\begin{equation}
\begin{multlined}
f'(T_i) = c_V (\rho^{n+1}, T_i)  \\
-\frac{\left ( \rho^{n+1} - \rho^{n} \right )}{ \left( \rho^{n+1} - \rho^{n} \right ) ^2}  \left [ \frac{ \partial \, p}{\partial T} \bigg \rvert_{\rho} (\rho^{n+1}, T_i) \right ] \;  .
\end{multlined}
\end{equation}
Note we have dropped the spatial subscripts for clarity as all values required are for the zone under consideration. 
The values of; $\epsilon$, $p$, $c_V $ and $ \partial \, p / \partial T \rvert_{\rho} $ are found by linear interpolation of the tabulated values.
The derivatives are pre-calculated at each point in the table using a second order modified central difference method for unevenly spaced points. The initial value for the iteration is determined by assuming an isentropic volume change from the previous time step.
The change in temperature, $\Delta T$, for a given density change, $\Delta \rho$, is given by
\begin{equation}
\Delta T =\frac{ \partial \, T}{\partial \rho} \bigg \rvert_{S} \Delta \rho \; ,
\end{equation}
which can be written
\begin{equation}
\Delta T =  \frac{T}{c_V \rho^2}  \frac{ \partial \, p}{\partial T} \bigg \rvert_{\rho} \Delta \rho \; ,
\end{equation}
which is straightforward to calculate from the tabulated values. The solution is considered converged when
\begin{equation}
\bigg | \frac{T_{i+1}-T_{i}}{T_{i+1}}  \bigg | < 10^{-10} \; .
\end{equation}
A maximum of 1000 iterations are allowed but the routine typically converges within $2$-$3$ iterations.

%xxxxxxxxxxxxxxxxxxxxxxxxxxx
\subsubsection{IAPWS Water Equation of State}
\label{sup:sec:IAPWS_EOS}

The IAPWS water EOS \citep{Wagner2002} is a high quality empirically fitted EOS widely used in industrial applications.
The advantages of this EOS is that it provides a smooth form for the material properties, however the EOS is limited in its pressure range and therefore it can only be used for lower values of $u_{\rm G}$. The solution to the energy equation using this EOS is found in the same way as for the tabulated EOS (Appendix~\ref{sup:sec:tab_EOS}), except with the thermodynamic parameters and their derivatives calculated using the fortran subroutines provided by the IAPWS rather than by interpolation of an EOS table.

%xxxxxxxxxxxxxxxxxxxxxxxxxxx
\subsubsection{Tillotson Equation of State}

The Tillotson EOS \citep{Tillotson1962} has been widely used in shock physics and the study of planetary collisions. 
The EOS is based on empirical fits along the hugoniot forced to converge to the Thomas-Fermi limit at high pressures. 
Although the EOS is designed for use primarily in compression it includes parameters to approximate the behaviour of material upon release.
For this work we use the parameters from \cite{O'Keefe1982} for water.

Following \cite{Melosh1989book} the EOS implemented here has four different regimes: compressed states ($\rho/\rho_0 \geq 1$, where $\rho_0$ is the reference density), cold expanded states ($\rho/\rho_0 < 1$ and $\epsilon < \epsilon_{iv}$, where $\epsilon_{iv}$ is the energy of incipient vaporisation), hot expanded states ($\rho/\rho_0  \leq 1$ and $\epsilon > \epsilon_{cv}$, where $\epsilon_{cv}$ is the energy of complete vaporisation), and a transition region between the hot and cold expanded states (where $\rho/\rho_0 < 1$ and $\epsilon_{iv} < \epsilon < \epsilon_{cv}$).
Firstly, for compressed states and for cold expanded states the pressure is given by
\begin{equation}
\label{sup:eqn:tillot1}
p=\left [ a + \frac{b}{\frac{\epsilon}{\epsilon_0 \eta^2} +1} \right ] \rho \epsilon + A \mu +B \mu^2 \; ,
\end{equation}
where: $a$, $b$, $A$, $B$ and $\epsilon_0$ are fitted parameters; $\eta = \rho/\rho_0 $; $\mu = \eta -1 $; and $\rho_0$ is the reference density of the material. In order to ensure numerical stability, a low density cutoff is applied in the cold expanded state when $\rho/\rho_0 < 0.8$, where a nominal pressure of $10$~Pa is given to a zone.
In hot expanded states:
\begin{equation}
\begin{multlined}
\label{sup:eqn:tillot2}
p=a \rho \epsilon \\
+ \left [ \frac{b \rho \epsilon}{\frac{\epsilon}{\epsilon_0 \eta^2} +1} + A \mu \exp \left \{-\beta \left ( \frac{1}{\eta}-1 \right ) \right \} \right ] \\
\exp \left \{-\alpha \left ( \frac{1}{\eta}-1 \right )^2 \right \} \;,
\end{multlined}
\end{equation}
where $\alpha$ and $\beta$ are two more fitted parameters that control the rate at which the EOS tends to the ideal gas law.
For expanded states with internal energies between that of incipient and complete vaporization a hybrid formula is used to transition smoothly between the cold and hot expanded states,
\begin{equation}
\label{sup:eqn:tillot3}
p=\frac{\left(\epsilon-\epsilon_{iv} \right ) p_E + \left ( \epsilon_{cv} - \epsilon \right ) p_{C}}{\epsilon_{cv} - \epsilon_{iv}} \;  ,
\end{equation}
where $p_C$ is the pressure calculated using Equation~\ref{sup:eqn:tillot1} and $p_E$ is the pressure calculated using Equation~\ref{sup:eqn:tillot2}.

Using the Tillotson EOS the energy equation can be solved analytically. However, identifying the relevant regime to use in our code requires knowledge of the solution. We therefore find the possible solutions for $\epsilon$ in all four regimes and use the physically meaningful solution. 
Subbing Equations~\ref{sup:eqn:tillot1}, \ref{sup:eqn:tillot2} and \ref{sup:eqn:tillot3} into the energy equation in finite difference form (Equation~\ref{sup:eqn:E_conc_fd}) produces three separate quadratic equations.
These can then be solved analytically giving two solutions for each equation.
In the case that the material is in compression we choose the solution of the equation derived from Equation~\ref{sup:eqn:tillot1} that: 1) gives the correct sign of the change in $\epsilon$ for the density change; and 2) is closest to the previous value of $\epsilon$.
If the material is in an expanded state solutions from all three equations are potentially valid and so we choose the solution that: 1) is in the correct energy range for its regime; 2) gives the correct sign of the change in $\epsilon$ for the density change; and 3) is closest to the previous value of $\epsilon$.
The relevant equation can then be used to find the pressure. 
Note that care must be taken in averaging properties in the cross over region.

In order to calculate artificial viscosity we also need to calculate the adiabatic sound speed. 
The sound speed is defined as
\begin{equation}
c^2=\frac{\partial p }{\partial \rho}  \bigg \rvert_{S} \; ,
\end{equation}
where $S$ is specific entropy.
Using standard differential relations this can be expressed as
\begin{equation}
c^2=\frac{\partial p }{\partial \rho}  \bigg \rvert_{\epsilon} + \frac{\partial p }{\partial \epsilon}  \bigg \rvert_{\rho} \frac{\partial \epsilon }{\partial \rho}  \bigg \rvert_{S} \;.
\end{equation}
Substituting for the final term from the fundamental thermodynamic relation,
\begin{equation}
\label{sup:eqn:c2}
c^2=\frac{\partial p }{\partial \rho}  \bigg \rvert_{\epsilon} + \frac{p}{\rho^2} \frac{\partial p }{\partial \epsilon}  \bigg \rvert_{\rho} \;  .
\end{equation}
All the terms in this equation are known or can be found by differentiating the relevant equation for pressure in each regime.
Equation~\ref{sup:eqn:c2} is used to calculate the sound speed in each regime with the exception of the low density cutoff where a nominal sound speed of $0.5u_{\rm G}$ is used. 

Due to the limitations of the Tillotson EOS, WONDY can encounter unphysical solutions in expanded states. In such cases, the code is unable to continue and most of our calculations using Tillotson did not complete the full model run.
However, unphysical states are usually encountered late in the simulation when the loss has plateaued and so the total loss calculated is only slightly underestimated.

%xxxxxxxxxxxxxxxxxxxxxxxxxxxxxxxxxxxxxxxxxxxxxxxxxxxxxxxxxxxxxxxx
%xxxxxxxxxxxxxxxxxxxxxxxxxxxxxxxxxxxxxxxxxxxxxxxxxxxxxxxxxxxxxxxx
\subsection{Boundary Conditions}
\label{sup:sec:boundary}

The breakout of the shock at the base of the atmosphere/ocean is simulated by giving the lower boundary an initial velocity, $u_{\rm G}$, and then allowing the boundary to evolve ballistically. The lower boundary evolves following the momentum equation (Equation~\ref{sup:eqn:mom_conc_fd}) ignoring forces other than gravity, which in finite difference form is expressed as: 
\begin{equation} 
\begin{multlined}
\label{sup:eqn:a_ground} 
a^n_0 = - \frac{G M_p}{\left (x_0^n \right ) ^2} \; .
\end{multlined}
\end{equation}

In this framework, $u_{\rm G}$ is the particle velocity of the surface upon release of the shock in the planet to the base of the atmosphere or ocean. As for the case of the ocean/atmosphere interface, $u_{\rm G}$ can be determined using an impedance-match calculation. Due to the lower impedance of gases as opposed to water, release of the shock directly to an atmosphere results in higher ground velocities than in the presence of an ocean and in both cases the ground velocity is lower than if the shock released to vacuum. When calculating loss by convolving 1D results with the ground velocity calculated in 2D or 3D impact simulations it is imperative that corrections are made to translate the ground velocity.

Driving the lower boundary in this way creates some numerical complications. 
The sudden acceleration of the boundary generates excessively high artificial viscosities in the lowermost zones. 
This causes these zones to have artificially high temperatures and low densities which can be outside the range of the EOS. 
We overcome this by limiting the artificial viscosity in the first four zones to be less than $q_{max}$.
The exact value of $q_{max}$, within a reasonable range, has little effect on the amount of loss so we set it to $0.2$ of that calculated for the first time step for the first zone using Equation~\ref{sup:eqn:q_fd}.
We also initialise the first zone with $q^0_1=q_{max}$ and the second and third zones with $q^0_{j-\frac{1}{2}}=0.01 q^0_{j-\frac{3}{2}}$.
This helps to maintain reasonable values for variables in the lowermost zones with no affect on the final results.

In some simulations (see Section~\ref{sec:discussion:issues}) the velocity of the ground is forced to increase linearly over some finite rise time, $t_{\rm rise}$, until the specified ground velocity is reached. The specification of the artificial viscosity in cells just above the boundary is the same as in the standard case.

The lower boundary is assumed to stop when it returns to its initial position.
However it is necessary to slow the boundary gradually to avoid numerical errors.
The boundary is prescribed to fall exponentially after it returns to a position $x_0<R_t+0.15 \left (x_{0}^{max}-R_t \right )$ where $x_{0}^{max}$ is the maximum height reached by the boundary.
The rate is chosen so that the velocity of the boundary is continuous from before being slowed.
In finite difference form
\begin{equation}
u_0^{n+\frac{1}{2}}=u_0^{n-\frac{1}{2}} \exp \left \{ \frac{u_{slow} \Delta t ^{n-\frac{1}{2}}}{0.15 \left (x_{0}^{max}-R_t \right )} \right \} \;  ,
\end{equation}
where $u_{slow}$ is the velocity of the boundary at the point the boundary starts to be slowed.
Since the slow down happens late in the simulation the time step can be quite large by the time the slow down begins.
We therefore artificially reduce the time step when the boundary is $R_t+0.15 \left (x_{0}^{max}-R_t \right ) < x_0<R_t+0.3 \left (x_{0}^{max}-R_t \right )$ to $\leq 0.1 $ s so that the slow down can be properly calculated. 
After slow down begins the time step is allowed to evolve normally.

In some simulations (see Section~\ref{sec:discussion:issues}) the ground is stopped at some time, $t_{\rm stop}$. This is achieved by immediately setting the velocity of the boundary to zero at the specified time. Once the boundary has stopped, it begins to fall under gravity and its evolution continues as described above.

The boundary at the top of the atmosphere is treated as a stress free boundary i.e. a vacuum. 
This is implemented by having an additional zone at the top of the atmosphere with $( \rho, p ) = 0$ at all time steps.
The finite difference equations are then applied as normal.

%xxxxxxxxxxxxxxxxxxxxxxxxxxxxxxxxxxxxxxxxxxxxxxxxxxxxxxxxxxxxxxxx
%xxxxxxxxxxxxxxxxxxxxxxxxxxxxxxxxxxxxxxxxxxxxxxxxxxxxxxxxxxxxxxxx
\subsection{Atmosphere and Ocean Initialisation}
\label{sup:sec:init}

The initial conditions for our simulations are a hydrostatic atmosphere and, optionally, an ocean. 
We set the extent of the atmosphere by setting the pressure and temperature at the base of the atmosphere.
The atmosphere is then assumed to be polytropic,
\begin{equation}
\left ( \frac{p}{p_0} \right ) ^{\frac{\gamma -1 }{\gamma}} = \left ( \frac{\rho}{\rho_0} \right ) ^{\gamma -1 } = \frac{T}{T_0} \; ,
\end{equation}
where $p_0$, $\rho_0$ and $T_0$ are the pressure, density and temperature at the base of the atmosphere.
We then integrate up from the base of the atmosphere using a 4th order Runge-Kutta method to find the top of the atmosphere (in this model the pressure drops to zero at a finite height).
The atmosphere is then divided into the specified number of zones, $N_{\rm atm}$, each of equal thickness.
For the ideal gas EOS the structure of the atmosphere is determined by $p_0$, $T_0$, $\gamma$ and $m_{\rm a}$ (the molar mass of the gas) which sets the density at a given $p$, $T$.
The values for $\gamma$ and  $m_{\rm a}$ for each of the atmospheres used here are given in Table~\ref{sup:tab:atmo_params}.

\begin{table}
\centering
\begin{tabular}{l  c  c}
Atmosphere & $\gamma$ &  $m_{\rm a}$ [g]\\
\hline
H$_2$ & $1.4$ & $2$\\
H$_2$O & 1.25 & 18 \\
CO$_2$ & $1.29$ & $44$\\
N$_2$ & $1.4$ & $28$\\
Earth-like & 1.4 & 29 \\
\end{tabular}
\caption{The ideal gas parameters (ratio of specific heat capacities, $\gamma$, and molar mass, $m_{\rm a}$)for the atmospheres used in this paper.}
\label{sup:tab:atmo_params}
\end{table}

The extent of the ocean is set by specifying the depth of the ocean, $H_{\rm oc}$.
The temperature and pressure of the bottom of the atmosphere are assumed to be continuous with the ocean.
The exact structure of the ocean however depends on the EOS being used.
For the tabulated EOS an isothermal ocean is initialised with temperature $T_0$.
In the case of the IAPWS EOS an isentropic ocean is used with the entropy set by $p_0$ and $T_0$.
In order to compare directly with the work of \cite{Genda2005} when using the Tillotson EOS we initialise a constant specific internal energy ocean with $\epsilon = 120$ J kg$^{-1}$.
In each case a 4th order Runge-Kutta method is used to integrate down from the surface to find the properties for each of the $N_{\rm oc}$ ocean cells. Due to the limited compressibility of water, the ocean is close to isothermal in both the isentropic and isoenergetic initializations.
Cells are of equal thickness. 

The mass, radii, and escape velocity of the planets used in this study are given in Table~\ref{sup:tab:planets}. For Earth and Mars -mass planets the radii were taken as the present-day equatorial radii of those planets. For intermediate mass planets, the radii were calculated using the HERCULES planetary structure code \citep{Lock2017,Lock2019HERCULES} using the present-day Earth's core mass fraction of 0.323 \citep{Yoder1995}.  In HERCULES, a body is modeled as consisting of a series of nested concentric layers of constant density and a potential field method is used to calculate the equilibrium structure of the body with a given thermal state, composition, mass and angular momentum, using realistic equations of state. For this work, the mantle and core of each planet were assumed to be forsterite and pure iron, respectively, and were modeled using the the ANEOS equation of state (EOS) model \citep{thompson1972,Canup2012,Melosh2007}. The EOS are documented as `aneos-T70' (iron) and `aneos-gadget' (forsterite) respectively in two Zenodo repositories \citep{Stewart2020ironEOS,Stewart2019forsteriteEOS}. The mantle was assumed to be isentropic with a specific entropy of 3.2~kJ~K$^{-1}$~kg$^{-1}$ corresponding to a mantle potential temperature of around 1900~K. This thermal state approximates that of the Hadean mantle. The core was also assumed to be isentropic and have a thermal state similar to the present day. The core specific entropy was set at 1.5~kJ~K$^{-1}$~kg$^{-1}$, corresponding to a temperature of 3800~K at the pressure of the present-day core-mantle boundary. We used the same HERCULES parameters as in previous studies, that have been shown to accurately model the structure of Earth-like planets \citep{Lock2017,Lock2019pressure,Lock2019HERCULES,Lock2020}. The planets were modeled by $N_{\rm lay}^{\rm core}=20$ evenly spaced layers in the core and $N_{\rm lay}^{\rm mantle}=80$ layers in the mantle, with $N_{\mu}=1000$ points describing the shape of each layer. The expression for the gravity field was truncated at order $2k_{\rm max}=12$. The minimum pressure at the surface of the planet was set to 10~bar. The tolerance for the convergence of the shape of equipotential layers was $\xi_{\rm toll}^{\mu}=10^{-10}$ and the tolerance for the convergence of the mass of the planet was $\xi_{\rm toll}=10^{-8}$. The step used for calculating gradients in the solution algorithm was $\delta\xi=10^{-2}$. For further details of the definitions of these parameters the reader is referred to the HERCULES user manual \citep{Lock2019HERCULES}.

\begin{table}
\centering
\begin{tabular}{c c c}
Mass [$M_{\rm Earth}$]	 & Radius [km]	& $v_{\rm esc}$ [km~s$^{-1}$] \\
\hline
0.107 &	3396	 & 5.02 \\
0.3	& 4607	& 7.20 \\
0.5	& 5366	& 8.62 \\
0.7	& 5907	& 9.72 \\
0.9	& 6337	& 10.64 \\
1.0	& 6371	& 11.18 \\
\end{tabular}
\caption{Properties of planets used for hydrodynamic simulations in this study}
\label{sup:tab:planets}
\end{table}

%xxxxxxxxxxxxxxxxxxxxxxxxxxxxxxxxxxxxxxxxxxxxxxxxxxxxxxxxxxxxxxxxxxxxxxxxxxxxxxxxxxxxxxxxxxxxxxxxxxxxxxxxxxxxxxxxxxxxxxxxxxxxxxxxxxxxxxxx
%xxxxxxxxxxxxxxxxxxxxxxxxxxxxxxxxxxxxxxxxxxxxxxxxxxxxxxxxxxxxxxxxxxxxxxxxxxxxxxxxxxxxxxxxxxxxxxxxxxxxxxxxxxxxxxxxxxxxxxxxxxxxxxxxxxxxxxxx
%xxxxxxxxxxxxxxxxxxxxxxxxxxxxxxxxxxxxxxxxxxxxxxxxxxxxxxxxxxxxxxxxxxxxxxxxxxxxxxxxxxxxxxxxxxxxxxxxxxxxxxxxxxxxxxxxxxxxxxxxxxxxxxxxxxxxxxxx
\section{Sensitivity Tests for 1D Model}

\subsection{Sensitivity to Code Parameters}
\label{sup:sec:1Dsensitivity_tests}

\begin{figure}
\centering
\includegraphics[width=\columnwidth]{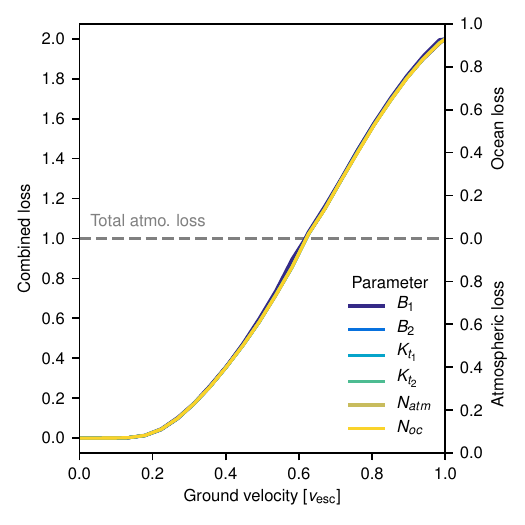}
\caption{The results of our 1D calculations are not sensitive to the intrinsic parameters in the code. Shown is the loss as a function of ground velocity for an example simulation ($H_{\rm oc} = 3 $~km, $p_0 = 100 $~bar and a CO$_2$ atmosphere at $300$~K) calculated varying run parameters $B_1$ (1-10), $B_2$ (0.1-5), $K_{\rm t_1}$ (1.05-1.2), $K_{\rm t_2}$ (0.6-0.9), $N_{\rm oc}$ (250-750), and $N_{\rm atm}$ (250-750). Each parameter was varied while holding all others constant at the values given in Table~\ref{sup:tab:code_constants}. }
\label{sup:fig:1Dsensitivity}
\end{figure}

We have tested the sensitivity of our 1D models to the intrinsic code parameters ($B_1$, $B_2$, $K_{\rm t_1}$ and $K_{\rm t_2}$) as well as the number of zones in the ocean and atmosphere ($N_{\rm atm}$ and $N_{\rm oc}$).
To do this we ran a series of calculations varying each of the parameters in turn and comparing the calculated loss. 
The final results showed little variation with any of the parameters over the ranges we explored (e.g., Figure~\ref{sup:fig:1Dsensitivity}), although some runs with lower $B_1$ values failed. For some combinations of parameters ringing was observed around the shock front indicating insufficient numerical viscosity, but such effects did not affect the final loss. We are therefore confident that our results are not significantly affected by our choice of these parameters.

\begin{table}
\centering
\begin{tabular}{l  c}
Parameter & Standard Value \\
\hline
$B_1$ &  $2$\\
$B_2$ & $0.1$\\
$K_{\rm t_1}$ & $0.75$ \\
$K_{\rm t_2}$ & $1.05$ \\
$N_{\rm oc}$ & $500$\\
$N_{\rm atm}$ & $500$\\
$t_{\rm final}$ & $5\times 10^4 $ s\\
\end{tabular}
\caption{Values for constants used in all 1D model runs for which results are presented here (details in text). $B_1$ and $B_2$ are used to determine the strength of the artificial viscosity. $K_{\rm t_1}$ and $K_{\rm t_2}$ control the size of each time step. $N_{\rm oc}$ and $N_{\rm atm}$ are the number of zones used for the ocean and atmosphere respectively. $t_{\rm final}$ is the total runtime for the simulations.}
\label{sup:tab:code_constants}
\end{table}

%xxxxxxxxxxxxxxxxxxxxxxxxxxxxxxxxxxxxxxxxxxxxxxxxxxxxxxxxxxxxxxxxxxxxxxxxxxxxxxxxxxxxxxxxxxxxxxxxxxxxxxxxxxxxxxxxxxxxxxxxxxxxxxxxxxxxxxxx
%xxxxxxxxxxxxxxxxxxxxxxxxxxxxxxxxxxxxxxxxxxxxxxxxxxxxxxxxxxxxxxxxxxxxxxxxxxxxxxxxxxxxxxxxxxxxxxxxxxxxxxxxxxxxxxxxxxxxxxxxxxxxxxxxxxxxxxxx
%xxxxxxxxxxxxxxxxxxxxxxxxxxxxxxxxxxxxxxxxxxxxxxxxxxxxxxxxxxxxxxxxxxxxxxxxxxxxxxxxxxxxxxxxxxxxxxxxxxxxxxxxxxxxxxxxxxxxxxxxxxxxxxxxxxxxxxxx
\section{Parameterization of loss as a function of ground velocity}
\label{sec:results:fit}

In this section, we present a parameterization of the loss due to a given ground motion from a body with a given escape velocity and atmosphere to ocean mass ratio. This parameterization is not intended to give physical insight, but rather to provide expressions for calculating global loss from 3D impact simulations \citep[e.g.,][]{Kegerreis2019,Denman2020,Denman2022_atmoloss_sE}. We will first describe the parameterization for the limiting case of no ocean and then for the general scenario including oceans and atmospheres of varying masses. Python functions and an interactive widget that implement this parameterization are available through GitHub. 

%xxxxxxxxxxxxxxxxxxxxxxxxxxxxxxxx
\subsection{The no-ocean case}

For the no-ocean case, we describe the loss using a modified logistics function,
\begin{equation}
 f_{\rm NO} \left (\frac{u_{\rm G}}{v_{\rm esc}} \right ) = 
 		\alpha_1 \left [1+\exp{\left \{\alpha_2 \left ( \frac{u_{\rm G}}{v_{\rm esc}} \right ) + \alpha_3 \right \}} \right ]^{\alpha_4} +\alpha_5 \; ,
\end{equation}
where $u_{\rm G}$ is the ground velocity, $v_{\rm esc}$ is the escape velocity, and $\alpha_i$ are constants. By requiring loss to be zero in the case of zero ground velocity, i.e., $f_{\rm NO}=0$ when $u_{\rm G}=0$, we find
\begin{equation}
    \alpha_5=-\alpha_1 [1+\exp{(\alpha_3)}]^{\alpha_4} \; .
\end{equation}
Similarly, by requiring loss to be complete when $u_{\rm G}=v_{\rm esc}$ we find
\begin{equation}
    \alpha_1=\frac{1}{[1+\exp{(\alpha_2 + \alpha_3}) ]^{\alpha_4} -[1+\exp{(\alpha_3)}]^{\alpha_4}} \; .
\end{equation}
In addition, we force $f_{\rm NO} =0$ when $u_{\rm G}/v_{\rm esc} <0$ and $f_{\rm NO} =1$ when $u_{\rm G}/v_{\rm esc} >1$. 

In the no-ocean case we set $\alpha_3=-1$ and performed a least squares fit to find $\alpha_2=-4.32$ and $\alpha_4=-3.77$. For the fit we used the calculations shown in Figure~\ref{fig:NO}A: 0.1, 0.5, 1, 5, 10, 50, 100, and 500~bar atmospheres on planets with mass 0.107 ($M_{\rm Mars}$), 0.3, 0.5, 0.7, 0.9 and 1~$M_{\rm Earth}$. The surface temperature was 300~K, and the atmosphere was \co2 with a mean molecular weight of $m_{\rm a}=44$ and a ratio of specific heat capacities of $\gamma=1.29$. Loss was calculated at 0.05~$v_{\rm esc}$ increments from 0.05 to 0.95~$v_{\rm esc}$. 

This parameterization offers a very accurate fit to our model runs. Figure~\ref{fig:NO}C shows the misfit between our parameterization and the fitted calculations. The root mean square (RMS) misfit for all runs is 0.0046 and the maximum misfit is 0.0089.

%xxxxxxxxxxxxxxxxxxxxxxxxxxxxxxxx
\subsection{Loss in the presence of an ocean}

We chose to parameterize loss in the $\mathcal{R}$-$u_{\rm G}$-loss space (Figure~\ref{fig:Rloss}). Loss is described by a modified logisitics function
\begin{equation}
 f (u_{\rm G}, v_{\rm esc}, \mathcal{R})= 
 		\alpha_1 [1+\exp{(\alpha_2 \log_{10}{\left(\mathcal{R}\right )} + \alpha_3}) ]^{\alpha_4} +\alpha_5  \; ,
 		\label{eqn:paramf}
\end{equation}
where $\alpha_i$ are functions of the ground velocity and the escape velocity, $v_{\rm esc}$. We find that loss due to a given ground velocity tends to the no-ocean case as $\mathcal{R}\rightarrow \infty$ (Figure~\ref{fig:Rloss}) and so we enforce $f\rightarrow f_{\rm NO}$ as $\mathcal{R}\rightarrow \infty$ by setting
\begin{equation}
    \alpha_1=f_{\rm NO} - \alpha_5 \; ,
\end{equation}
and requiring that $\alpha_2\le0$. 

The velocity dependence of $\alpha_i$ are given by arbitrary function chosen as they provided a good fit to our simulation results. $\alpha_2$ and $\alpha_3$ are described by 3$^{\rm rd}$-order polynomials:
\begin{equation}
 \alpha_{i} = a^i_1+a^i_2 \frac{u_{\rm G}}{v_{\rm esc}} +a^i_3 \left(\frac{u_{\rm G}}{v_{\rm esc}} \right )^2 +a^i_4 \left(\frac{u_{\rm G}}{v_{\rm esc}} \right )^3 \; ,
\end{equation}
and $\alpha_4$ is given by
\begin{align}
\begin{aligned}
 \alpha_4 = & a^4_1 \exp\left\{ \sin{\left (\frac{u_{\rm G}}{v_{\rm esc}} \pi \right)} \right. \\
 & \left. \times \left [ a^4_2 \frac{u_{\rm G}}{v_{\rm esc}} +a^4_3 \left(\frac{u_{\rm G}}{v_{\rm esc}} \right )^2 +a^4_4 \left(\frac{u_{\rm G}}{v_{\rm esc}} \right )^3 \right ]\right\} \; ,
 \end{aligned}
\end{align}
where $a^i_j$ are coefficients that can depend on $v_{\rm esc}$. The functional form for the ground velocity dependence of $\alpha_5$ is more complex. The value of $\alpha_5$ is the asymptote for loss as $\mathcal{R} \rightarrow 0$ and is close to the lowest $\mathcal{R}$ loss curve we calculate (e.g., the yellow line in Figure~\ref{fig:ug_loss}). To describe a loss curve in $u_{\rm G}$-loss space, we divide the functional form for $\alpha_5$ into two parts for the atmospheric loss ($g_1$) and ocean loss ($g_2$) regimes respectively: 
\begin{equation}
 \alpha_5 = \begin{dcases*}
        0 & \; $u_{\rm G} \leq 0$ \\
        g_1 & \; $0< u_{\rm G} <u_{\rm tal}$\\
        1 & \; $u_{\rm tal} \leq u_{\rm G} < v_{\rm esc}$ \;  \&  \;  $g_2\leq1$ \\
        g_2 & \;  $u_{\rm tal} \leq u_{\rm G} < v_{\rm esc}$ \;  \&  \; $g_2>1$  \\
        2 &  \; $u_{\rm G}\geq v_{\rm esc}$ \; ,
        \end{dcases*}
\end{equation}
where
\begin{equation}
    g_1= a^5_1 \left ( \frac{u_{\rm G}}{v_{\rm esc}} \right )^{a^5_3} \exp{ \left ( a^5_2 \frac{u_{\rm G}}{v_{\rm esc}} \right ) } \; ,
\end{equation}
\begin{align}
\begin{aligned}
    g_2= 2 & + a^5_4 \left [\frac{u_{\rm G}-u_{\rm tal}}{v_{\rm esc}-u_{\rm tal}} -1 \right ] \\
    &+ a^5_5 \left [\left (\frac{u_{\rm G}-u_{\rm tal}}{v_{\rm esc}-u_{\rm tal}} \right )^2-1 \right ] \; ,
\end{aligned}
\end{align}
and $u_{\rm tal}$ is the velocity at which total atmospheric loss is achieved (i.e., when $g_1=1$) given by
\begin{equation}
    u_{\rm tal}=v_{\rm esc} \left (\frac{a^5_3}{a^5_2} \right )  \min_{k=-1,0} W_k \left \{  \frac{a^5_3}{a^5_2}  \left ( a^5_1 \right )^{-1/a^5_3} \right \} \; ,
    \label{eqn:utal}
\end{equation}
where $W_k$ is the Lambert $W$ function of order $k$. The velocity regime between the atmospheric and ocean loss regimes where $\alpha_5=1$ loss emulates a feature seen in our low $\mathcal{R}$ simulations where significant ocean loss is delayed for a short range of $u_{\rm G}$ after total atmospheric loss is achieved (Figure~\ref{fig:ug_loss}). 

Dependence on planetary mass is captured by a linear dependence of the coefficients for $\alpha_2$ ($a^2_j$), $\alpha_4$ ($a^4_j$), and $\alpha_5$ ($a^5_j$) on $v_{\rm esc}$ such that
\begin{equation}
    a^i_j=a^i_{j,1} + a^i_{j,2} \frac{v_{\rm esc}}{v_{\rm esc}^{\rm Earth}} \; ,
    \label{eqn:param_vesc_dep}
\end{equation}
where $v_{\rm esc}^{\rm Earth}=11.2$~km~s$^{-1}$ is the escape velocity of Earth. Note that it is possible to parameterize the $v_{\rm esc}$ dependence on $\alpha_5$ by using a function for the absolute value of $u_{\rm G}$ convolved with the no-ocean loss function, but we found that, because of the complications of the varying pressure of release and the highly non-linear no-ocean loss function, a simple escape velocity scaling of the parameters was able to give a more accurate fit.  

To determine the parameters $a^i_{j,n}$ we performed a least squares fit on the results of the simulations shown in Figure~\ref{fig:Rloss}: all combinations of 1, 5, 10, 50, 100, 300 and 500~bar atmospheres and oceans of depths 0.1, 0.5, 1, 2, 3, 5, 10, 20 and 30~km, on planets with mass $M_{\rm Mars}$ (0.107~$M_{\rm Earth}$) and 0.3, 0.5, 0.7, 0.9 and 1~$M_{\rm Earth}$. Additional simulations with 900~bar atmospheres and oceans of 0.1~km depth were also included. The surface temperature was assumed to be 300~K, and the atmosphere was CO$_2$ ($m_{\rm a}=44$, $\gamma=1.29$). Loss was calculated at 0.05~$v_{\rm esc}$ increments from 0.05 to 0.95~$v_{\rm esc}$. The best-fit parameters are given in Table~\ref{tab:fits}.

\begin{table}
\centering
\setlength{\tabcolsep}{12pt}
\begin{tabular}{c c  c c}
 \multicolumn{1}{c}{$i$} &  \multicolumn{1}{c}{$j$} &  \multicolumn{1}{c}{$a_{j,1}^i$} &  \multicolumn{1}{c}{$a_{j,2}^i$} \vspace{0.1cm} \\
 \hline
$2$ & 1 & -16.2584384 &  14.2828078 \\
2 & 2 &  71.5029708 & -86.4465281  \\
2 & 3 & -117.862430 &  164.871407 \\
2 & 4 & 60.1395645 & -95.6142191 \\
3 & 1 &  -7.66878923 &   \multicolumn{1}{c}{-} \\
3 & 2 & 43.9524753 &  \multicolumn{1}{c}{-} \\
3 & 3 & -105.616622 &  \multicolumn{1}{c}{-} \\
3 & 4 & 77.4557563 &  \multicolumn{1}{c}{-} \\
4 & 1 & -0.767532594 & -0.263620386 \\
4 & 2 & -12.0443695 & -11.5024085 \\
4 & 3 & 44.7091911e &  39.8696345 \\
4 & 4 & -38.8346991 & -28.7008315 \\
5 & 1 & 10660.1668 & 13897.9517 \\
5 & 2 & -12.9490638 &  0.640100268 \\
5 & 3 & 7.35306571 & -0.384763832 \\
5 & 4 & 2.12776033 & -0.324256357 \\
5 & 5 &  -1.14723912 &  0.418100341 \\
\end{tabular}
\caption{Fitted coefficients for each of the parameters $a_{j,k}^i$ (see Equation~\ref{eqn:param_vesc_dep}). }
\label{tab:fits}
\end{table}

We find that our parameterization offers an accurate representation of the dependence of loss on $\mathcal{R}$, $u_{\rm G}$ and $M_{\rm p}$. Figure~\ref{fig:Rloss} shows the fit in $\mathcal{R}$-loss space for example ground velocities. The global RMS is 0.041, the RMS at any given ground velocity does not exceed 0.062, and the maximum misfit of any simulation result is 0.25. Unsurprisingly, the misfit is largest in the transition between the low and high $\mathcal{R}$ regimes where the gradient of the function in the $\mathcal{R}$-loss regime is greatest and the slight sensitivity to parameters such as absolute velocity, atmospheric pressure, ocean height etc. are greatest. 

\begin{figure}
    \centering
    \includegraphics[width=\columnwidth]{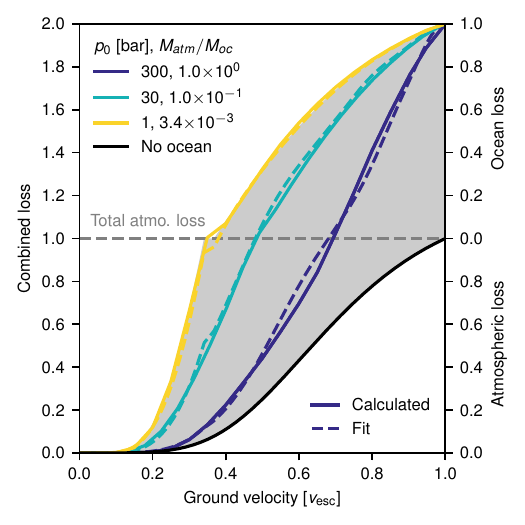}
    \caption{Our parameterization can accurately determine loss as a function of ground velocity. Shown are our simulation results for the same example surface conditions as in Figure~\ref{fig:ug_loss}A but for \co2 atmospheres (solid lines) and loss curves calculated using our parameterization (dashed lines). Grey band shows the full range of possible loss from Earth-mass bodies determined using our parameterization.}
    \label{fig:model-uG-loss}
\end{figure}

The parameterization also reproduces realistic loss curves in $u_{\rm G}$-loss space. Figure~\ref{fig:model-uG-loss} shows our simulation results for the same four example loss curves as in Figure~\ref{fig:ug_loss}A, but for \co2 atmospheres, (solid lines), with corresponding curves calculated using our parameterization (dashed lines). The grey band gives the full range of possible loss for an Earth-mass planet for any value of $\mathcal{R}$ using our parameterization.

%xxxxxxxxxxxxxxxxxxxxxxxxxxxxxxxx
\subsection{Connecting shock strength to ground velocity}

As discussed in Section~\ref{sec:results:up_ug_relation_NO}, the relationship between the strength of the shock in the planet and the ground velocity varies depending on the properties of the ocean and/or atmosphere in contact with the ground. It is also worth noting that the relationship could be affected somewhat by changing the shock properties of the ground. When calculating the loss due to a given shock, for example when combining 1D and 3D simulation (Section~\ref{sec:discussion:coupling}), it is therefore necessary to calculate the relationship between the properties of the shock in the planet (here we use the particle velocity, $u_{p}$, as the reference variable) and the ground velocity for the specific combination of ground and ocean and/or atmosphere that is under consideration. As this relationship can vary substantially, we do not attempt to provide a comprehensive set of $u_p$-$u_{\rm G}$ relations. Instead, we provide a python package and a widget that workers can use to calculate individual or sets of $u_p$-$u_{\rm G}$ relations (see data availability) for different surface conditions and a range of EOS.  

We do make one exception. The $u_p$-$u_{\rm G}$ relationship for breakout of a shock into the base of the ocean from a forsterite mantle is relatively constant over the range of conditions we have considered (oceans of depths 0.1-30~km, and atmospheric pressure up to 900~bar). In this case, the $u_p$-$u_{\rm G}$ relationship can be well described by
\begin{equation}
    u_{\rm G} =\beta_1 u_{p}^{\beta_2} \; ,
\end{equation}
where $\beta_1=2.6441$, and $\beta_2=0.9252$.

% \section*{Supplementary References}
% \bibliographystyle{elsarticle-harv} 
% \bibliography{References_processed}

\bibliography{References_processed}

\begin{thebibliography}{}
\expandafter\ifx\csname natexlab\endcsname\relax\def\natexlab#1{#1}\fi
\providecommand{\url}[1]{\href{#1}{#1}}
\providecommand{\dodoi}[1]{doi:~\href{http://doi.org/#1}{\nolinkurl{#1}}}
\providecommand{\doeprint}[1]{\href{http://ascl.net/#1}{\nolinkurl{http://ascl.net/#1}}}
\providecommand{\doarXiv}[1]{\href{https://arxiv.org/abs/#1}{\nolinkurl{https://arxiv.org/abs/#1}}}

\bibitem[{Abe \& Matsui(1988)}]{Abe1988}
Abe, Y., \& Matsui, T. 1988, Journal of the Atmospheric Sciences, 45, 3081,
  \dodoi{10.1175/1520-0469(1988)045<3081:EOAIGH$>$2.0.CO;2}

\bibitem[{Asphaug {et~al.}(2021)Asphaug, Emsenhuber, Cambioni, Gabriel, \&
  Schwartz}]{Asphaug2021}
Asphaug, E., Emsenhuber, A., Cambioni, S., Gabriel, T.~S., \& Schwartz, S.~R.
  2021, Planetary Science Journal, 2, 200, \dodoi{10.3847/PSJ/ac19b2}

\bibitem[{Bower {et~al.}(2019)Bower, Kitzmann, Wolf, Sanan, Dorn, \&
  Oza}]{Bower2019}
Bower, D.~J., Kitzmann, D., Wolf, A.~S., {et~al.} 2019, Astronomy \&
  Astrophysics, 631, A103, \dodoi{10.1051/0004-6361/201935710}

\bibitem[{Cameron \& Ward(1976)}]{Cameron1976}
Cameron, A. G.~W., \& Ward, W.~R. 1976, in Lunar and {Planetary} {Science}
  {Conference} {Abstracts}, Vol.~7, 120

\bibitem[{Canup(2004)}]{Canup2004anrev}
Canup, R.~M. 2004, Annual Review of Astronomy and Astrophysics, 42, 441,
  \dodoi{10.1146/annurev.astro.41.082201.113457}

\bibitem[{Canup(2012)}]{Canup2012}
---. 2012, Science, 338, 1052, \dodoi{10.1126/science.1226073}

\bibitem[{Canup \& Asphaug(2001)}]{Canup2001}
Canup, R.~M., \& Asphaug, E. 2001, Nature, 412, 708, \dodoi{10.1038/35089010}

\bibitem[{Caracas \& Stewart(2023)}]{Caracas2023}
Caracas, R., \& Stewart, S.~T. 2023, Earth and Planetary Science Letters, 608,
  118014, \dodoi{10.1016/j.epsl.2023.118014}

\bibitem[{Carter {et~al.}(2018)Carter, Leinhardt, Elliott, Stewart, \&
  Walter}]{Carter2018}
Carter, P.~J., Leinhardt, Z.~M., Elliott, T., Stewart, S.~T., \& Walter, M.~J.
  2018, Earth and Planetary Science Letters, 484, 276,
  \dodoi{10.1016/j.epsl.2017.12.012}

\bibitem[{Carter {et~al.}(2020)Carter, Lock, \& Stewart}]{Carter2020}
Carter, P.~J., Lock, S.~J., \& Stewart, S.~T. 2020, Journal of Geophysical
  Research: Planets, 125, \dodoi{10.1029/2019JE006042}

\bibitem[{Chen \& Ahrens(1997)}]{Chen1997}
Chen, G.~Q., \& Ahrens, T.~J. 1997, Physics of the Earth and Planetary
  Interiors, 100, 21, \dodoi{10.1016/S0031-9201(96)03228-1}

\bibitem[{Denman {et~al.}(2022)Denman, Leinhardt, \&
  Carter}]{Denman2022_atmoloss_sE}
Denman, T.~R., Leinhardt, Z.~M., \& Carter, P.~J. 2022, Monthly Notices of the
  Royal Astronomical Society, 513, 1680, \dodoi{10.1093/mnras/stac923}

\bibitem[{Denman {et~al.}(2020)Denman, Leinhardt, Carter, \&
  Mordasini}]{Denman2020}
Denman, T.~R., Leinhardt, Z.~M., Carter, P.~J., \& Mordasini, C. 2020, Monthly
  Notices of the Royal Astronomical Society, 496, 1166,
  \dodoi{10.1093/mnras/staa1623}

\bibitem[{Elkins-Tanton(2012)}]{Elkins-Tanton2012}
Elkins-Tanton, L.~T. 2012, Annual Review of Earth and Planetary Sciences, 40,
  113, \dodoi{10.1146/annurev-earth-042711-105503}

\bibitem[{Fegley {et~al.}(2023)Fegley, Lodders, \&
  Jacobson}]{Fegley2023_BSE_cond}
Fegley, B., Lodders, K., \& Jacobson, N.~S. 2023, Geochemistry, 125961,
  \dodoi{10.1016/j.chemer.2023.125961}

\bibitem[{Genda \& Abe(2003)}]{Genda2003}
Genda, H., \& Abe, Y. 2003, Icarus, 164, 149,
  \dodoi{10.1016/S0019-1035(03)00101-5}

\bibitem[{Genda \& Abe(2005)}]{Genda2005}
---. 2005, Nature, 433, 842, \dodoi{10.1038/nature03360}

\bibitem[{Grady {et~al.}(1987)Grady, Aylmer, Kurat, Ntaflos, Ott, Palme, \&
  Spettel}]{Grady1987}
Grady, M., Aylmer, D., Kurat, G., {et~al.} 1987, Memoirs of National Institute
  of Polar Research. Special Issue, 46, 162

\bibitem[{Grady {et~al.}(1986)Grady, Wright, Carr, \&
  Pillinger}]{Grady1986_enstatite}
Grady, M.~M., Wright, I.~P., Carr, L.~P., \& Pillinger, C.~T. 1986, Geochimica
  et Cosmochimica Acta, 50, 2799, \dodoi{10.1016/0016-7037(86)90228-0}

\bibitem[{Halliday(2013)}]{Halliday2013}
Halliday, A.~N. 2013, Geochimica et Cosmochimica Acta, 105, 146,
  \dodoi{10.1016/j.gca.2012.11.015}

\bibitem[{Hartmann \& Davis(1975)}]{Hartmann1975}
Hartmann, W.~K., \& Davis, D.~R. 1975, Icarus, 24, 504,
  \dodoi{10.1016/0019-1035(75)90070-6}

\bibitem[{Henderson(1989)}]{Henderson1989_refraction_shock}
Henderson, L.~F. 1989, Journal of Fluid Mechanics, 198, 365,
  \dodoi{10.1017/S0022112089000170}

\bibitem[{Hicks(1978)}]{Hicks1978}
Hicks, D.~L. 1978, Mathematics of Computation, 32, 1123,
  \dodoi{10.1090/S0025-5718-1978-0483944-0}

\bibitem[{Hubbard \& MacFarlane(1980)}]{Hubbard1980}
Hubbard, W.~B., \& MacFarlane, J.~J. 1980, Journal of Geophysical Research:
  Solid Earth, 85, 225, \dodoi{10.1029/JB085IB01P00225}

\bibitem[{Ikoma \& Genda(2006)}]{Ikoma2006}
Ikoma, M., \& Genda, H. 2006, The Astrophysical Journal, 648, 696,
  \dodoi{10.1086/505780}

\bibitem[{Jontof-Hutter(2019)}]{Jontof-Hutter2019_exo_comp}
Jontof-Hutter, D. 2019, Annual Review of Earth and Planetary Sciences, 47, 141,
  \dodoi{10.1146/annurev-earth-053018-060352}

\bibitem[{Kasting(1988)}]{Kasting1988_runaway_gh}
Kasting, J.~F. 1988, Icarus, 74, 472, \dodoi{10.1016/0019-1035(88)90116-9}

\bibitem[{Kegerreis {et~al.}(2020{\natexlab{a}})Kegerreis, Eke, Catling,
  Massey, Teodoro, \& Zahnle}]{Kegerreis2020}
Kegerreis, J.~A., Eke, V.~R., Catling, D.~C., {et~al.} 2020{\natexlab{a}}, The
  Astrophysical Journal, 901, L31, \dodoi{10.3847/1538-4357/ab9810}

\bibitem[{Kegerreis {et~al.}(2019)Kegerreis, Eke, Gonnet, Korycansky, Massey,
  Schaller, \& Teodoro}]{Kegerreis2019}
Kegerreis, J.~A., Eke, V.~R., Gonnet, P., {et~al.} 2019, Monthly Notices of the
  Royal Astronomical Society, 487, 5029, \dodoi{10.1093/mnras/stz1606}

\bibitem[{Kegerreis {et~al.}(2020{\natexlab{b}})Kegerreis, Eke, Massey, \&
  Teodoro}]{Kegerreis2020a}
Kegerreis, J.~A., Eke, V.~R., Massey, R.~J., \& Teodoro, L. F.~A.
  2020{\natexlab{b}}, The Astrophysical Journal, 897, 161

\bibitem[{Kegerreis {et~al.}(2018)Kegerreis, Teodoro, Eke, Massey, Catling,
  Fryer, Korycansky, Warren, \& Zahnle}]{Kegerreis2018}
Kegerreis, J.~A., Teodoro, L. F.~A., Eke, V.~R., {et~al.} 2018, The
  Astrophysical Journal, 861, 52, \dodoi{10.3847/1538-4357/aac725}

\bibitem[{Kerridge(1985)}]{Kerridge1985}
Kerridge, J.~F. 1985, Geochimica et Cosmochimica Acta, 49, 1707

\bibitem[{Kipp \& Lawrence(1982)}]{Kipp1982}
Kipp, M.~E., \& Lawrence, R.~J. 1982, NASA STI/Recon Technical Report N, 83,
  10396

\bibitem[{Lammer {et~al.}(2014)Lammer, Stokl, Erkaev, Dorfi, Odert, Gudel,
  Kulikov, Kislyakova, \& Leitzinger}]{Lammer2014}
Lammer, H., Stokl, A., Erkaev, N.~V., {et~al.} 2014, Monthly Notices of the
  Royal Astronomical Society, 439, 3225, \dodoi{10.1093/mnras/stu085}

\bibitem[{Lemmon {et~al.}(2000)Lemmon, Jacobsen, Penoncello, \&
  Friend}]{Lemmon2000}
Lemmon, E.~W., Jacobsen, R.~T., Penoncello, S.~G., \& Friend, D.~G. 2000,
  Journal of Physical and Chemical Reference Data, 29, 331,
  \dodoi{10.1063/1.1285884}

\bibitem[{Lock(2023{\natexlab{a}})}]{LockStewart2023_zenodo_v1_forced_general}
Lock, S. 2023{\natexlab{a}},
  sjl499/{Lock}\_Stewart\_2023\_PSJ\_atmospheric\_loss: {Supporting} resources
  for {Lock} \& {Stewart} (2023),  Zenodo, \dodoi{10.5281/zenodo.8368279}

\bibitem[{Lock(2023{\natexlab{b}})}]{simples_Zenodo_v1_general}
---. 2023{\natexlab{b}}, sjl499/simples: {Shock} {Impedance} {Match} {Package}
  and {Loss} {Event} {Simulator},  Zenodo, \dodoi{10.5281/zenodo.8367762}

\bibitem[{Lock(2019)}]{Lock2019HERCULES}
Lock, S.~J. 2019, Zenodo, \dodoi{10.5281/zenodo.3509365}

\bibitem[{Lock \& Stewart(2017)}]{Lock2017}
Lock, S.~J., \& Stewart, S.~T. 2017, Journal of Geophysical Research: Planets,
  122, 950, \dodoi{10.1002/2016JE005239}

\bibitem[{Lock \& Stewart(2019)}]{Lock2019pressure}
---. 2019, Science Advances, 5, eaav3746, \dodoi{10.1126/sciadv.aav3746}

\bibitem[{Lock {et~al.}(2018)Lock, Stewart, Petaev, Leinhardt, Mace, Jacobsen,
  \& Ćuk}]{Lock2018moon}
Lock, S.~J., Stewart, S.~T., Petaev, M.~I., {et~al.} 2018, Journal of
  Geophysical Research: Planets, 123, 910, \dodoi{10.1002/2017JE005333}

\bibitem[{Lock {et~al.}(2020)Lock, Stewart, \& Ćuk}]{Lock2020}
Lock, S.~J., Stewart, S.~T., \& Ćuk, M. 2020, Earth and Planetary Science
  Letters, 530, 115885, \dodoi{10.1016/j.epsl.2019.115885}

\bibitem[{Lodders(2003)}]{Lodders2003}
Lodders, K. 2003, The Astrophysical Journal, 591, 1220, \dodoi{10.1086/375492}

\bibitem[{Marty(2012)}]{Marty2012}
Marty, B. 2012, Earth and Planetary Science Letters, 313-314, 56,
  \dodoi{10.1016/j.epsl.2011.10.040}

\bibitem[{Marty {et~al.}(2016)Marty, Avice, Sano, Altwegg, Balsiger, Hässig,
  Morbidelli, Mousis, \& Rubin}]{Marty2016}
Marty, B., Avice, G., Sano, Y., {et~al.} 2016, Earth and Planetary Science
  Letters, 441, 91 , \dodoi{http://dx.doi.org/10.1016/j.epsl.2016.02.031}

\bibitem[{Melosh(1989)}]{Melosh1989book}
Melosh, H.~J. 1989, Impact cratering: {A} geologic process (Oxford University
  Press)

\bibitem[{Melosh(2003)}]{Melosh2003}
---. 2003, Journal of Applied Physics, 94, 4320, \dodoi{10.1063/1.1602566}

\bibitem[{Melosh(2007)}]{Melosh2007}
---. 2007, Meteoritics \& Planetary Science, 42, 2079,
  \dodoi{10.1111/j.1945-5100.2007.tb01009.x}

\bibitem[{Moore \& Gibson(1969)}]{Moore1969_N_chondrites}
Moore, C.~B., \& Gibson, E.~K. 1969, Science, 163, 174,
  \dodoi{10.1126/science.163.3863.174}

\bibitem[{Moore \& Lewis(1966)}]{Moore1966_E_C}
Moore, C.~B., \& Lewis, C.~F. 1966, Earth and Planetary Science Letters, 1,
  376, \dodoi{10.1016/0012-821X(66)90029-X}

\bibitem[{Mukhopadhyay \& Parai(2019)}]{Mukhopadhyay2019}
Mukhopadhyay, S., \& Parai, R. 2019, Annual Review of Earth and Planetary
  Sciences, 47, 389, \dodoi{10.1146/annurev-earth-053018-060238}

\bibitem[{Nakajima \& Stevenson(2015)}]{Nakajima2015}
Nakajima, M., \& Stevenson, D.~J. 2015, Earth and Planetary Science Letters,
  427, 286, \dodoi{10.1016/j.epsl.2015.06.023}

\bibitem[{Odert {et~al.}(2018)Odert, Lammer, Erkaev, Nikolaou, Lichtenegger,
  Johnstone, Kislyakova, Leitzinger, \& Tosi}]{Odert2018_Mars_sized}
Odert, P., Lammer, H., Erkaev, N.~V., {et~al.} 2018, Icarus, 307, 327,
  \dodoi{10.1016/j.icarus.2017.10.031}

\bibitem[{O'Keefe \& Ahrens(1982)}]{O'Keefe1982}
O'Keefe, J.~D., \& Ahrens, T.~J. 1982, Journal of Geophysical Research, 87,
  6668, \dodoi{10.1029/JB087iB08p06668}

\bibitem[{Olson \& Sharp(2019)}]{Olson2019}
Olson, P.~L., \& Sharp, Z.~D. 2019, Physics of the Earth and Planetary
  Interiors, 294, 106294, \dodoi{10.1016/j.pepi.2019.106294}

\bibitem[{O’Brien {et~al.}(2014)O’Brien, Walsh, Morbidelli, Raymond, \&
  Mandell}]{OBrien2014}
O’Brien, D.~P., Walsh, K.~J., Morbidelli, A., Raymond, S.~N., \& Mandell,
  A.~M. 2014, Icarus, 239, 74, \dodoi{10.1016/j.icarus.2014.05.009}

\bibitem[{Quintana {et~al.}(2016)Quintana, Barclay, Borucki, Rowe, \&
  Chambers}]{Quintana2016}
Quintana, E.~V., Barclay, T., Borucki, W.~J., Rowe, J.~F., \& Chambers, J.~E.
  2016, The Astrophysical Journal, 821, 126,
  \dodoi{10.3847/0004-637X/821/2/126}

\bibitem[{Raymond \& Izidoro(2017)}]{Raymond2017a}
Raymond, S.~N., \& Izidoro, A. 2017, Icarus, 297, 134,
  \dodoi{10.1016/j.icarus.2017.06.030}

\bibitem[{Raymond {et~al.}(2007)Raymond, Quinn, \& Lunine}]{Raymond2007}
Raymond, S.~N., Quinn, T., \& Lunine, J.~I. 2007, Astrobiology, 7, 66,
  \dodoi{10.1089/ast.2006.06-0126}

\bibitem[{Reufer {et~al.}(2012)Reufer, Meier, Benz, \& Wieler}]{Reufer2012}
Reufer, A., Meier, M.~M., Benz, W., \& Wieler, R. 2012, Icarus, 221, 296,
  \dodoi{10.1016/j.icarus.2012.07.021}

\bibitem[{Robert(2003)}]{Robert2003_DH}
Robert, F. 2003, Space Science Reviews, 106, 87,
  \dodoi{10.1023/A:1024629402715}

\bibitem[{Rufu {et~al.}(2017)Rufu, Aharonson, \& Perets}]{Rufu2017}
Rufu, R., Aharonson, O., \& Perets, H.~B. 2017, Nature Geoscience, 10, 89,
  \dodoi{10.1038/ngeo2866}

\bibitem[{Schaefer \& Fegley(2007)}]{Schaefer2007_ord_outgas}
Schaefer, L., \& Fegley, B. 2007, Icarus, 186, 462,
  \dodoi{10.1016/j.icarus.2006.09.002}

\bibitem[{Schaefer \& Fegley(2017)}]{Schaefer2017}
---. 2017, The Astrophysical Journal, 843, 120,
  \dodoi{10.3847/1538-4357/aa784f}

\bibitem[{Schaller {et~al.}(2018)Schaller, Gonnet, Draper, Chalk, Bower,
  Willis, \& Hausammann}]{Schaller2018}
Schaller, M., Gonnet, P., Draper, P.~W., {et~al.} 2018, Astrophysics Source
  Code Library, ascl:1805.020

\bibitem[{Schlichting \& Mukhopadhyay(2018)}]{Schlichting2018}
Schlichting, H.~E., \& Mukhopadhyay, S. 2018, Space Science Reviews, 214, 34,
  \dodoi{10.1007/s11214-018-0471-z}

\bibitem[{Schlichting {et~al.}(2015)Schlichting, Sari, \&
  Yalinewich}]{Schlichting2015}
Schlichting, H.~E., Sari, R., \& Yalinewich, A. 2015, Icarus, 247, 81,
  \dodoi{10.1016/j.icarus.2014.09.053}

\bibitem[{Senft \& Stewart(2008)}]{Senft2008}
Senft, L.~E., \& Stewart, S.~T. 2008, Meteoritics \& Planetary Science, 43,
  1993, \dodoi{10.1111/j.1945-5100.2008.tb00657.x}

\bibitem[{Sossi {et~al.}(2020)Sossi, Burnham, Badro, Lanzirotti, Newville, \&
  O'Neill}]{Sossi2020}
Sossi, P.~A., Burnham, A.~D., Badro, J., {et~al.} 2020, Science advances, 6,
  eabd1387, \dodoi{10.1126/sciadv.abd1387}

\bibitem[{Span {et~al.}(2000)Span, Lemmon, Jacobsen, Wagner, \&
  Yokozeki}]{Span2000}
Span, R., Lemmon, E.~W., Jacobsen, R.~T., Wagner, W., \& Yokozeki, A. 2000,
  Journal of Physical and Chemical Reference Data, 29, 1361,
  \dodoi{10.1063/1.1349047}

\bibitem[{Span \& Wagner(1996)}]{Span1996}
Span, R., \& Wagner, W. 1996, Journal of Physical and Chemical Reference Data,
  25, 1509, \dodoi{10.1063/1.555991}

\bibitem[{Stewart {et~al.}(2020)Stewart, Davies, Duncan, Lock, Root, Townsend,
  Kraus, Caracas, \& Jacobsen}]{Stewart2020_key_req_EOS}
Stewart, S., Davies, E., Duncan, M., {et~al.} 2020, AIP Conference Proceedings,
  2272, 080003, \dodoi{10.1063/12.0000946}

\bibitem[{Stewart(2020)}]{Stewart2020ironEOS}
Stewart, S.~T. 2020, Zenodo, \dodoi{10.5281/ZENODO.3866507}

\bibitem[{Stewart {et~al.}(2018)Stewart, Lock, \& Caracas}]{Stewart2018LPSC}
Stewart, S.~T., Lock, S.~J., \& Caracas, R. 2018, in Lunar and {Planetary}
  {Science} {Conference} {Abstracts}, Vol.~49, 1708

\bibitem[{Stewart {et~al.}(2019)Stewart, Davies, Duncan, Lock, Root, Townsend,
  Kraus, Caracas, \& Jacobsen}]{Stewart2019forsteriteEOS}
Stewart, S.~T., Davies, E.~J., Duncan, M.~S., {et~al.} 2019, Zenodo,
  \dodoi{10.5281/ZENODO.3478631}

\bibitem[{Thompson \& Lauson(1972)}]{thompson1972}
Thompson, S.~L., \& Lauson, H.~S. 1972, Improvements in the {CHART} {D}
  radiation-hydrodynamic code {III}: {Revised} analytic equations of state,
  Tech. rep., Sandia Laboratories

\bibitem[{Tillotson(1962)}]{Tillotson1962}
Tillotson, J.~H. 1962, Report No. GA-3216, General Atomic, San Diego, CA

\bibitem[{Tucker \& Mukhopadhyay(2014)}]{Tucker2014}
Tucker, J.~M., \& Mukhopadhyay, S. 2014, Earth and Planetary Science Letters,
  393, 254, \dodoi{10.1016/j.epsl.2014.02.050}

\bibitem[{Wagner(2002)}]{Wagner2002}
Wagner, W. 2002, Journal of Physical and Chemical Reference Data, 31, 387,
  \dodoi{10.1063/1.1461829}

\bibitem[{Wiik(1956)}]{Wiik1956_stony}
Wiik, H.~B. 1956, Geochimica et Cosmochimica Acta, 9, 279,
  \dodoi{10.1016/0016-7037(56)90028-X}

\bibitem[{Yalinewich \& Schlichting(2018)}]{Yalinewich2018}
Yalinewich, A., \& Schlichting, H.~E. 2018, Monthly Notices of the Royal
  Astronomical Society, 486, 2780, \dodoi{10.1093/mnras/stz1049}

\bibitem[{Yoder(1995)}]{Yoder1995}
Yoder, C.~F. 1995, in {AGU} {Reference} {Shelf} 1, ed. T.~J. Ahrens (American
  Geophysical Union), 1--31

\bibitem[{Young {et~al.}(2019)Young, Shahar, Nimmo, Schlichting, Schauble,
  Tang, \& Labidi}]{Young2019}
Young, E.~D., Shahar, A., Nimmo, F., {et~al.} 2019, Icarus, 323, 1,
  \dodoi{10.1016/j.icarus.2019.01.012}

\bibitem[{Zahnle {et~al.}(2010)Zahnle, Schaefer, \& Fegley}]{Zahnle2010}
Zahnle, K., Schaefer, L., \& Fegley, B. 2010, Cold Spring Harbor Perspectives
  in Biology, 2, a004895, \dodoi{10.1101/cshperspect.a004895}

\bibitem[{Zel'Dovich \& Raizer(2002)}]{ZelDovich&Raizer2002book}
Zel'Dovich, Y.~B., \& Raizer, Y.~P. 2002, Physics of shock waves and
  high-temperature hydrodynamic phenomena (Dover Publications Inc)

\bibitem[{Ćuk \& Stewart(2012)}]{Cuk2012}
Ćuk, M., \& Stewart, S.~T. 2012, Science, 338, 1047,
  \dodoi{10.1126/science.1225542}

\end{thebibliography}
\bibliographystyle{aasjournal} 

%\section*{Supplemental References}
%\bibliographystyle{../elsarticle-harv} 
%\bibliography{../References}
\end{document}